\documentclass[12pt,a4paper]{article}
\pdfoutput=1

\usepackage{jheppub}


\usepackage{graphicx}
\usepackage{amsmath}
\usepackage{amssymb}
\usepackage{marginnote}
 
\title{A chiral magnetic spiral in the holographic Sakai-Sugimoto model}

\author{Alfonso Ballon-Bayona, Kasper Peeters and Marija Zamaklar}

\affiliation{Department of Mathematical Sciences,\\
Durham University,\\
South Road,\\
Durham DH1 3LE,\\
United Kingdom.}

\emailAdd{c.a.m.ballonbayona@durham.ac.uk}
\emailAdd{kasper.peeters@durham.ac.uk} 
\emailAdd{marija.zamaklar@durham.ac.uk}

\date{April 11th, 2011}
\preprint{DCPT-12/33}

\abstract{We investigate the effect of a magnetic field on the vacuum
  of low-temperature QCD at large-$N_c$ in the presence of a chiral
  chemical potential, using the holographic Sakai-Sugimoto model.
  Above some critical chemical potential we find an instability, which
  triggers a decay of the homogeneous vacuum to a non-homogeneous
  configuration with a spiral form, which we construct explicitly. We
  find that this decay is suppressed for sufficiently large magnetic
  field, and determine the critical strength. We find that the new
  vacuum does not exhibit the chiral magnetic effect.  This is related
  to the way the chiral chemical potential is introduced. We discuss
  an alternative way of introducing the chiral chemical potential that
  leads to a nonzero chiral magnetic effect.}

\begin{document}
\maketitle

\section{Introduction and discussion}

Recent work by various authors has provided evidence that in QCD, a
magnetic field leads to several interesting phenomena, for values of
the magnetic field which are substantially smaller than anticipated
in earlier studies. The origin of these effects lies in the chiral
axial anomaly and the additional couplings it provides. For
instance, it has been shown \cite{Son:2007ny} that the axial anomaly
is responsible for the appearance of \emph{pion domain walls}, which
carry baryon charge. These are stable for sufficiently large baryon
chemical potentials. Importantly, they form for relatively small
critical values of the magnetic field, as this critical value scales
with the pion mass.

In the presence of an axial chemical potential
\cite{Kharzeev:2007jp,Fukushima:2008xe}, an external magnetic field
can trigger the appearance of a vectorial current in the direction
of this field,
\begin{equation}
\vec{j}_V \sim e \mu_A \vec{B}\,.
\end{equation}
The origin of this so called \emph{chiral magnetic effect} (CME) is
the asymmetry between left- and right-handed fermions (parametrised by
the axial potential $\mu_A$), which leads to separation of electric
charge. Related to this is the \emph{chiral separation effect}
\cite{Son:2004tq,Metlitski:2005pr,Son:2009tf}, where instead of an
axial potential, a baryonic quark potential is introduced. The
magnetic field now leads to a separation of chiral charge, and an
\emph{axial} current is generated,
\begin{equation}
\vec{j}_A \sim e \mu_B \vec{B}\,.
\end{equation}
Combining these two effects leads to evidence for the existence of a
so-called \emph{chiral magnetic wave} \cite{Kharzeev:2010gd}.  All
these effects suggest that the consequence of a magnetic field may be
much more important (and potentially easier to see in experiments)
than previously thought.  

These effects were originally derived using a quasi-particle picture
of chiral charge carriers, which is only valid at weak coupling. They
have meanwhile also been confirmed in a number of holographic
descriptions of the strong coupling regime. The pion gradient walls
were found in the Sakai-Sugimoto model in
\cite{Bergman:2008qv,Thompson:2008qw} (see
also~\cite{Bergman:2012na}), and there are even abelian analogues of
this solution \cite{Kim:2010pu} which involve an $\eta'$-meson gradient
(and of course no baryon number in this case). Whether or not
holographic models exhibit the CME or CSE is subject of more
controversy, as it requires a careful definition of the holographic
currents. An analysis of the D3/D7 system was given in
\cite{Hoyos:2011us,Gynther:2010ed} and we will comment extensively on
the Sakai-Sugimoto model later in this paper.

A major question is what happens when the chemical potentials are
large enough so that they trigger a condensation of bound states. Even
without a magnetic field and without an axial anomaly, various models
predict the formation of a non-isotropic (although homogeneous) vector
meson condensate, once the axial chemical potential becomes of the
order of the meson mass (see \cite{Aharony:2007uu} for an analysis in
the Sakai-Sugimoto model relevant here, and a list of references to
other works). When the axial anomaly is present, the condensate is
typically no longer even homogeneous, but forms a spiral structure
\cite{Bayona:2011ab}. A lot of work in this direction focuses on
high temperature models, but in fact such condensation already happens
at low and zero temperatures~\cite{Aharony:2007uu,Bayona:2011ab}.

\medskip

In the present paper, we therefore examine the Sakai-Sugimoto model in
the presence of a magnetic field and an abelian chiral chemical
potential, and set out to determine whether there is an instability
against decay to a non-homogeneous chiral spiral also in this case. We
will also analyse whether the ground state exhibits a chiral magnetic
effect, and whether there is an $\eta'$-gradient in this case. We will
only consider the confined chirally broken phase of this model,
i.e.~the low-temperature behaviour.

Our main results are as follows. First, we determine the location of
the phase boundary between the homogeneous solution of
\cite{Bergman:2008qv,Thompson:2008qw} and a new chiral spiral like
solution\footnote{The stability analysis of the homogeneous solution
  presented in \cite{Kim:2010pu} is incorrect in some crucial aspects;
  we will comment on the details and the consequences later.}. We
construct the non-homogeneous chiral spiral like solution
explicitly.\footnote{The chiral magnetic wave mentioned earlier is
  different from the chiral magnetic spiral, because the chiral
  magnetic wave involves the density and the component of the current
  parallel to the magnetic field, while the chiral magnetic spiral
  involves the components of the current transverse to the magnetic
  field.} We establish that the latter still does not exhibit the CME,
but point out that this conclusion relies crucially on subtleties
related to the precise way in which the chemical potential is
introduced. Finally, we show that a magnetic field tends to stabilise
the homogeneous solution, and there is a critical value beyond which
the chiral spiral ceases to exist. The parameter space in which we
need to scan for non-linear spiral solutions is rather large and the
numerical analysis is consequently sometimes tricky; we comment on
details of this procedure in the appendix.

An important issue in this analysis is the precise way in which the
chemical potential is included in the holographic setup. It was
recently pointed out \cite{Landsteiner:2011tf,Gynther:2010ed} that the
usual approach, which roughly amounts to identifying the chemical
potential with the asymptotic value of the gauge field, may be
incorrect in case the associated charge is not conserved (like, in our
case, the axial charge). However, we can define $\mu_A$ as the
integrated radial electric flux between the two boundaries of the
brane. Following the field theory arguments
of~\cite{Landsteiner:2011tf} we will argue that there are then two
natural formalisms for introducing the chemical potential.  In
formalism A we introduce $\mu_A$ through boundary conditions in the
component ${\cal A}_0$ of the gauge field. In formalism B the chemical
potential instead sits in ${\cal A}_z$. In the presence of the axial
anomaly these two formalisms are inequivalent.

In formalism A, we find that the non-homogeneous phase characterised
by wave number $k$, is dominant and there is a preferred value of this
wave number $k = k_{\text{g.s.}}$. When the magnetic field is small
compared to the chiral chemical potential $B \ll \mu_A$,
$k_{\text{g.s.}}$ depends only very weakly on the value of
$\mu_A$. This result is consistent with our previous
work~\cite{Bayona:2011ab}. For sufficiently large magnetic field we
find that the non-homogeneous ground state is suppressed.  In
formalism B, on the other hand, the homogeneous state always has lower
energy than the non-homogeneous one, and there is hence no chiral
magnetic spiral.

As far as the chiral magnetic effect is concerned, we find that in
formalism A the effect is absent. This result is consistent with
previous calculations on the Sakai-Sugimoto model
\cite{Rebhan:2009vc}.  In formalism B, on the other hand, there exists
a non-zero chiral magnetic effect. This seems to be more in line with
recent lattice calculations~\cite{Yamamoto:2012bi}, chiral model
calculations~\cite{Fukushima:2012fg} and holographic bottom-up
models~\cite{Gorsky:2010xu,Rubakov:2010qi} for the confined chirally
broken phase of QCD. However, a main shortcoming of formalism~B is
that the inhomogeneous phase, whose existence is indicated by
our perturbative analysis, is absent in the full nonlinear theory. Since
the perturbative analysis is blind to the subtle issues on how one
introduces the chemical potential in the holographic setup, we are
inclined to think that, as it stands, formalism B is
incorrect. Whether formalism A is correct on the other hand, or also
needs to be altered, is an open question at the moment, and needs a
separate study, which we leave for future work.

\section{Chemical potentials, currents and anomalies}

\subsection{Effective five-dimensional action}

In order to set the scene and to introduce our conventions, let us
start by giving a brief review of the basics of the Sakai-Sugimoto
model~\cite{Sakai:2004cn,Sakai:2005yt}. For a more detailed
description of the features of this model which are relevant here, we
refer the reader to the original papers or to~\cite{Bayona:2011ab}.

The Sakai-Sugimoto model at low temperature consists of $N_f$ flavour
D8-branes and $N_f$ anti-D8-branes which are placed in the
gravitational background sourced by a large number $N_c$ of D4-branes
which are compactified on a circle of radius~$R$. In the simplest set
up, the probes are asymptotically positioned at the antipodal points
of a circle, while in the interior of the geometry they merge together
into one U-shaped object. The gauge theory on the world-volume of the
probe brane is nine-dimensional, containing a four-dimensional sphere,
the holographic direction and four directions parallel to the
boundary. By focusing on the sector of the theory that does not
include excitations over the $S^4$, one can integrate the probe brane
action over this sub-manifold and end up with an effective five
dimensional DBI action on the probe brane world-volume. By expanding
this action to the leading order in the string tension, one ends up
with a five dimensional Yang-Mills theory with a Chern-Simons
term. For $N_f = 1$ we can write the action as
\begin{equation} 
\label{fullaction}
S= S_{\text{YM}} + S_{\text{CS}} =
- \frac{\kappa}{2} \int\! {\rm d}^4 x {\rm d}z \sqrt{-g} \, {\cal F}^{mn} {\cal
  F}_{mn} + \frac{\alpha}{4} \, \epsilon^{\ell m n p q} \int\! {\rm
  d}^4x {\rm d}z {\cal A}_{\ell} {\cal F}_{mn} {\cal F}_{pq} \, . 
\end{equation}
where the indices are now raised or lowered using the effective five
dimensional metric $g_{mn}$. This metric is defined as
\begin{equation} 
{\rm d}s_{(5)}^2 = g_{mn} {\rm d}x^m {\rm d}x^n =
M_{\text{KK}}^2 K_z^{2/3} \, \eta_{\mu \nu} {\rm d}x^\mu {\rm d}x^\nu \,+\,
K_z^{-2/3} {\rm d}z^2 \, , \label{gmn}
\end{equation}
where $K_z \equiv (1 + z^2)$. The $x^\mu$ directions are parallel to
the boundary and $z$ is the holographic direction orthogonal to the
boundary. The coupling constants $\kappa$ and $\alpha$ are given in
terms of the number of colours $N_c$, the compactification mass scale
$M_{\text{KK}}$ and the 't~Hooft coupling $\lambda$ by
\begin{equation} 
\kappa= \frac{\sqrt{\alpha '} g_s N_c^2 M_{\text{KK}}}{108 \pi^2} = \frac{\lambda N_c}{216 \pi^3} \,,  \qquad
\alpha = \frac{N_c}{24 \pi^2} \, .
\end{equation}
The Chern-Simons term written in~\eqref{fullaction} is valid only for
a single D8 probe, i.e.~for a $U(1)$ gauge theory on the brane
world-volume.

\subsection{Symmetries and chemical potentials}

Holographic models encode \emph{global} symmetries of the dual gauge
theory in the form of \emph{gauge} symmetries in the bulk theory, and
these relations hold both for the closed as well as the open sectors
of the string theory. The same is true for the Sakai-Sugimoto model at
hand, where there are two independent gauge theories living near the
two boundaries of the flavour D8-$\overline{{\rm D}8}$ brane
system. These $U(N_f)_L$ and $U(N_f)_R$ gauge symmetries correspond to
the global $U(N_f)\times U(N_f)_R$ flavour symmetries of the dual
gauge theory.  In the low-temperature phase of the Sakai-Sugimoto
model that we are interested in, the two branes are connected in the
interior of the bulk space, and thus the gauge fields ${\cal A}_{M}^L$
and ${\cal A}_{M}^R$ are limits of a single gauge field living on the
two connected branes. Therefore, one cannot independently perform
gauge transformations on these two gauge fields, but is constrained to
gauge transformations which, as $z\rightarrow \pm \infty$, act in a
related way.  Specifically, since near the boundary a large bulk gauge
transformation acts as ${\cal A}_{L/R} \rightarrow g_{L/R} {\cal
  A}_{L/R} g^{-1}_{L/R} $, then clearly any state (and in particular
the trivial vacuum ${\cal A}=0$) is invariant under the vectorial
transformations, i.e.~those transformations for which $g_{L} = g_{R}$.
This means that the vector-like symmetry is unbroken in this model. On
the other hand, the fact that the branes are joined into one U-shaped
object means that the axial symmetry is broken, which corresponds to
the \emph{spontaneous} breaking of the axial symmetry in the dual
gauge theory. The corresponding Goldstone bosons can be seen
explicitly in the spectrum of the fluctuations on the brane
world-volume.

The relationship between the bulk gauge field and the source and
global symmetry current of the dual gauge theory is encoded in the
asymptotic behaviour of the former.  More precisely, the bulk gauge
field ${\cal A}_M(z,x^\mu)$ behaves, near the boundary and in the
${\cal A}_z=0$ gauge, as
\begin{equation}
\label{asymptotic}
{\cal A}_\nu(x^\mu,z) \rightarrow 
  a_{\nu}(x^\mu)\Big(1 + \mathcal{O}(z^{-2/3})\Big) 
   + \rho_\nu(x^\mu) \frac{1}{z}\Big(1 + \mathcal{O}(z^{-2/3}) \Big)\,.
\end{equation}
Here $\rho_\nu$ parametrises the normalisable mode, while
$a_\nu(x^\mu)$ describes the non-norm\-alisable behaviour of the
field. The latter is interpreted as a source in the dual field theory
action, where it appears as
\begin{equation}
\label{e:boundaryaJ}
\int\!{\rm d}^4x\, a_\nu(x^\mu) J^\nu(x^\mu)\, .
\end{equation}
Hence the expectation value of the current corresponding to the global
symmetry in the gauge theory is given by
\begin{equation}
\label{currentsholography}
J^\mu(x)_{\pm} = \frac{\delta S}{\delta A_{\mu}(x,z \rightarrow \pm
  \infty)} \, .
\end{equation}
When the bulk action is just the ordinary Yang-Mills action (in curved
space), this expectation value is the same as the coefficient $\rho_\nu$
in the expansion \eqref{asymptotic}. However, in the presence of the
Chern-Simons term, the coefficient $\rho_\nu$ is different from the
current, as we will explicitly demonstrate for the system at hand in
section~\ref{s:anomalies}. This difference has been a source of some
of confusion in the literature. Its importance has recently been
emphasised in the context of the chiral magnetic effect in the
Sakai-Sugimoto model in~\cite{Rebhan:2010ax}.

From~\eqref{e:boundaryaJ} we also see that adding a chemical
potential to the field theory corresponds to adding a source for
$J^0$, which implies the boundary condition for the holographic gauge
field ${\cal A}_\nu(x)= \mu \delta_{\nu 0}$.

For the Sakai-Sugimoto model, the bulk field ${\cal A}_m$ living on
the D8-branes has \emph{two} asymptotic regions, corresponding to each
brane, and hence there are two independent chemical potentials $\mu_L$
and $\mu_R$ which can be separately turned on. Instead of left and
right chemical potentials one often introduces vectorial and axial
potentials, defined respectively as $\mu_B=\frac{1}{2}(\mu_R +\mu_L)$
and $\mu_A=\frac{1}{2}(\mu_R-\mu_L)$. For the $N_f=2$ case, the
vectorial and axial chemical potentials for the $U(1)$ subgroup of the
$U(2)$ gauge group on the two D8-branes correspond to the baryonic and
axial chemical potential in the dual gauge theory, while the
non-abelian $SU(2)$ chemical potentials are mapped to the vectorial
and axial isospin potentials.

In what follows we will also be interested in studying the system in
the presence of an \emph{external} (non-dynamical) magnetic source,
which will be introduced by turning on a nontrivial profile for the
non-normalisable component $a_\nu(x)$ of the bulk field.

\subsection{The Chern-Simons term, anomalies and the Bardeen counter term}
\label{s:anomalies}

The symmetries and currents of the model discussed in the previous
section are, however, valid only at leading order in $\lambda^{-1}$. At
the next order in $\lambda^{-1}$, the Yang-Mills action on the brane
world-volume receives many corrections, among which the Chern-Simons
term. This term turns out to be crucial for the existence of a
nontrivial ground state of the system in the presence of the external
magnetic field, a situation which we will study in the following
sections. On manifolds with boundaries, however, the Chern-Simons term
is manifestly gauge non-invariant and it also spoils conservation of
the vectorial and axial currents. Both of these reflect the fact that
the Chern-Simons term is the holographic manifestation of the vector
and axial gauge anomalies in the dual theory.

Apart from the Chern-Simons term there are of course also various
other corrections. One important class of terms is coming from the
expansion of the DBI action. In previous work~\cite{Bayona:2011ab} we
have seen that the qualitative picture of chiral spiral formation at
vanishing magnetic field does not change when these corrections are
taken into account. We will assume that this holds true here as
well. For the other higher derivative corrections, it is important to
note that we will find that our ground state has a momentum scale of
order $M_{\text{KK}}$, not $1/l_s$, so higher derivatives can typically
be ignored as long as $M_{\text{KK}} l_s$ is small.

Returning to the role of the Chern-Simons term in describing
anomalies, let us start by decomposing the gauge field in terms of the
axial and vectorial components, as
\begin{equation}
{\cal A}_m (x,z)= {\cal A}_m^V (x,z) + {\cal A}_m^A (x,z) \, . \label{VA}
\end{equation}
These two components transform under under an inversion of the
holographic coordinate $z \to - z$ as
\begin{equation}
{\cal A}_\mu^{V/A}(-z,x) = \pm {\cal A}_\mu^{V/A}(z,x) \,, \quad 
{\cal A}_z^{V/A}(-z,x) = \mp {\cal A}_z^{V/A}(z,x) \, . \label{mirror}
\end{equation}
Furthermore, the $\mu$ component of these fields are related to the
dual gauge fields as
\begin{equation}
{\cal A}^V_\mu ( z = \pm \infty , x ) = A^V_\mu (x) \quad , \quad {\cal A}^A_\mu (z = \pm \infty , x) = \mp  A^A_\mu (x) \, , \label{bdyfields}
\end{equation}
where the non-calligraphic $A^{V/A}_\mu(x)$ are the boundary vector and
axial-vector gauge fields. When written in terms of these vectorial and
axial potentials the action reads
\begin{multline}
\label{fullactionAV}
S =  \int\! {\rm d}^4 x {\rm d}z \Big \{ 
- \frac{\kappa}{2} \sqrt{-g} \left [ {\cal F}^{mn}_V {\cal F}_{mn}^V + {\cal F}^{mn}_A {\cal F}_{mn}^A \right ] \\[1ex]
+ \frac{\alpha}{4} \epsilon^{\ell m n p q} \left [ 2 {\cal A}_\ell^V {\cal F}_{mn}^A {\cal F}_{pq}^V 
+ {\cal A}_\ell^A ( {\cal F}_{mn}^V {\cal F}_{pq}^V + {\cal F}_{mn}^A {\cal F}_{pq}^A ) \right ] \Big \} \,.
\end{multline}
We now want to compute the currents following the prescription
\eqref{currentsholography}. The variation of the action can be written as
\begin{multline}
\delta S = \int\! {\rm d}^4 x {\rm d}z  
\Big \{ \left [ \frac{ \partial {\cal L}}{ \partial {\cal A}^V_\ell} - \partial_m {\cal P}^{m \ell}_V  \right ] \delta {\cal A}^V_\ell  
+  \left [ \frac{ \partial {\cal L}}{ \partial {\cal A}^A_\ell} - \partial_m {\cal P}^{m \ell}_A  \right ] \delta {\cal A}^A_\ell \\[1ex]
 + \partial_m \left[ {\cal P}^{m \ell}_V   \delta {\cal A}^V_\ell
 + {\cal P}^{m \ell}_A  \delta {\cal A}^A_\ell \right ] \Big \} \, , \label{varYMCS} 
\end{multline}
where the derivatives of the Lagrangian are given by
\begin{equation}
\begin{aligned}
\frac{ \partial {\cal L}}{ \partial {\cal A}^V_\ell} &= \frac{\alpha}{2} \epsilon^{\ell m n p q} {\cal F}_{mn}^A {\cal F}_{pq}^V \, , \\[1ex]
\frac{ \partial {\cal L}}{ \partial {\cal A}^A_\ell} &= \frac{\alpha}{4} \epsilon^{\ell m n p q} ( {\cal F}_{mn}^V {\cal F}_{pq}^V + {\cal F}_{mn}^A {\cal F}_{pq}^A )\, , \\[1ex]
{\cal P}^{m \ell}_V &\equiv \frac{ \partial {\cal L}}{ \partial (\partial_m {\cal A}^V_\ell) }  
= - 2 \kappa \sqrt{-g} {\cal F}^{m \ell}_V + \alpha \epsilon^{m \ell n p q} ({\cal A}_n^V {\cal F}_{pq}^A + {\cal A}_n^A {\cal F}_{pq}^V ) \, , \\[1ex]
{\cal P}^{m \ell}_A &\equiv\frac{ \partial {\cal L}}{ \partial (\partial_m {\cal A}^A_\ell) }  
= - 2 \kappa \sqrt{-g} {\cal F}^{m \ell}_A + \alpha \epsilon^{m \ell n p q} ({\cal A}_n^V {\cal F}_{pq}^V + {\cal A}_n^A {\cal F}_{pq}^A ) \,. \label{genmom}
\end{aligned}
\end{equation}
Note that ${\cal P}^{m \ell}_{V/A}$ are antisymmetric in $m
\leftrightarrow \ell$. Imposing that the bulk term in the variation
(the first line in \eqref{varYMCS}) vanishes gives the equations of
motion,
\begin{equation}
\begin{aligned}
 2 \kappa \partial_m ( \sqrt{-g} {\cal F}^{m \ell}_V ) + \frac32 \alpha \epsilon^{\ell m n p q} {\cal F}_{mn}^A {\cal F}_{pq}^V &= 0 \, , \\[1ex]
 2 \kappa \partial_m ( \sqrt{-g} {\cal F}^{m \ell}_A ) + \frac34
 \alpha \epsilon^{\ell m n p q} ({\cal F}_{mn}^V {\cal F}_{pq}^V +
 {\cal F}_{mn}^A {\cal F}_{pq}^A )  &= 0  \, . \label{eomYMCS} 
\end{aligned}
\end{equation}
We will need these shortly to show that the currents are not
conserved. The boundary term in the action variation can be written as
\begin{equation}
  \delta
  S_{\text{bdy}} = \int\! {\rm d}^4x {\rm d}z \, \Big \{ \partial_z
  \left [ {\cal P}^{z \mu}_V \delta {\cal A}^V_\mu + {\cal P}^{z \mu}_A
    \delta {\cal A}^A_\mu \right ] + \partial_\mu \left [ {\cal P}^{\mu
      \ell}_V \delta {\cal A}^V_\ell + {\cal P}^{\mu \ell}_A \delta
    {\cal A}^A_\ell \right ] \Big \} \,. \label{decompbdyterm} 
\end{equation}
This implies that the holographic vector and axial currents are given
by the expressions
\begin{equation}
\begin{aligned}
\label{currentss}
J^\mu_V(x) &= - 4 \kappa \lim_{z \to \infty} \left [ \sqrt{-g} {\cal F}^{z \mu}_V \right ] 
 - 2  \alpha \epsilon^{\mu \nu \rho \sigma} ( A_\nu^V F_{\rho \sigma}^A + A_\nu^A F_{\rho \sigma}^V ) \, ,  \\[1ex]
J^\mu_A(x) &= 4 \kappa \lim_{z \to \infty} \left [ \sqrt{-g} {\cal F}^{z \mu}_A \right ] 
 - 2 \alpha \epsilon^{\mu \nu \rho \sigma} ( A_\nu^V F_{\rho \sigma}^V + A_\nu^A F_{\rho \sigma}^A ) \, ,
\end{aligned}
\end{equation}
where we have used the boundary conditions (\ref{bdyfields}) and $F^{V/A}_{\mu \nu} = \partial_\mu A_\nu^{V/A} - \partial_\nu A_\mu^{V/A}$ are the boundary field strengths. 

Using the equations of motion \eqref{eomYMCS} we can explicitly show that these two
currents are not conserved due to the presence of the Chern-Simons
term. One finds
\begin{equation}
\begin{aligned}
\partial_\mu J^\mu_V (x) &=   2 \lim_{z \to \infty} \partial_\mu \left [ \sqrt{-g} {\cal P}^{z \mu}_V \right ]  
\equiv - 2  \lim_{z \to \infty} \left [ \sqrt{-g} \frac{ \partial {\cal L} }{\partial {\cal A}^V_z}  \right ] \\[1ex]
&=  - \alpha \, \epsilon^{\mu \nu \rho \sigma} F_{\mu \nu}^A F_{\rho \sigma}^V \,,\\[1ex]
\partial_\mu J^\mu_A (x) &=   - 2 \lim_{z \to \infty} \partial_\mu \left [ \sqrt{-g} {\cal P}^{z \mu}_V \right ]  \equiv  2  \lim_{z \to \infty} \left [ \sqrt{-g} \frac{ \partial {\cal L} }{\partial {\cal A}^A_z}  \right ] \\[1ex]
&=  \frac{\alpha}{2} \, \epsilon^{\mu \nu \rho \sigma} ( F_{\mu \nu}^V F_{\rho \sigma}^V + F_{\mu \nu}^A F_{\rho \sigma}^A )\,,
\end{aligned}
\end{equation}
We see two things here. Firstly, the anomaly (i.e.~the right-hand side
of above equations) is indeed sourced by the Chern-Simons
term. Secondly, the anomaly is present only if the \emph{boundary}
value of the bulk gauge \emph{field strength} is non-vanishing. In
other words, only in the presence of an \emph{external} field does the
anomaly show up.

While the axial anomaly is not problematic (it just reflects the fact
that this symmetry is no longer present in the dual quantum field
theory), this is not the case with the vectorial symmetry. In QED
coupled to chiral fermions one has to require that the vector current
is strictly conserved, since its non-conservation would imply a gauge
anomaly. It is possible to make the vectorial current conserved by
adding extra boundary terms to the action, as was first shown
by Bardeen in~\cite{Bardeen:1969md}. 
 
In the holographic setup, one expects that such a Bardeen-type counter
term will appear, but this type of term should come from the
requirement that the full theory in the bulk \emph{in the presence of
  the boundary} is gauge invariant under vectorial gauge
transformations. Let us therefore consider a generic \emph{vectorial} gauge
transformation in the bulk
\begin{equation} 
\delta {\cal A}_\ell^V = \partial_\ell \Lambda_V (x,z) \,, \quad \delta {\cal A}_\ell^A = 0
\, , 
\label{vecgauge} 
\end{equation}
where $\Lambda_V(x,z)$ is a function even in $z$. Under this
transformation, the action \eqref{fullactionAV} is invariant up to
a boundary term,
\begin{equation}
\begin{aligned}
\delta S^{(\Lambda_V)}_{\text{bdy}} &= \int\! {\rm d}^4x {\rm d}z   (\partial_\ell \Lambda_V) \partial_m {\cal P}^{m \ell}_V \\[1ex]
&= \frac{\alpha}{2} \epsilon^{\ell m n p q} \int\! {\rm d}^4x {\rm d}z (\partial_\ell \Lambda_V) {\cal F}_{mn}^A {\cal F}_{pq}^V \\[1ex]
&= \alpha \epsilon^{\ell m n p q} \int\! {\rm d}^4x {\rm
  d}z \partial_m \left [ (\partial_\ell \Lambda_V) {\cal A}^A_n {\cal
    F}_{pq}^V \right ] \,.
\end{aligned}
\end{equation}
Therefore, if we want to impose invariance of the action under
\eqref{vecgauge} we need to add the anomaly counter term correction
\begin{equation}
\begin{aligned}
S_{\text{an.}} &= - \alpha \epsilon^{\ell m n p q} \int\! {\rm d}^4x {\rm d}z \partial_m \left [ {\cal A}^V_\ell {\cal A}^A_n {\cal F}_{pq}^V \right ] \\[1ex]
&= \frac{\alpha}{2} \epsilon^{\ell m n p q} \int\! {\rm d}^4x {\rm d}z \left [ - {\cal A}^V_\ell {\cal F}_{mn}^A {\cal F}_{pq}^V + {\cal A}_\ell^A {\cal F}_{mn}^V {\cal F}_{pq}^V \right ] \,.
\end{aligned}
\end{equation}
We see that there are two contributions of this surface term: at the
``holographic'' boundaries $z\rightarrow \pm \infty$ and at the
boundary at spatial infinity $|\vec{x}| \rightarrow \infty$. Its
contribution at holographic infinity is indeed the Bardeen counter
term as derived in quantum field theory~\cite{Bardeen:1969md}, and
which was, in the holographic setup, initially postulated (added by
hand) in \cite{Rebhan:2010ax}. We see that its presence automatically
follows from the requirement of classical gauge invariance of the bulk
theory in the presence of the boundary.

The contribution at spatial infinity, would typically vanish, as all
physical states in the system are localised in the interior. However,
in the presence of external sources which generate an external
magnetic field that fills out the whole four-dimensional space, like
the one we will be considering, it is not a priori clear if this is
true, and one has to be careful about possible extra contributions to
the action and currents. It will however turn out that for our
non-homogeneous ansatz, these extra terms are irrelevant and that only
the Bardeen counter term is non-vanishing.

Let us continue by showing the effect of adding the $S_{\text{an.}}$
term on the currents. The total action now reads
\begin{multline}
\tilde S = S + S_{\text{an.}} = \int\! {\rm d}^4 x {\rm d}z \Big \{ 
- \frac{\kappa}{2} \sqrt{-g} \left [ {\cal F}^{mn}_V {\cal F}_{mn}^V + {\cal F}^{mn}_A {\cal F}_{mn}^A \right ] \\[1ex]
+ \frac{\alpha}{4} \epsilon^{\ell m n p q}  {\cal A}_\ell^A ( \, 3 \, {\cal F}_{mn}^V {\cal F}_{pq}^V + {\cal F}_{mn}^A {\cal F}_{pq}^A ) \Big \}  \equiv \int\! {\rm d}^4 x {\rm d}z {\cal \tilde L} \,. \label{newYMCS}
\end{multline}
The variation of this new action can again be written in the form
\eqref{varYMCS}, with $\tilde {\cal L}$ and $\tilde P$, instead of ${\cal L}$ and $P$, i.e.  
\begin{equation}
\begin{aligned}
\frac{\partial {\cal \tilde L}}{\partial {\cal A}_\ell^V} &= 0 \, , \cr 
\frac{\partial {\cal \tilde L}}{\partial {\cal A}_\ell^A} &= \frac{\alpha}{4} \epsilon^{\ell m n p q} ( \, 3 \, {\cal F}_{mn}^V {\cal F}_{pq}^V + {\cal F}_{mn}^A {\cal F}_{pq}^A ) \, , \cr
{\tilde P}_V^{m \ell} &\equiv \frac{ \partial {\cal \tilde L}}{ \partial (\partial_m {\cal A}^V_\ell) } 
= - 2 \kappa \sqrt{-g} {\cal F}^{m \ell}_V + 3 \alpha   \epsilon^{m \ell n p q} {\cal A}_n^A {\cal F}_{pq}^V   \, , \cr
{\tilde P}_A^{m \ell} &\equiv \frac{ \partial {\cal \tilde L}}{ \partial (\partial_m {\cal A}^A_\ell) } 
= - 2 \kappa \sqrt{-g} {\cal F}^{m \ell}_A +  \alpha \epsilon^{m \ell n p q} {\cal A}_n^A {\cal F}_{pq}^A \, .
\end{aligned}
\end{equation}
The equations of motion are of course unchanged (as the Bardeen
counter term is only a surface term), while the new currents are
obtained as
\begin{equation}
\tilde J^\mu_{V/A}(x) = \frac{ \delta \tilde S}{ \delta A^{V/A}_\mu (x)} = \pm \frac{ \delta \tilde S}{ \delta {\cal A}^{V/A}_\mu (x, z = \infty)} = \pm 2 \lim_{z \to  \infty} {\cal \tilde P}^{z \mu}_{V/A}   \,.
\end{equation}
Explicitly, the expressions read
\begin{equation}
\label{extracurrents}
\begin{aligned}
\tilde J^\mu_V(x) &= - 4 \kappa \lim_{z \to \infty} \left [ \sqrt{-g} {\cal F}^{z \mu}_V \right ] 
 - 6 \alpha \epsilon^{\mu \nu \rho \sigma} A_\nu ^A F_{\rho \sigma}^V  \, ,  \\[1ex]
\tilde J^\mu_A(x) &= 4 \kappa \lim_{z \to \infty} \left [ \sqrt{-g} {\cal F}^{z \mu}_A \right ] 
 - 2 \alpha \epsilon^{\mu \nu \rho \sigma}  A_\nu^A F_{\rho \sigma}^A
 \, .
\end{aligned}
\end{equation}
The divergences of these currents are 
\begin{equation}
\begin{aligned}
\partial_\mu \tilde J^\mu_V (x) &= 0 \, ,  \\[1ex]
\partial_\mu \tilde J^\mu_A (x) &=  \frac{\alpha}{2} \epsilon^{\mu \nu \rho \sigma} 
\left [ 3  F_{\mu \nu}^V  F_{\rho \sigma}^V +  F_{\mu \nu}^A F_{\rho \sigma}^A \right ] \,,
\end{aligned}
\end{equation}
where we used again the equations of motion \eqref{eomYMCS}. This
clearly shows that the vector current is now conserved, while the
anomaly is seen only in the axial sector. When the coefficient
$\alpha$ is taken from string theory, the non-conservation of the axial
current is exactly the same (including the numerical factor) as in QED
coupled to external fermions~\cite{Peskin:1995a}. In what follows we
will work with these renormalised currents and action.  We should also
emphasise that when one considers chemical potential for a
symmetry which is anomalous (as it is case here), just knowing the corrected
action \eqref{newYMCS} may not be enough. Also, one has to be careful
about the boundary conditions one has to impose on the states, as not
all boundary conditions are allowed. We discuss these subtle issues in
section~\ref{s:chempots_nonconserved}.

\subsection{The corrected Hamiltonian}

The Lagrangian density of the action after inclusion of the anomaly
counter term can be written as
\begin{multline}
\tilde {\cal L} = - \kappa \sqrt{-g} \left [ {\cal F}^{0a}_V {\cal F}_{0a}^V + \frac12 {\cal F}^{ab}_V {\cal F}_{ab}^V  
+ {\cal F}^{0a}_A {\cal F}_{0a}^A + \frac12 {\cal F}^{ab}_A {\cal F}_{ab}^A \right ] \\[1ex]
+ \frac{\alpha}{4} \epsilon^{0abcd} \Big [ {\cal A}_0^A ( 3 {\cal F}_{ab}^V {\cal F}_{cd}^V + {\cal F}_{ab}^A {\cal F}_{cd}^A ) 
+ 4 {\cal A}_b^A ( 3 {\cal F}_{0a}^V {\cal F}_{cd}^V + {\cal F}_{0a}^A {\cal F}_{cd}^A ) \Big ] \,.
\end{multline}
The conjugate momenta associated with the vector and axial gauge
fields thus take the form 
\begin{equation}
\begin{aligned}
\tilde \Pi^a_V &= \frac{ \partial \tilde L}{ \partial (\partial_0 {\cal A}_a^V )} 
= \tilde P^{0a}_V = - 2 \kappa \sqrt{-g} {\cal F}^{0a}_V + 3 \alpha \epsilon^{0abcd} {\cal A}_b^A {\cal F}_{cd}^V \,, \\[1ex]
\tilde \Pi^a_A &= \frac{ \partial \tilde L}{ \partial (\partial_0 {\cal A}_a^A )} 
= \tilde P^{0a}_A = - 2 \kappa \sqrt{-g} {\cal F}^{0a}_A +  \alpha \epsilon^{0abcd} {\cal A}_b^A {\cal F}_{cd}^A \,.
\end{aligned}
\end{equation}
We then obtain the on-shell Hamiltonian as $\tilde H = \tilde
H_{\text{Bulk}} + \tilde H_{\text{Bdy}}$, where the two contributions read
\begin{equation}
\begin{aligned}
\tilde H_{\text{Bulk}} &= \kappa \int {\rm d}^3x {\rm d}z    \sqrt{-g} \left [ - {\cal F}^{0a}_V {\cal F}_{0a}^V + \frac12 {\cal F}^{ab}_V {\cal F}_{ab}^V  
- {\cal F}^{0a}_A {\cal F}_{0a}^A + \frac12 {\cal F}^{ab}_A {\cal F}_{ab}^A \right ]  \cr 
&= \kappa \int\! {\rm d}^3x {\rm d}z    \sqrt{-g} \left [ - {\cal
    F}^{0a} {\cal F}_{0a} + \frac12 {\cal F}^{ab} {\cal F}_{ab}
\right ] \, , \label{tHBulk}
\end{aligned}
\end{equation}
\begin{equation}
\tilde H_{\text{Bdy}} =  \int {\rm d}^3 x {\rm d}z \, \partial_a \left [ \tilde \Pi^a_V {\cal A}_0^V + \tilde \Pi^a_A {\cal A}_0^A \right ] \label{tHBdy} \, .
\end{equation}
Here we have used the gauge field equations (\ref{eomYMCS}) for the
time component $\ell = 0$ (generalised Gauss law).

\section{The spatially modulated phase}

Having settled the issue of how to deal with the Chern-Simons term in
the presence of external fields for the Sakai-Sugimoto model, we now
want to find the ground state of the system, at strong coupling, in
the presence of external magnetic field and non-vanishing axial and
baryon chemical potentials. Based on the weak coupling, partonic, arguments
which were mentioned in the introduction, we expect that the ground
state should be a chiral magnetic spiral like configuration. In
particular, we expect it to be \emph{non-homogeneous}.

So far, for Sakai-Sugimoto model a non-homogeneous ground state was
constructed in the presence of large enough axial chemical potential
but with no external fields in \cite{Bayona:2011ab} (at low
temperature; see \cite{Ooguri:2010xs} for a high-temperature analysis).
The main reason why such a state appeared was due to the nontrivial
Chern-Simons term. We now want to see if this state persists and how
it is modified once the external magnetic field is introduced.

\subsection{Magnetic ground state ansatz in the presence of chemical potentials}

We are interested in studying the ground state of the system at non-zero
axial chemical potential $\mu_A$ and non-zero baryon chemical potential
$\mu_B$.  In addition, we will turn on a constant magnetic field in
the $x^1$ direction $\vec{B}=B \hat x_1 $.  The boundary conditions
associated with this physical scenario are
\begin{equation}
\label{boundarystuff5d}
\vec{{\cal A}}( z = \pm \infty ) \,=\, \frac12 \vec{B} \times \vec{x}  , 
\quad \quad {\cal A}_0( z = \pm \infty ) = \mp  \mu_A + \mu_B \,.
\end{equation}
We will see that in fact our ansatz is insensitive to the baryon
chemical potential, but we will keep it in the formulas for a little
while longer.

Let us now consider our particular ansatz, in the gauge ${\cal A}_z
=0$. First, we want to introduce the vectorial and axial chemical
potentials, hence we turn on ${\cal A}_0= f(z)$ with the above
boundary conditions. Second, we need to introduce the (constant)
magnetic field in the boundary, hence we turn on a component of ${\cal
  A}$ transverse to the direction of $\vec{B}$: $\vec{\cal
  A}_B^T(x_2,x_3,z)$. In principle this function could depend on $z$,
but the Bianchi identity tells us that in fact the magnetic field is
\emph{independent} of~$z$.  Therefore, we have for $\vec{\cal A}_B^T$ 
\begin{equation}
\vec{{\cal A}}_B^T ( x_2,x_3 ) \,=\, \frac12 \vec{B} \times \vec{x} =
\frac{B}{2} \left [ - x_3 \hat x_2 + x_2 \hat x_3 \right
]\,. \label{newansatzABT} 
\end{equation}
Next by looking at the equations of motion for the $\vec{\cal A}_B^L $
component parallel to the magnetic field, we see that this component
also has to be turned on, and is only a function of $z$, 
\begin{equation} 
\vec{\cal
  A}_B^L= a(z) \hat x_1 \, , \label{newansatzABL} 
\end{equation}
Finally, we expect that a chiral magnetic spiral will appear in the
direction transverse to the external magnetic field 
\begin{equation} 
\vec{\cal A}_W^T (x_1,z) = h (z) \left [ \cos(k x_1) \hat x_2 -\sin(k x_1)
  \hat x_3 \right ] \, , \label{newansatzAWT} 
\end{equation}
i.e.~it represents the chiral wave transverse to the boundary magnetic
field satisfying 
\begin{equation}
\nabla \times \vec{\cal A}_W^T \,=\,k \,
\vec{\cal A}_W^T \, , 
\end{equation} 
where $k=\pm |\vec{k}|$ and $\vec{k}$ is the spatial momentum. So in
summary our ansatz is given by
\begin{equation}
\begin{aligned}
{\cal A}_0^V &= f_V(z) \quad , \quad  {\cal A}_0^A = f_A (z) \, , \\[1ex]
\vec{\cal A}_V &= \frac{B}{2} \left [ - x_3 \hat x_2 + x_2 \hat x_3 \right ] + h_V(z) \left [ \cos (k x_1) \hat x_2 - \sin (k x_1) \hat x_3 \right ] + a_V(z) \hat x_1 \, ,\\[1ex]
\vec{ \cal A}_A &= a_A (z) \hat x_1 +  h_A(z) \left [ \cos (k x_1) \hat x_2 - \sin (k x_1) \hat x_3 \right ] \, , \label{newansatz}
\end{aligned}
\end{equation}
where the fields satisfy the boundary conditions
\begin{equation}
\begin{aligned}
f_V (z \to \pm \infty) &= \mu_B\,, \quad& f_A (z \pm \infty) &= \mp \mu_A \, , \\[1ex]
h_V( z \to \pm \infty) &= 0 \quad , \quad& h_A( z \to \pm \infty) &= 0 \, , \\[1ex]
a_V (z \to \pm \infty) &= 0 \quad , \quad& a_A( z \to \pm \infty ) &= \mp j \,  \label{boundarystuff} \, .
\end{aligned}
\end{equation}

The Gauss law, i.e.~the zeroth component of the equation of motion
~\eqref{eomYMCS}, is automatically satisfied for our ansatz. The
remaining equations reduce, after integrating one of them with
integration constant $\tilde\rho$, to
\begin{align}
&  \sqrt{- g} \, g^{zz} g^{00} \partial_z f
=  3 \frac{\alpha}{\kappa} \left [  B a + \frac{k}{2}  h^2   \right ] - \tilde \rho \, , \label{feq} \\[1ex]
& \partial_z \left [ \sqrt{- g} \, g^{zz} g^{xx} \partial_z a \right ] 
+ 3 \frac{\alpha}{\kappa}  B \partial_z f   = 0  \,, \label{aeq}\\[1ex]
&  \partial_z \left [\sqrt{- g} \, g^{zz} g^{xx} (\partial_z h)  \right ]
- \sqrt{-g} \,  (g^{xx})^2  k^2  h 
+ 3 \frac{\alpha}{\kappa}    k  \partial_z f h
 = 0 \label{heq} \,.
\end{align}
Restricting now to the metric~\eqref{gmn}, and substituting
eq.~\eqref{feq} into ~\eqref{aeq} and \eqref{heq} we obtain our master
equations,
\begin{align}
& K_z \partial_z \hat f = - \hat b 
- \frac{1}{2}  \hat k \hat h^2  \, , \label{hatfequation} \\[1ex]
& K_z \partial_z \left [ K_z \partial_z \hat b \right ] 
-  \hat B^2 \left [ \hat b  +  \frac{1}{2} \hat k \hat h^2  \right ] 
= 0 \, , \label{hataequation} \\[1ex]
& K_z \partial_z \left [ K_z \partial_z \hat h \right ] 
  -  K_z^{2/3} \hat k^2 \hat h -   \hat k \hat h \left [ 
\hat b + \frac{1}{2} \hat k \hat h^2 \right ] = 0 \, , 
\label{hathequation} 
\end{align}
where 
\begin{equation}
\label{e:bversusa}
\hat b \equiv  \hat B \hat a - \hat \rho \, , 
\end{equation}
and we have also introduced a set of dimensionless variables
 $\hat f$, $\hat h$, $\hat a$, $\hat k$, $\hat \rho$ and 
$\hat B$ defined by
\begin{align}
 &  f =  \bar \lambda \, M_{\text{KK}} \hat f \quad ,  \quad 
h = \bar \lambda \, M_{\text{KK}} \, \hat h \quad , \quad 
 a =  \bar \lambda \, M_{\text{KK}} \hat a \, ,\\[1ex]
& k= M_{\text{KK}} \hat k \quad , \quad 
\tilde \rho = \bar \lambda \, M^3_{\text{KK}} \, \hat \rho \quad , \quad 
B = \bar \lambda M^2_{\text{KK}}  \hat B \,,  \label{hatdimless} 
\end{align}
with $\bar\lambda = \lambda/(27\pi)$. These coupled equations are in
general not solvable analytically, except in the special case $k=0$,
which we will review next.

\subsection{Review of the homogeneous solution}

Before embarking on the full task of finding the non-homogeneous
solutions to the equations of motion, we will in this section first
review the homogeneous solution (i.e.~the solution for which $k=0$) in
the presence of a constant magnetic field. This is an abelian version
of the solution first constructed
in~\cite{Bergman:2008qv,Thompson:2008qw}, see
also~\cite{Rebhan:2009vc}. In the homogeneous case there is no
transverse spiral, i.e.~$h=0$ and the equations of motion simplify to
\begin{equation}
\begin{aligned}
\partial_{\tilde{z}} \hat{f} =& \hat\rho - \hat B \hat{a} \, , \\[1ex]
\left ( \partial_{\tilde z}^2 - \hat B^2 \right ) \hat{a} &= -
\hat B \hat{\rho}\,,
\end{aligned}
\end{equation}
with $\tilde z = \arctan z$. These equations can be integrated exactly,
and the solution takes the form 
\begin{equation}
\begin{aligned}
\label{homogenious}
\hat{a}(z)&= \frac{\hat{C}_A}{\hat B} \cosh (\hat B \arctan z) +
\frac{\hat{C}_B}{\hat B} \sinh (\hat B \arctan z) + \frac{\hat
  \rho}{\hat B} \, ,\\
\hat{f}(z) &= - \frac{\hat{C}_A}{\hat
  B} \sinh ( \hat B \arctan z) - \frac{\hat{C}_B}{\hat B} \cosh (\hat B \arctan z) + \hat{f}_0 \,,
\end{aligned}
\end{equation}
where $\hat{C}_A,\hat{C}_B,\tilde{\rho}$ and $\hat{f}_0$ are four integration constants
for the two second order differential equations. The corresponding
field strengths take the form 
\begin{equation}
\begin{aligned}
\label{fieldstr}
{\cal F}_{z1} &=  \frac{1}{1 + z^2} 
\left [ C_A \sinh ( \hat B \arctan z) + C_B \cosh (\hat B \arctan z)  \right ]\,,  \\
{\cal F}_{z0} &= - \frac{1}{1 + z^2} 
\left [ C_A \cosh ( \hat B \arctan z) + C_B \sinh (\hat B \arctan z)  \right ] \,,
\end{aligned}
\end{equation}
and $C_{A/B} = \bar \lambda M_{KK} \hat C_{A/B}$. In terms of the baryonic and
axial chemical potentials the boundary condition on $f(z)$ reads
\begin{equation}
f (z \to \pm \infty) = \mu_{L/R} = \mu_B \mp \mu_A \, .
\end{equation}
From the expression for $f(z)$ these potentials are related to $\hat{C}_A$ and $\hat{C}_B$ by 
\begin{equation}
\hat{C}_A =  \frac{\hat B }{\sinh \left ( \frac{\pi}{2} \hat B \right )} \hat{\mu}_A \, , \quad \quad 
\hat{C}_B = - \frac{\hat B } { \cosh \left ( \frac{\pi}{2} \hat B \right )} ( \hat{\mu}_B - \hat{f}_0)  \,,
\end{equation}
where $\mu_{A/B} = \bar \lambda M_{KK} \hat
\mu_{A/B}$. In analogy with the analysis
of \cite{Son:2007ny} we are here interested in configurations in which
there is a non-vanishing pion (or rather, $\eta'$) gradient in the
direction of the external field. Since we are working in the ${\cal
  A}_z=0$ gauge, a pion field will appear as part of the axial,
non-normalisable component of all ${\cal A}_\mu$'s (see
\cite{Sakai:2004cn,Sakai:2005yt}). Hence, we impose the boundary
condition\footnote{We stick to this notation, introduced
  in~\cite{Rebhan:2009vc}, but want to emphasise that even though
  using the symbol $j$ suggests that the asymptotic value of $a$ is a
  current, it is not.}
\begin{equation}
a (z \to \pm \infty) = \mp j \, .
\end{equation} 
We should note here that this is not the most general boundary
condition for the field $a(z)$, since we have set the even part to
zero. Having thus reduced the parameter space to the set $\{\mu_A,\mu_B,
j\}$, we have the relations
\begin{equation}
\begin{aligned}
\hat{C}_A &= \frac{\hat
  B }{\sinh \left ( \frac{\pi}{2} \hat B \right )} \hat{\mu}_A\,,
\quad&   
\hat \rho &= - \hat B \coth
\left ( \frac{\pi}{2} \hat B \right ) \hat{\mu}_A \,, \\[1ex]
\hat{C}_B &= - \frac{\hat B} { \sinh \left ( \frac{\pi}{2} \hat B
  \right )} \hat{j}\,,\quad & \hat{f}_0 &= \hat{\mu}_B
 - \coth \left ( \frac{\pi}{2} \hat B \right ) \hat{j} \,, 
\end{aligned}
\end{equation}
where $j = \bar \lambda M_{KK} \hat j$.
Using these relations we can rewrite $\hat f(z)$ and $\hat a(z)$ as
\begin{equation}
\begin{aligned}
\label{homosolu}
\hat{f}(z) &=\hat\mu_B - \hat\mu_A \frac{ \sinh (\hat B \arctan z)}{\sinh \left
  ( \frac{\pi}{2} \hat B \right )} + \hat j \left [ \frac{ \cosh ( \hat
    B \arctan z) }{ \sinh \left ( \frac{\pi}{2} \hat B \right )} -
  \coth \left ( \frac{\pi}{2} \hat B \right ) \right ] \, , \\[1ex]
\hat{a}(z) &= \hat\mu_A \left [ \frac{ \cosh ( \hat B \arctan z) }{ \sinh \left
    ( \frac{\pi}{2} \hat B \right )} - \coth \left ( \frac{\pi}{2}
  \hat B \right ) \right ] - \hat j \frac{ \sinh (\hat B \arctan z)}{
  \sinh \left ( \frac{\pi}{2} \hat B \right ) } \, . 
 \end{aligned}
\end{equation}
Since the constant $\mu_B$ does not appear in the Hamiltonian, it is
effectively a free parameter, which we are free to set to zero. As
expected, we find that the baryon chemical potential has no effect on
this abelian system. In contrast, minimising the Hamiltonian
will impose a constraint on the axial chemical potential and $j$, as
expected for physical systems.

In summary, we have two physical boundary values, $\mu_A$ and $j$,
which are implicitly expressed in terms of the two parameters $C_A$
and $C_B$. At any given fixed values of $\mu_A$ and $j$, we will want
to compare the homogeneous solution given above to possible
non-homogeneous condensates and determine which of the two has lower
energy.

\subsection{Currents for the homogeneous and non-homogeneous ansatz}

The work of \cite{Rebhan:2010ax,Rebhan:2009vc}, which studied the
homogeneous solution discussed above, resulted in the interesting
conclusion that there is \emph{no} chiral magnetic effect present in
the holographic Sakai-Sugimoto model. There are some subtleties with
this which were pointed out in \cite{Landsteiner:2011tf}, to which we
will return shortly. However, their result also leaves the open
question as to whether there is a chiral magnetic effect for more
general solutions to the equations of motion, for instance the
non-homogeneous ones which we consider here.

So even before we find the full non-homogeneous solution, an important
lesson might be learnt from an evaluation of the corrected holographic
currents \eqref{extracurrents} for the non-homogeneous ansatz. For our
ansatz (\ref{newansatz}), the corrected currents become
\begin{equation}
\label{allcurrents}
\begin{aligned}
\tilde J^0_V &=  -  4 \kappa \lim_{z \to \infty} \left [ \sqrt{-g} g^{zz} g^{00} \partial_z f_V \right ] - 12 \alpha  B j  \\[1ex] 
&= - 12 \alpha  \lim_{z \to \infty} \left [ B a_A + k h_V h_A \right ] - 12 \alpha  B j = 0 \, , \\[1ex]
\tilde J^1_V &=  - 4 \kappa \lim_{z \to \infty} \left [ \sqrt{-g} g^{zz} g^{xx} \partial_z a_V \right ] + 12 \alpha B \mu_A   \\[1ex]
&= 12 \alpha B \lim_{z \to \infty} f_A + 12 \alpha B \mu_A = 0 \,  , \\[1ex]
\tilde J^2_V &=  - 4 \kappa \lim_{z \to \infty} \left [ \sqrt{-g} g^{zz} g^{xx} \partial_z {\cal A}_2^V \right ]  
= - 4 \kappa M_{\text{KK}}^2  \lim_{z \to \infty} \left [ K_z  \partial_z h_V \right ] \cos(kx_1) \, , \\[1ex]
\tilde J^3_V &=  - 4 \kappa \lim_{z \to \infty} \left [ \sqrt{-g} g^{zz} g^{xx} \partial_z {\cal A}_3^V \right ] 
=  4 \kappa M_{\text{KK}}^2  \lim_{z \to \infty} \left [ K_z  \partial_z h_V \right ] \sin(k x_1) \, , \\[1ex]
\tilde J^0_A &=  4 \kappa \lim_{z \to \infty} \left [ \sqrt{-g} g^{zz} g^{00} \partial_z f_A \right ] \\[1ex]
&=  \lim_{z \to \infty} \left [ 12 \alpha  \left ( B a_V + \frac{k}{2} h_V^2 + \frac{k}{2} h_A^2  \right ) - 4 \kappa \tilde \rho \right ] = - 4 \kappa \tilde \rho \, , \\[1ex]
\tilde J^1_A &=  4 \kappa \lim_{z \to \infty} \left [ \sqrt{-g} g^{zz} g^{xx} \partial_z a_A  \right ] 
=  4 \kappa M_{\text{KK}}^2  \lim_{z \to \infty} \left [ K_z  \partial_z a_A \right ] \, , \\[1ex]
\tilde J^2_A &=   4 \kappa \lim_{z \to \infty} \left [ \sqrt{-g} g^{zz} g^{xx} \partial_z {\cal A}_2^A \right ]  
=  4 \kappa M_{\text{KK}}^2  \lim_{z \to \infty} \left [ K_z  \partial_z h_A \right ] \cos(kx_1) \, , \\[1ex] 
\tilde J^3_A &=    4 \kappa \lim_{z \to \infty} \left [ \sqrt{-g} g^{zz} g^{xx} \partial_z {\cal A}_3^A \right ] 
= - 4 \kappa M_{\text{KK}}^2  \lim_{z \to \infty} \left [ K_z  \partial_z h_A \right ] \sin(k x_1) \, . 
\end{aligned}
\end{equation}
where we have used the gauge field equations for the components $\ell
= 0$ and $\ell =1$. 

From \eqref{allcurrents} we see two important facts. First, the
density of particles carrying baryonic charge is zero. This confirms
once more that there is nothing baryonic in the solutions under
consideration, in agreement with the fact that the baryon chemical
potential $\mu_B$ decouples completely.

The second observation is that the component of the vector current
\emph{in the direction of the external magnetic field} is zero. This is in
sharp contrast to what one would expect if there was a chiral
magnetic effect present.  We should, however, emphasise that it is
possible to define corrected currents which are different from those
above, if one decides to deal with the anomalous symmetry in a
different way, see section~\ref{s:chempots_nonconserved}. However,
this alternative method, although it produces the chiral magnetic
effect, suffers from other shortcomings, as we will explain.

The expressions  (\ref{allcurrents}) simplify further when we restrict to the
homogeneous ansatz, for which one obtains
\begin{equation}
\begin{aligned}
\label{homogeniouscurrents}
\tilde J^0_V &= \tilde J^1_V = \tilde J^2_V = \tilde J^3_V = 0 \, , \cr
\tilde J^0_A &=  4 \kappa M_{\text{KK}}^2 \tilde B \mu_A \coth( \frac{\pi}{2} \tilde B ) = 12 \alpha B \mu_A \coth( \frac{\pi}{2} \tilde B ) \, , \cr
\tilde J^1_A &=  - 4 \kappa M_{\text{KK}}^2 \tilde B j \coth( \frac{\pi}{2} \tilde B ) = - 12 \alpha B ( \mu_B - f_0 ) \, , \cr
\tilde J^2_A &= \tilde J^3_A = 0 \,.
\end{aligned}
\end{equation}
We should emphasise once more that the vector (baryonic) currents
vanish \emph{only} when the contribution from the Bardeen term in the
action is properly taken into account.  One could in principle
evaluate an ``abelianised'' version of the baryon number (second Chern
class), $\sim \int F_{3z}F_{12}$. However, this expression is strictly
speaking valid only in a \emph{nonabelian} system, and in this
situation one expects that the charge computed using the corrected
conserved current $\tilde{J}_V^0$ should coincide with this
topological number.

Our analysis shows that the fact that there is no chiral magnetic
effect in the Sakai-Sugimoto model is not due to the simplified
homogeneous ansatz, and that it is also not a consequence of the
details of any numerical solution which we will present later; rather,
the chiral magnetic effect is absent for the entire class of solutions
captured by the ansatz~\eqref{newansatz}.  The contribution of the
Bardeen terms required to make the vector currents conserved is
crucial for the absence of the chiral magnetic effect.

We finally see that, in contrast to the homogeneous solution,
where no vector currents are present at all, the non-homogeneous system
exhibits transverse vector currents. This shows some resemblance to the
chiral magnetic spiral of \cite{Basar:2010zd}.  

\subsection{Chemical potentials for non-conserved charges}
\label{s:chempots_nonconserved}

Given the rather convincing quasi-particle picture of the origin of
the chiral magnetic effect at weak coupling
\cite{Kharzeev:2007jp,Fukushima:2008xe}, it is somewhat surprising
that it is not present in the model at hand at strong coupling. A
reason for this discrepancy has been suggested in
\cite{Landsteiner:2011tf}, in which it was emphasised that one should
be careful in computing the effects of a chemical potential in
theories for which the associated charge is not conserved.

The main observation made in \cite{Landsteiner:2011tf} is that there
are two ways to introduce a chemical potential into a thermal quantum
system. One is to twist the fermions along the thermal circle, i.e.~to
impose
\begin{equation}
\label{e:Bformalism}
\Psi_{L,R}(\tau) = - e^{\pm\beta\mu_A} \Psi_{L,R}(\tau-\beta)\,,
\end{equation}
instead of the usual anti-periodic boundary condition. This is what
one would do at weak coupling. It was called the ``B-formalism'' in
\cite{Landsteiner:2011tf}. The other way is to keep anti-periodic
boundary conditions, but instead use a shifted Hamiltonian,
\begin{equation}
\label{e:Aformalism}
\tilde H = H - \mu_A Q_L + \mu_A Q_R\,.
\end{equation}
This coupling can be viewed as a coupling to a gauge field for which
$\tilde A_0=\mu_A$. We will refer to this as the ``A-formalism'' (in
which we will temporarily put tildes on all objects, as above, for
clarity). For a non-anomalous symmetry, these two formalisms are
equivalent, and one can go from the B-formalism to the A-formalism
using a gauge transformation involving the external gauge field, with
parameter $\theta_A = -\mu_A t$, relating the gauge fields in the two
formalisms according to
\begin{equation}
A_\mu = \tilde{A}_\mu + \partial_\mu \theta_A\,.
\end{equation}

In terms of our holographic picture, this gauge transformation acts
directly on the 5d gauge field, with a parameter $\Theta_A(z)$ that is
$z$-dependent, 
\begin{equation}
\label{e:ThetaAdef}
\Theta_A(z) = t g_A(z)\,,\qquad g_A(z\rightarrow\pm\infty) = \pm\mu_A\,.
\end{equation}
which again acts on the gauge field as\footnote{The boundary gauge
  transformation parameter $\theta_A(x)$ is related to the 5d
  parameter by
\begin{equation}
  \theta_A(x) \equiv \mp \Theta_A(x, z \to \pm \infty) = -\mu_A t  \,, \label{thetadef}
\end{equation}
with the perhaps somewhat inconvenient signs following from our relation between the bulk and
boundary gauge fields~\eqref{bdyfields}.}
\begin{equation}
{\cal A}_m = \tilde{\cal A}_m + \partial_m \Theta_A\,.
\end{equation}
The difference between the two formalism can thus be formulated in a
clear holographic language as well: in the A-formalism the ansatz has
$\tilde{\cal A}_0$ asymptoting to the chemical potential, whereas in the
B-formalism the ansatz instead has this chemical potential stored in
the ${\cal A}_z$ component. The chemical potential is then best
written as
\begin{equation}
\mu_A  = - \frac12 \int_{-\infty}^{\infty}\!{\rm d}z\, {\cal F}_{z0} \, , \label{chiralmu}
\end{equation}
which is nicely gauge invariant and independent of the formalism used.

This is all clear and unambiguous when the symmetry is
non-anomalous. However, in the presence of an anomaly, one cannot pass
from the formalism defined by \eqref{e:Bformalism} to that defined by
\eqref{e:Aformalism}. This is what happens in our system: as we
discussed above, the 5d action after the anomaly correction
\eqref{newYMCS} is not invariant under a chiral gauge transformation,
\begin{multline}
 \tilde S [\tilde{\cal A }_\ell + \partial_\ell \Theta_A] = 
 \tilde S[\tilde{\cal A}_\ell] + \frac{\alpha}{4} \epsilon^{\ell m n p q}
 \int\!{\rm d}^4 x {\rm d}z\, (\partial_\ell \Theta_A) 
\left ( 3 \tilde{\cal F}_{mn}^V \tilde{\cal F}_{pq}^V + \tilde{\cal F}_{mn}^A \tilde{\cal
    F}_{pq}^A \right )  \\[1ex] =:
 \tilde S[\tilde{\cal A}_\ell] + \tilde S^\Theta[\tilde{\cal A}_\ell]\,.
\end{multline}
The anomaly implies that one of the two formalisms is incorrect.

The point of view of \cite{Landsteiner:2011tf} is that there are
strong indications that the B-formalism is the correct one in field
theory. If one insists on computing with untwisted fermions, one needs
to perform a gauge transformation, which not just introduces the
chemical potential into $A_0$, but also modifies the action and
Hamiltonian to correct for the fact that the action is not gauge
invariant. To be precise, when one uses untwisted fermions, the action
that one should use is the gauge-transformed action, which differs
from the original one by the anomaly. This was called the
``A'-formalism'' (note the prime) in \cite{Landsteiner:2011tf}.

Let us see how this logic works for the Sakai-Sugimoto system under
consideration here. The idea is thus that if we want to identify the
chemical potential with the asymptotic value of $\tilde{\cal A}_0$, we should
be working not with $\tilde S[\tilde {\cal A}_\ell]$ but rather with $\tilde
S[\tilde {\cal A}_\ell] + \tilde S^\Theta[\tilde{\cal A}_\ell]$. In order to
compute the currents, we need to compute the variation of
the $\Theta_A$, which takes the form
\begin{equation}
\delta  S^{(\Theta_A)} =  \alpha \epsilon^{m \ell n p q} \int\!{\rm
  d}^4x {\rm d}z\, \partial_m 
\left [ 3 (\partial_n \Theta_A) \tilde{\cal F}_{pq}^V \delta \tilde{\cal A}_\ell^V + (\partial_n \Theta_A) \tilde{\cal F}_{pq}^A \delta \tilde{\cal A}_\ell^A \right ] \,. 
\end{equation}
The contribution of the $\Theta_A$ term to the holographic currents can be obtained using the dictionary
\begin{equation}
\Delta \tilde J^\mu_{V/A} = \pm \frac{ \delta  S^{(\Theta_A)}} {\delta \tilde{\cal A}_\mu^{V/A} (x, z = \infty)} \,,
\end{equation}
We then obtain
\begin{equation}
\begin{aligned}
\Delta \tilde J^\mu_V &= - 6 \alpha \epsilon^{\mu \nu \rho \sigma}
(\partial_\nu \theta_A) \tilde F_{\rho \sigma}^V = 6 \alpha \mu_A
\epsilon^{\mu 0 \rho \sigma} \tilde F_{\rho \sigma}^V \\[1ex]
\Delta \tilde J^\mu_A &= - 2 \alpha \epsilon^{\mu \nu \rho \sigma}
(\partial_\nu \theta_A) \tilde F_{\rho \sigma}^A = 2 \alpha \mu_A
\epsilon^{\mu 0 \rho \sigma} \tilde F_{\rho \sigma}^A \,.
\end{aligned}
\end{equation}
These expressions are independent of the particular function $g_A(z)$
which one chooses in~\eqref{e:ThetaAdef}.  For our ansatz
\eqref{newansatz} with the boundary conditions \eqref{boundarystuff}
we get
\begin{equation}
\label{e:DeltaJ}
\begin{aligned}
\Delta \tilde J^0_V &= 0 \, , \\[1ex] 
\Delta \tilde J^1_V &= - 12 \alpha B \mu_A  \, , \\[1ex]
\Delta \tilde J^2_V &= \Delta \tilde J^3_V = 0 \, , \\[1ex] 
\Delta \tilde J^0_A &= \Delta \tilde J^1_A = \Delta \tilde J^2_A = \Delta \tilde J^3_A = 0 \, .
\end{aligned}
\end{equation}
This shows a very promising feature: any solution in this class will
now exhibit the chiral magnetic effect, as there is a non-vanishing
$\tilde J^1_V$ component.

Unfortunately, we will see later that things are more subtle at the
level of the Hamiltonian. There are two main problems when writing
down the $A'$-formalism for things more complicated than the
currents. Firstly, any bulk quantities such as the Hamiltonian will
typically depend on $g_A(z)$, not just on its asymptotic
values. Secondly, it will turn out that even for a `natural' choice of
$g_A(z)$, for instance $g_A(z)=-f_A(z)$, the new Hamiltonian has the
property that it does not lead to a minimum for non-homogeneous
configurations. To see this requires some more details about this
condensate, but let us here already present the expression for the
corrected Hamiltonian. It can be written as
\begin{equation}
H_{\text{Tot}}({\cal A_\ell},\Theta_A) =  \tilde H_{\text{Bulk}} ({\cal A_\ell}) + \tilde H_{\text{Bdy}} ({\cal A_\ell}) +  H^{(\Theta_A)} ({\cal A_\ell},\Theta_A) \label{Htot}
\end{equation}
where $\tilde H_{\text{Bulk}} ({\cal A_\ell})$ and $\tilde H_{\text{Bdy}} ({\cal A_\ell})$ are given by (\ref{tHBulk}), (\ref{tHBdy}) and the $\Theta_A$ term is 
\begin{multline}
H^{(\Theta_A)}({\cal A_\ell},\Theta_A) = \alpha \int {\rm d}^3 x {\rm d}z \Big \{  \epsilon^{0abcd} (\partial_b \Theta_A) 
\left [ 3 {\cal F}_{cd}^V (\partial_0 {\cal A}_a^V) + {\cal F}_{cd}^A (\partial_0 {\cal A}_a^A) \right ] \\[1ex]
- \frac14 \epsilon^{\ell m n p q} (\partial_\ell \Theta_A) ( 3 {\cal F}_{mn}^V {\cal F}_{pq}^V + {\cal F}_{mn}^A {\cal F}_{pq}^A ) 
\Big \}
\end{multline}
For our ansatz \eqref{newansatz} with the boundary conditions
\eqref{boundarystuff} the theta term in the Hamiltonian take the form
\begin{equation}
\begin{aligned}
H^{(\Theta_A)}({\cal A_\ell},\Theta_A) &= 2 \alpha \int {\rm d}^3 x {\rm d}z\, (\partial_0 \Theta_A) 
\left [ 3 B \partial_z a_V +  \frac{k}{2} \partial_z (3 h_V^2 + h_A^2) \right ] \cr
&=  - \frac23 {\cal H}_0  \int  {\rm d}z \, \partial_z \hat g_A   
\left [ 3 \hat B  \hat a_V + \frac{k}{2} (3 h_V^2 + h_A^2) \right ] \cr
&=  - \frac23 {\cal H}_0  \int  {\rm d}z \, \partial_z \hat g_A   
\left [ 3 ( \hat b_V + \frac{k}{2} h_V^2) + \frac{k}{2} h_A^2\right ] - 4 {\cal H}_0 \hat \rho \hat \mu_A \,. 
\end{aligned}
\end{equation}
Specialising to $g_A(z) = - f_A(z)$ we get 
\begin{multline}
\label{e:HTheta}
H^{(\Theta_A)}({\cal A_\ell},\Theta_A)
=  - \frac23 {\cal H}_0  \int  {\rm d}z \, \frac{1}{K_z} \left [ \hat b_V + \frac{ \hat k}{2} \hat h_V^2 + \frac{ \hat k}{2} \hat h_A^2 \right ]   
\left [ 3 ( \hat b_V + \frac{k}{2} h_V^2) + \frac{k}{2} h_A^2\right ] \\[1ex]
- 4 {\cal H}_0 \hat \rho \hat \mu_A \,. 
\end{multline}
In the homogeneous case we get 
\begin{equation}
\label{e:HTheta_hom}
H^{(\Theta_A)}({\cal A_\ell},\Theta_A)
=  - {\cal H}_0 \hat C_A^2 \left [ \frac{\sinh( \pi \hat B)}{\hat B} + \pi \right ]  - 4 {\cal H}_0 \hat \rho \hat \mu_A \,. 
\end{equation}
This boundary term can have drastic consequences for the phase
structure of the theory: we will see in section~\ref{s:inhomogeneous}
that it disfavours non-homogeneous configurations.

We should emphasise that the procedure for introducing the
A'-formalism in the holographic context is, unfortunately, rather
ambiguous. This is essentially because the required boundary condition
on the gauge transformation parameter $\Theta_A(z)$, given in
\eqref{thetadef}, does not uniquely specify the behaviour of
$\Theta_A(z)$ in the bulk. In contrast, this kind of ambiguity does not
appear in field theory, as gauge transformation which untwists
fermions, and ``moves'' chemical potential into temporal component of
the gauge potential, is unique.

Instead of using the A'-formalism one could consider doing the
holographic computation directly in the B-formalism. This would
require writing the solution in the $A_0=0$ gauge (the fermions are
not directly accessible so it is unclear whether additional changes
are required to implement the twisting). This does lead to a CME, but
the Hamiltonian again turns out to have no minimum for non-homogeneous
configurations. Some details are given in appendix~\ref{a:Bformalism}.

\subsection{Perturbative stability analysis of the homogeneous solution}
\label{s:perturbative_stability}

In this section we will perturbatively analyse the stability of the
homogeneous solution \eqref{homosolu}, in order to show that this
configuration is unstable and wants to ``decay'' to a non-homogeneous
solution~\eqref{newansatz} (whose explicit form will be found later). Our
analysis here is a revision of the work done in \cite{Kim:2010pu}, but
our findings differ from theirs in an important way which is crucial
for the remainder of our paper.

Our starting point is given by the equations of motion linearised around the
configuration \eqref{homosolu}. Following \cite{Kim:2010pu}, we will
look for fluctuations of the modes transverse to the direction of the
external field, i.e.~along $(A_1,A_2)$, since these should lead to the
formation of a chiral spiral. These fluctuations are given by
\begin{equation}
\delta A_{i} = \delta A_{i}(\omega,k) e^{-i \omega t + i k x_1}  \quad \quad (i=2,3) \, .
\end{equation}
The equations of motion for the fluctuations $(\delta A_1,\delta A_2)$
are coupled, but they diagonalise in the complex basis
\begin{equation} 
\delta A^{(\pm)}\equiv \delta A_2 \pm i \delta A_3 \, ,
\end{equation}
where the equations become
\begin{equation}
\label{perturba}
K_z^{-1/3}(\omega^2 - k^2) \delta A^{(\pm)} + M_{\text{KK}}^2 \partial_z(K_z \partial_z \delta A^{(\pm)}) \pm \frac{N_c}{8\pi^2 \kappa} (k \mathcal{F}_{0z} + \omega \mathcal{F}_{1z}) \delta A^{(\pm)} =0 \, .
\end{equation}
Here $\mathcal{F}_{1z},\mathcal{F}_{0z}$ are field strengths of the
background homogeneous solution \eqref{fieldstr}, and we will express
all results in terms of the constants $C_A$ and $C_B$ (instead
of~$\mu_A$ and~$j$) in order enable a simpler comparison with the
results of \cite{Kim:2010pu}.

Given values of $C_A$ and $C_B$, we numerically solve
equation~\eqref{perturba}. As usual in perturbation theory, solutions
with real $\omega$ represent fluctuations that are stable, while those
for which $\omega$ has a positive imaginary part correspond to
instabilities, since they are exponentially growing in
time. Fluctuations for which $\omega=0$ are marginal and have to be
analysed in the full nonlinear theory in order to see if they
correspond to unstable directions in configuration space. We will
argue now that, while \cite{Kim:2010pu} has correctly identified
the perturbatively unstable solutions with complex $\omega$, they have
missed the marginally unstable modes, which are actually unstable in
the full theory. In the following section we will then explicitly
construct the new vacua corresponding to these marginal modes.

Before presenting solutions to the equation \eqref{perturba}, observe
that this equation exhibits the symmetry $t\rightarrow
-t,\omega\rightarrow -\omega$, which means that all solutions will
come in pairs $(\omega, -\omega)$. Additionally, when solving this equation,
one looks for \emph{normalisable} solutions, i.e.~one looks for the
solutions that behave as $\delta A^{(\pm)} \sim 1/z$ near the boundaries.

\begin{figure}[t]
\includegraphics[width=.31\textwidth]{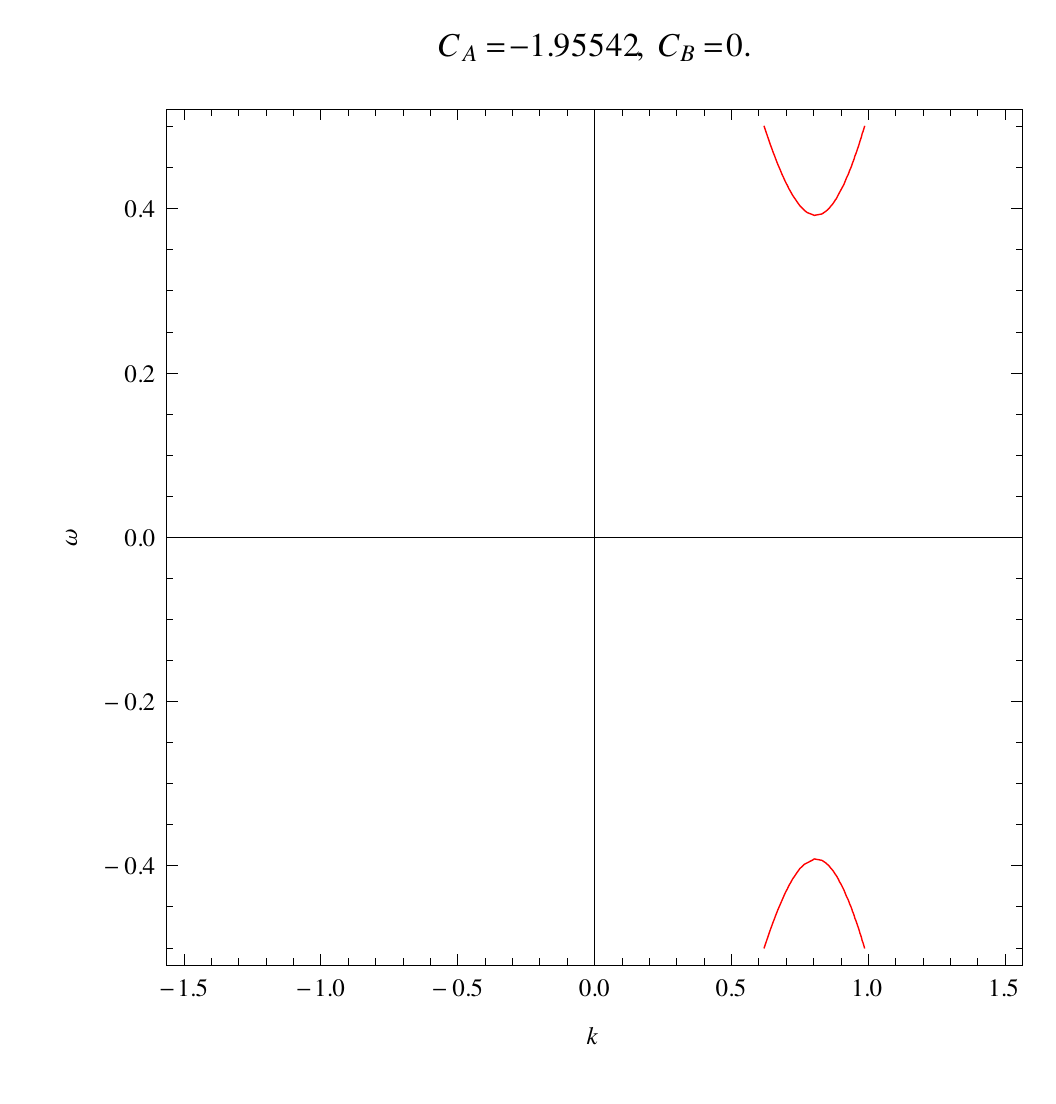}~~
\includegraphics[width=.31\textwidth]{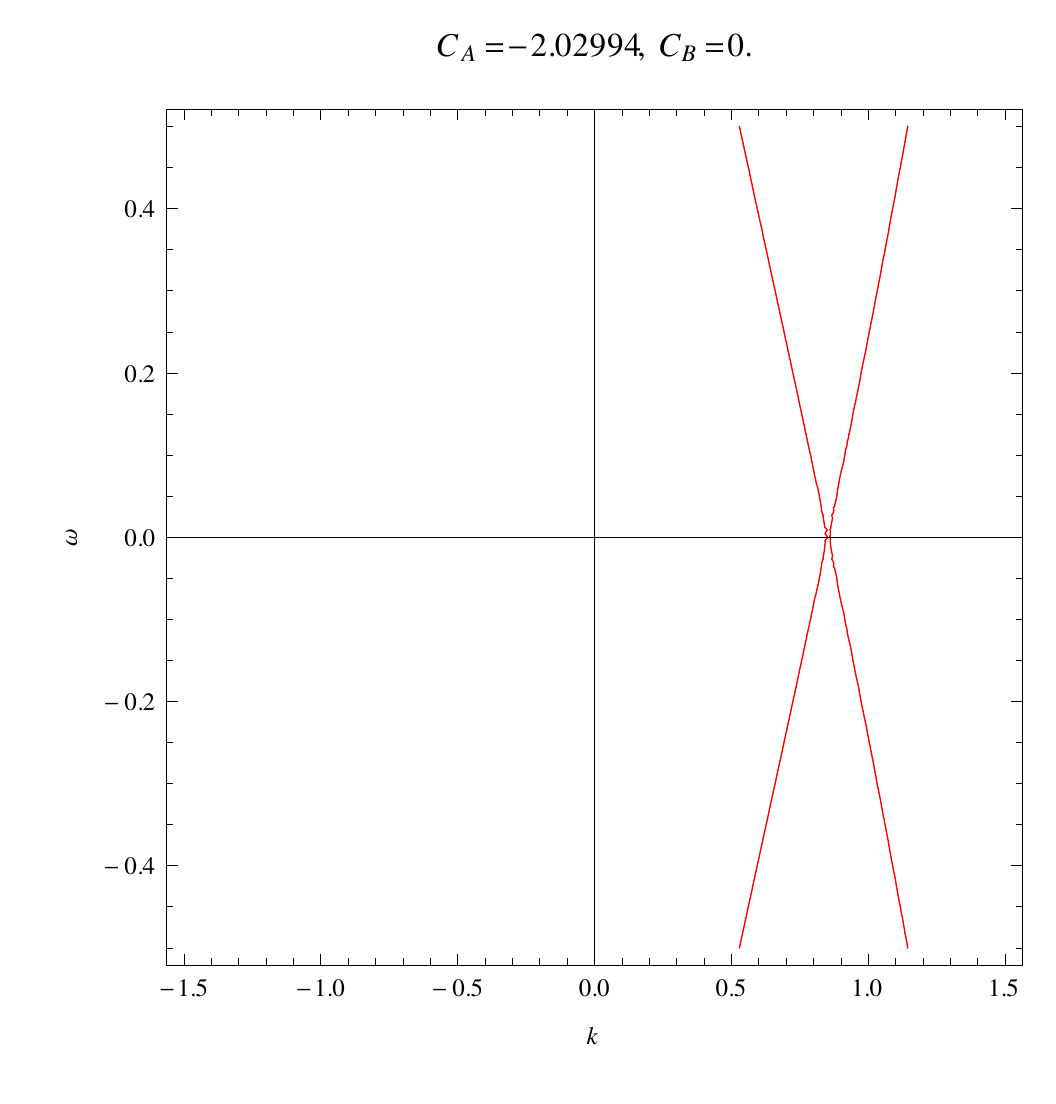}~~
\includegraphics[width=.31\textwidth]{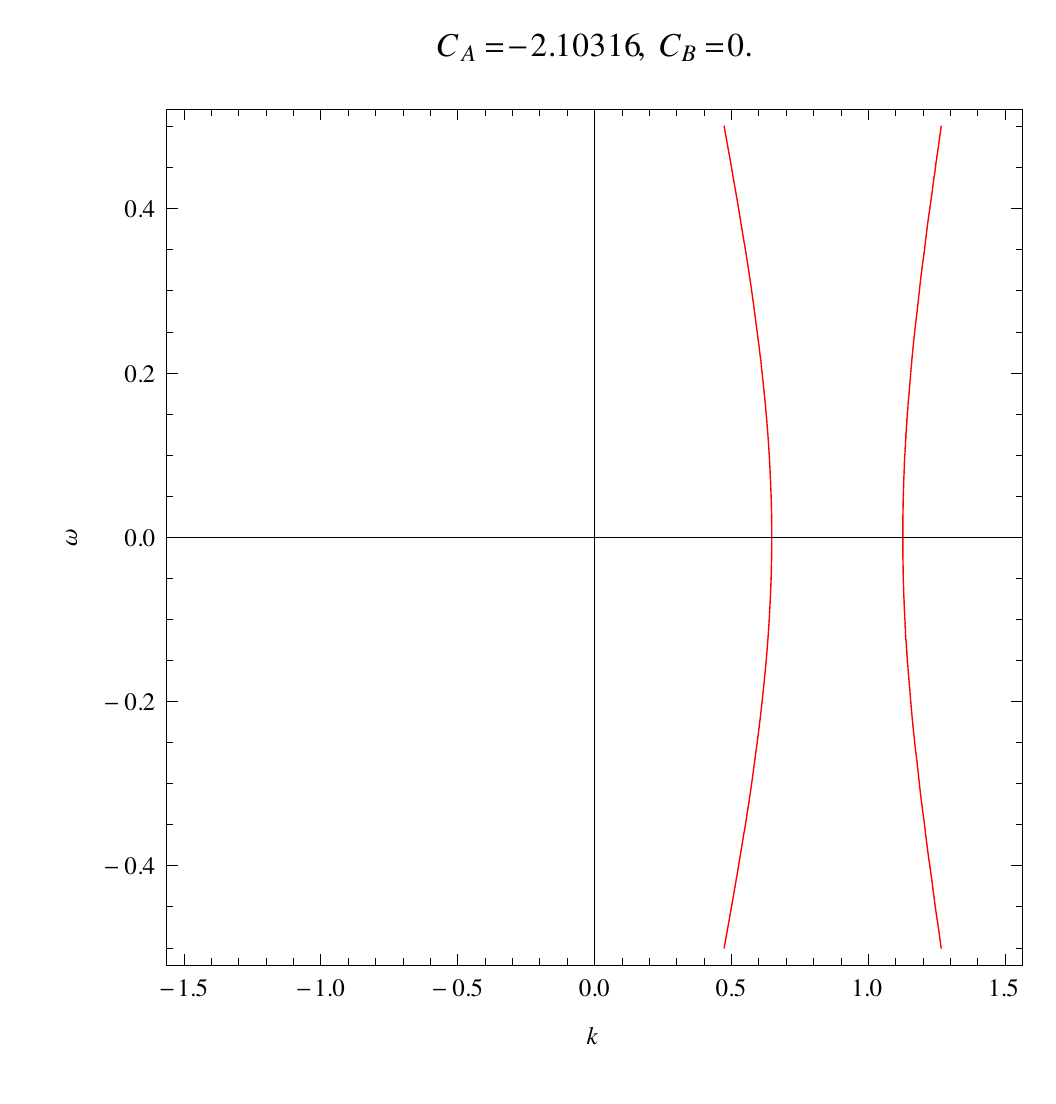}
\caption{\label{f:dispersion_relations_CB0} Dispersion relation for
  small fluctuations at vanishing $C_B$ and increasing values of
  $C_A$. In the last plot, the region between the two $\omega=0$ modes
  corresponds to modes with positive imaginary $\omega$, hence
  signalling proper instability. This follows Kim et al.~\cite{Kim:2010pu}.}
\end{figure}

Let us start by discussing the solutions with only $C_A$
non-vanishing. At $B=0$ these correspond to solutions with
$j=0$. Samples of these solutions are shown in
figure~\ref{f:dispersion_relations_CB0}. We see that as $|C_A|$ is
increased, the two branches of solutions come closer to the $\omega=0$
axis and then touch (the middle plot in figure
\ref{f:dispersion_relations_CB0}). For all those values of $|C_A|$,
given any momenta $k$, there is always real $\omega$
solution. However, as $|C_A|$ is increased even more, the two branches
of solutions separate along the $k$-axis, so that there is a region of
momenta for which there are no real $\omega$ solutions (see the third
plot of figure \ref{f:dispersion_relations_CB0}). For these
``forbidden'' values of the momenta, one can explicitly find
solutions with complex $\omega$, which clearly signal a proper
instability of the solution. These modes have previously been found in
\cite{Kim:2010pu}.

\begin{figure}[t]
\includegraphics[width=.31\textwidth]{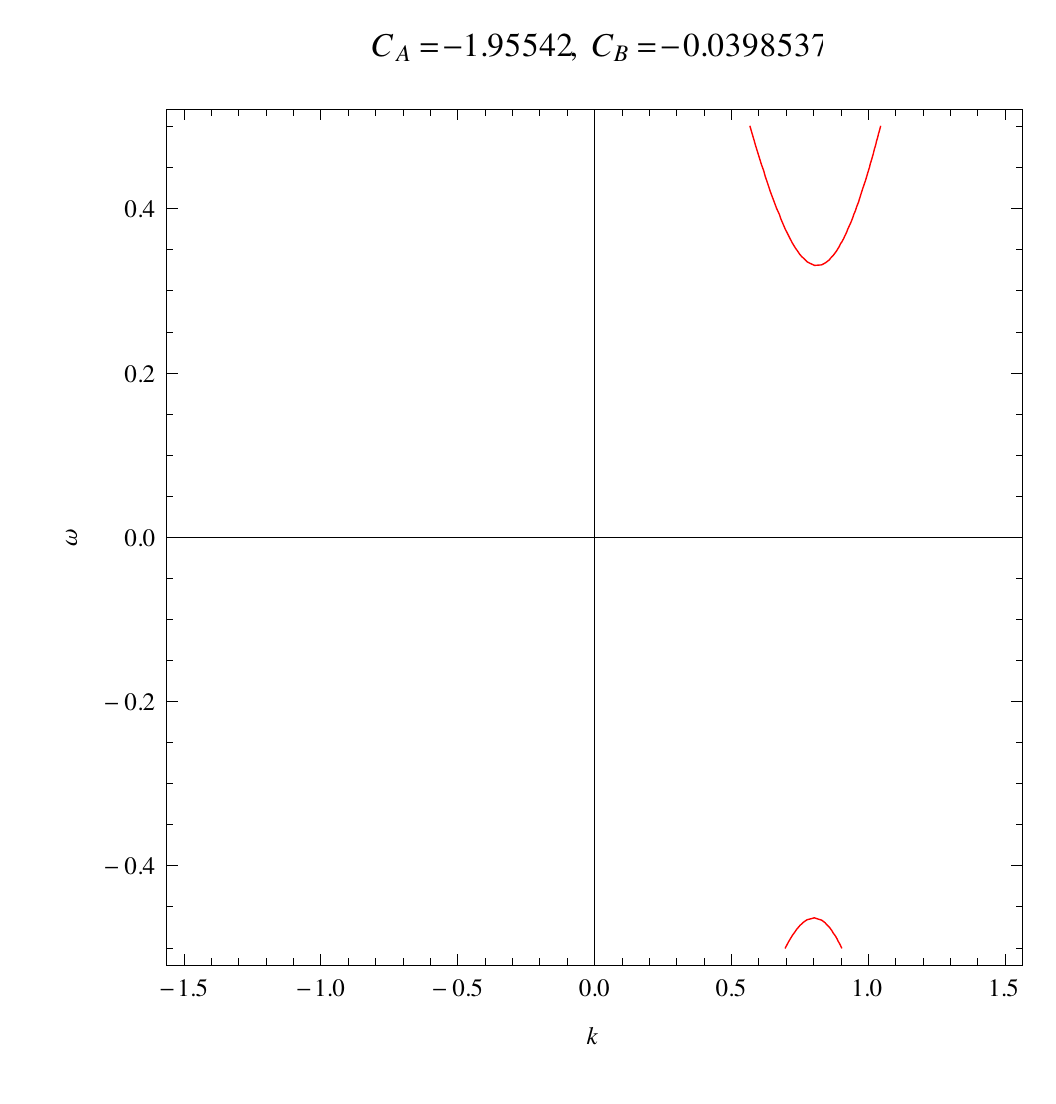}~~
\includegraphics[width=.31\textwidth]{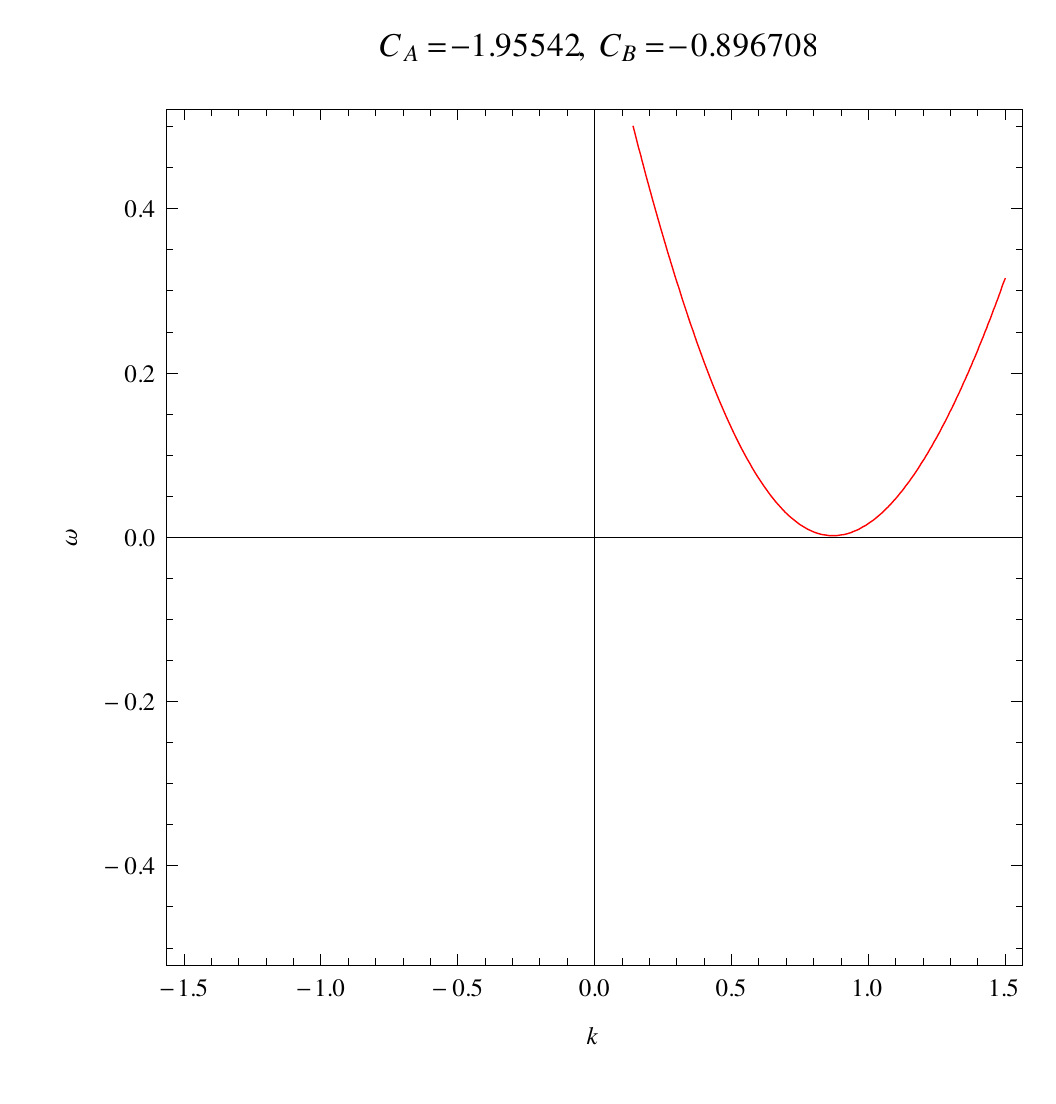}~~
\includegraphics[width=.31\textwidth]{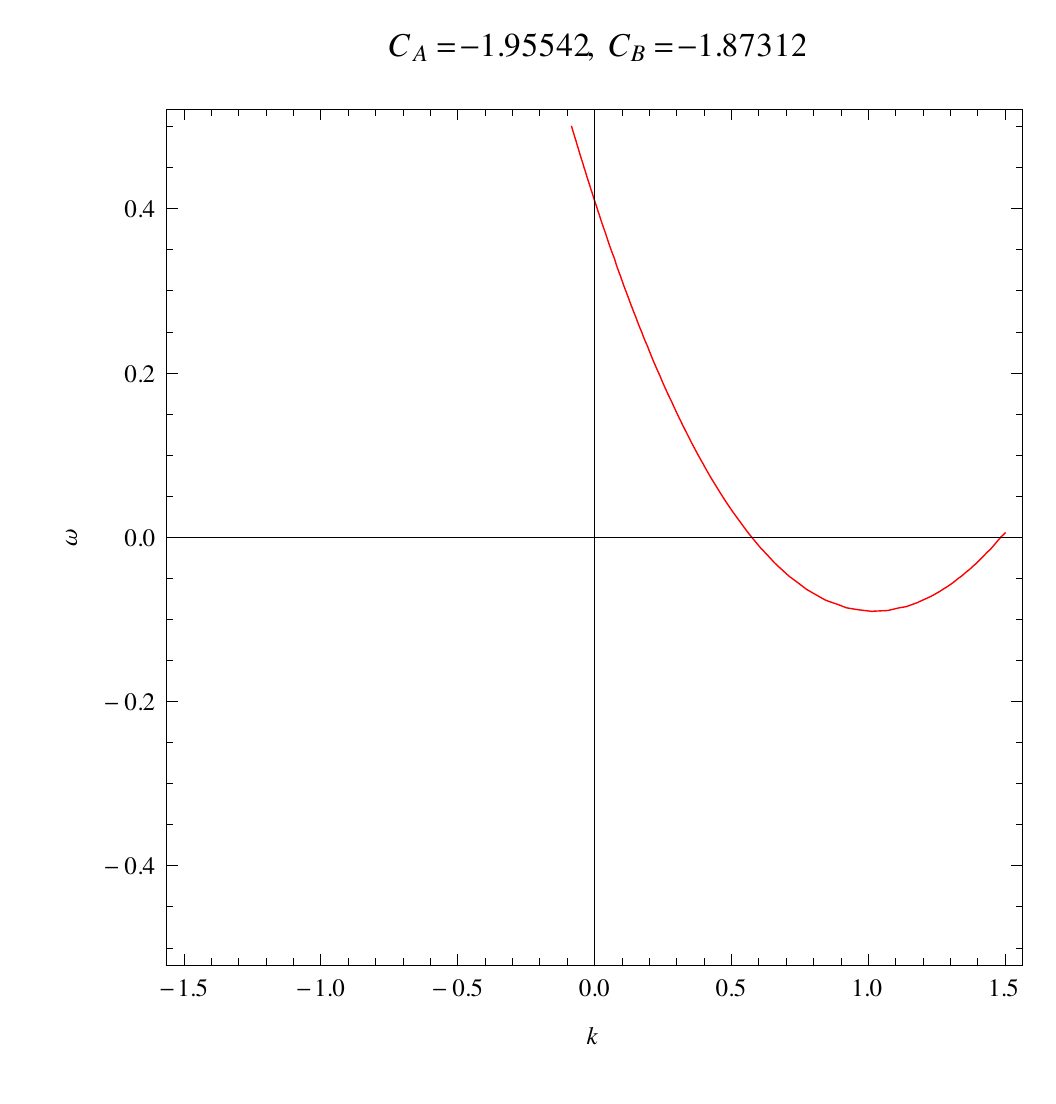}
\caption{\label{f:dispersion_relations_CBnon0} Dispersion relation at
  fixed value of $C_A$ and increasing values of $C_B$. Modes with
  $\omega=0$ occur for sufficiently large $C_B$, but in the region on
  the $\omega=0$ axis between these two roots, there are no strictly
  unstable modes. Nevertheless, the $\omega$ could signal that there
  appears a new ground state.}
\end{figure}

Let us now turn on $|C_B|$, while keeping $|C_A|$ fixed. Samples of
these solutions are shown in figure
\ref{f:dispersion_relations_CBnon0}.  We see that as $|C_B|$ is
increased, the two branches of solutions shift in the vertical
direction, while the distance between them remains non-vanishing (which
is the reason why this other branch is not visible in the second and
third plots). For some value of $|C_B|$, the upper branch crosses the
$\omega=0$ axis, and continues to go towards negative $\omega$. We
thus see that for large enough $|C_B|$, marginal $\omega=0$ modes are
always present in the spectrum. In the next section, we will show that
these modes, which were previously missed in the literature, are
actually unstable once non-linearities are taken into account.

\begin{figure}[t]
\begin{center}
\includegraphics[width=.7\textwidth]{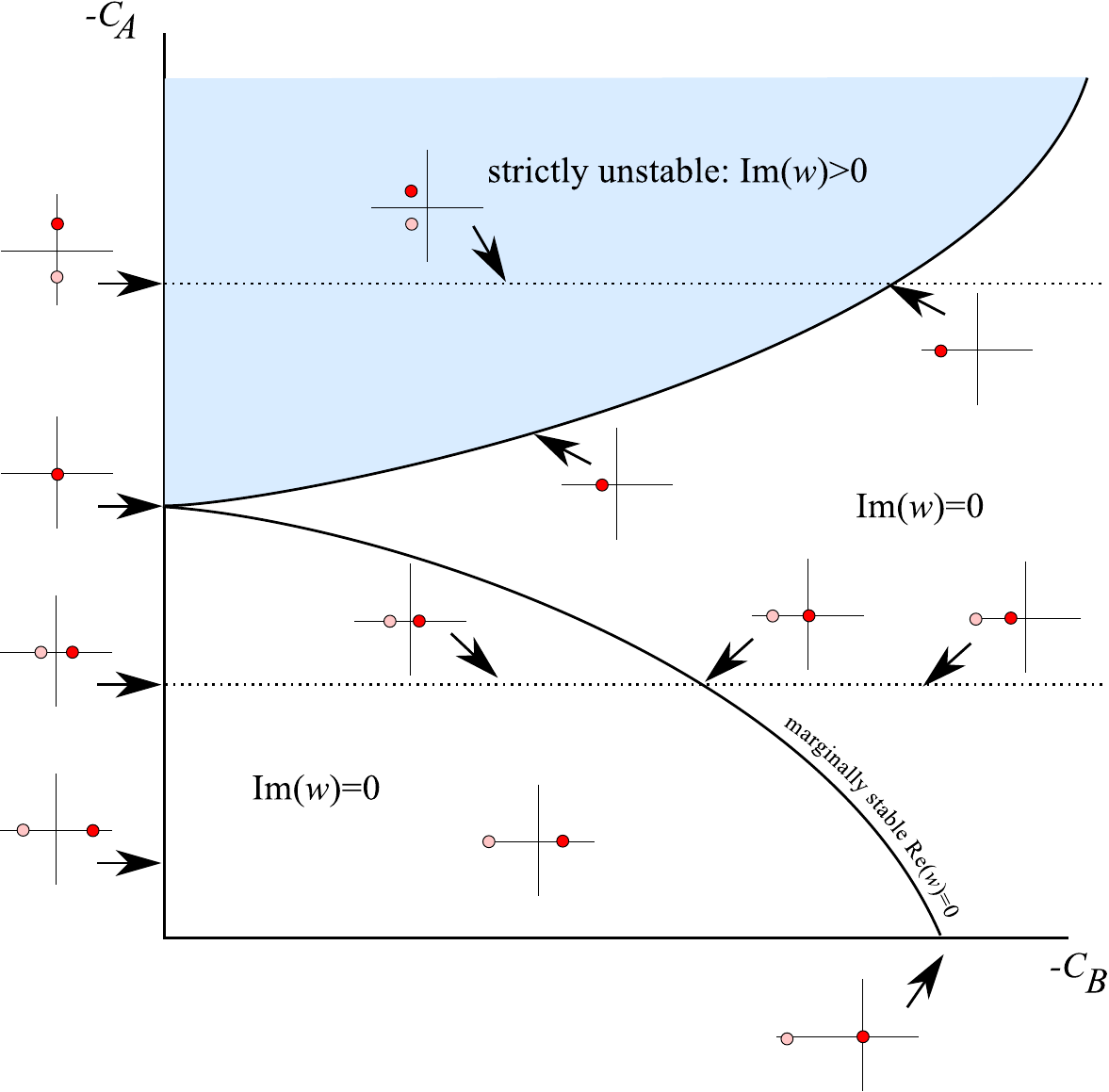}
\caption{\label{f:physical_modes} The physical picture (not to scale)
  that arises from the analysis of the dispersion relation. The plot
  displays the location of the physical modes at the extrema of the
  curves in figures~\protect\ref{f:dispersion_relations_CB0} and
  \protect\ref{f:dispersion_relations_CBnon0}, but now depicted in the complex
  $\omega$ plane.}
\end{center}
\end{figure}

These findings are summarised in
figure~\ref{f:physical_modes}. Depicted there is the behaviour of the
frequencies $\omega$ of the fluctuation modes at the `extrema' of the
dispersion relation plots, as a function of the two parameters $C_A$
and $C_B$.  For a generic value of $C_B$ and $C_A=0$, one root is
always positive.  As $C_A$ is increased from zero, the positive root
(red dot) moves towards the left, until it becomes zero. This defines
the lower, solid curve of marginal modes in
figure~\ref{f:physical_modes}.  Above this curve there is always a
marginally stable mode in the spectrum, and thus potentially an
instability. As $C_A$ is increased further, both roots eventually
develop a positive imaginary part, and we enter the region of strictly
unstable modes (the blue shaded region). This area was previously
discussed in \cite{Kim:2010pu}.\footnote{As we have already mentioned
  before, we disagree with their interpretation of $C_B$ being
  related to the baryon chemical potential; see also below.}

\vfill\eject

\subsection{The inhomogeneous solutions to the equations of motion}
\label{s:inhomogeneous}

Having established the location of the unstable and marginally stable
modes in the fluctuation spectrum of the homogeneous solution, we
would now like to explicitly construct the non-homogeneous solutions to
which they are expected to decay, and study their properties.  In our
previous work~\cite{Bayona:2011ab} we have explicitly constructed a
non-homogeneous solution in the presence of non-vanishing \emph{axial}
chemical potential, and in the absence of a magnetic field. We will
therefore first discuss solutions with nonzero $\mu_A$ and non-zero
$B$, which are a natural generalisation of those considered before. We
will then introduce a non-zero potential $j$ as well and discuss how
the solutions behave as a function of this parameter.

The upshot of our previous analysis \cite{Bayona:2011ab} was that for
large enough axial chemical potential, larger than some critical
value, a non-homogeneous solution is formed. The relation between the
chemical potential and axial particle density was almost linear,
similar to the homogeneous case. However, a particular density of
particles was achieved for a smaller value of the chemical potential
than in the homogeneous configuration. The wave number characterising
the period of the non-homogeneous solution turned out to be very
weakly sensitive to the actual value of the particle charge density in
the canonical ensemble (or to the value of the chemical potential in
the case of the grand canonical ensemble). The first things we would
like to know is whether any of these characteristics change in the
presence of an external magnetic field.

Our starting point is the nonlinear system of equations
\eqref{hatfequation}, \eqref{hataequation} and
\eqref{hathequation}. We first observe that the function $\hat{f}(z)$
appears only in the first equation \eqref{hatfequation}, and this
equation can be directly integrated once we determine the functions
$\hat{b}(z)$ and $\hat{h}(z)$, from the other two equations.

Hence, we first need to solve equations \eqref{hataequation} and
\eqref{hathequation}. The parameters $\hat{B}$ and $\hat{k}$ in these equations
correspond to the external magnetic field and the momentum of the transverse
spiral, and are fixed at this stage. There are then four undetermined
constants corresponding to non-normalisable and normalisable modes for
each of the functions $\hat{b}$ and $\hat{h}$. However, we solve equations
(\ref{hataequation}), (\ref{hathequation}) requiring that the transverse
spiral describes a normalisable mode,
\begin{equation}
\hat{h}(z) = \frac{h_0}{z} + \cdots \, ,
\end{equation}
i.e. we impose that the only external field is the magnetic field, and
that there are no transverse external fields.

\begin{figure}[t]
\begin{center}
\includegraphics[width=.45\textwidth]{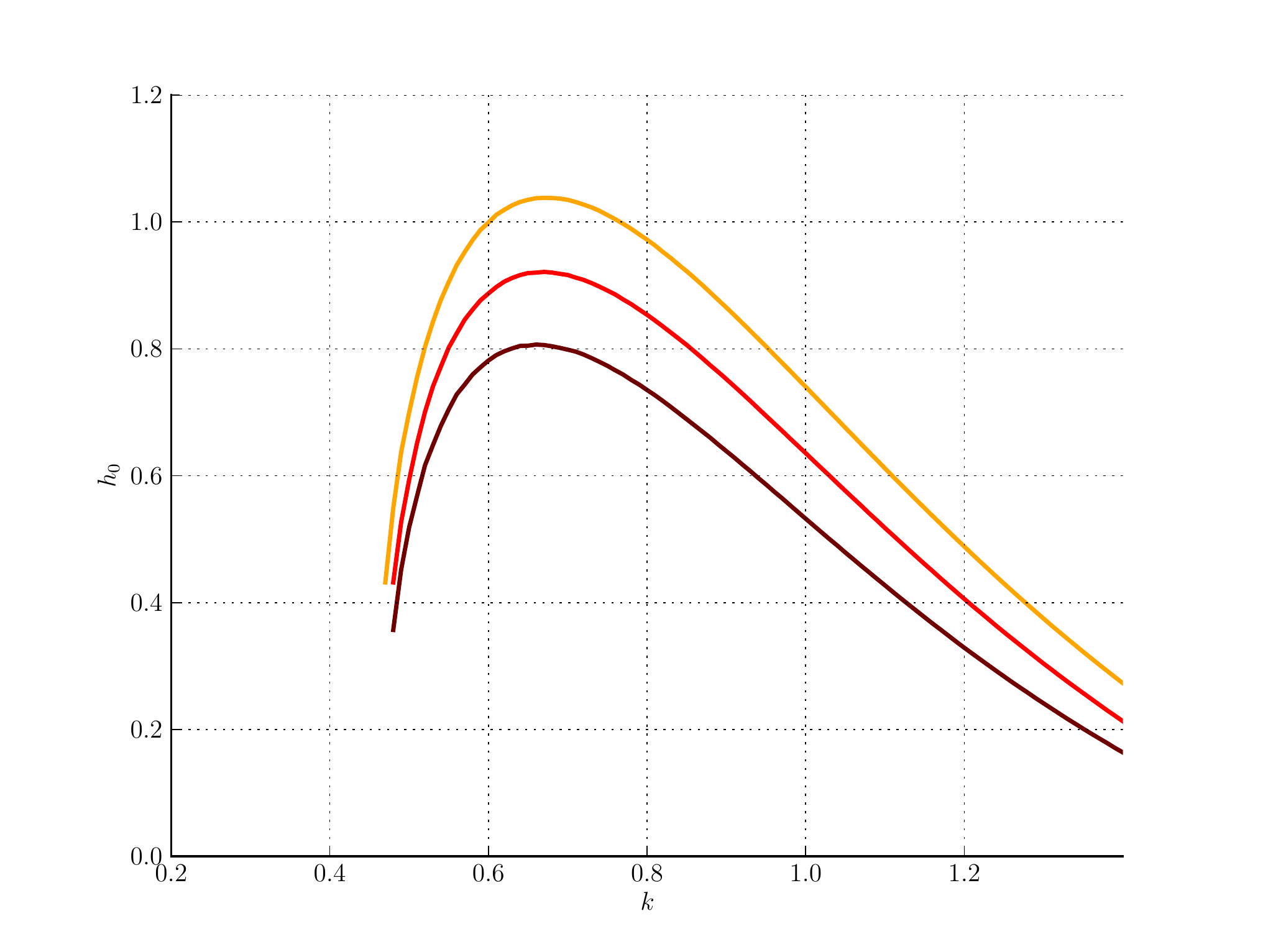}~~
\caption{Value of $h_0$ versus $k$ at fixed $B=1.0$, for increasing
  values of $|C_B|$. Increasing $|C_B|$ makes the curve go up, indicating
  that a full new non-linear ground state exists for increasing $|C_B|$
  and decreasing $|C_A|$. That is to say, the $\omega=0$ fluctuation
  modes are the indicators of the transition to a new ground state,
  not the ${\rm Im}(\omega)>0$ modes.\label{f:varyCB}}
\end{center}
\end{figure}

In contrast, the function $\hat{b}(z)$ (or alternatively,
$\hat{a}(z)$, see \eqref{e:bversusa}), which describes
the longitudinal field, does have a non-normalisable component. For
example, on the positive side of the U-shaped brane it generically
behaves as
\begin{equation}
\label{bexpansions}
 \hat{b}(z) = b_0 + \frac{b_1}{z} + \cdots \quad (z\rightarrow
 \infty)\,,
\end{equation}
where the coefficients $b_0$ and $b_1$ are given by
\begin{equation}
\begin{aligned}
 b_0 &= C_A \cosh\left(\frac{\pi B}{2}\right) + C_B
\sinh\left(\frac{\pi B}{2}\right)\,,  \\[1ex]
 b_1 &= B C_B \cosh\left(\frac{\pi
  B}{2}\right) -B C_A
\sinh\left(\frac{\pi B}{2} \right) \,.
\end{aligned}
\end{equation}
However, just like for the homogeneous solution, this non-normalisable
component of the solution corresponds to a gradient of the $\eta'$ field
and not to an external field (we will provide more evidence of this in
section~\ref{s:inhomogeneous}). Its unusual appearance as a
non-normalisable mode is a consequence of our choice of the $A_z=0$
gauge.

We should note that at this stage we do not impose that $\hat{b}$ or
$\hat{h}$ are of definite parity.  We numerically solve the equation of
motion using the shooting method, and to do this we only need to
specify three undetermined constants $h_0$, $C_A$ and $C_B$ on one
side of the U-shaped brane, as well as the parameter $B$ and
$k$ in the equations. We solve equations for various values of
constants and parameters, but keep only those for which $\hat{h}$ is
normalisable. 
\begin{figure}[t]
\begin{center}
\vspace{1cm}
\includegraphics[width=.43\textwidth]{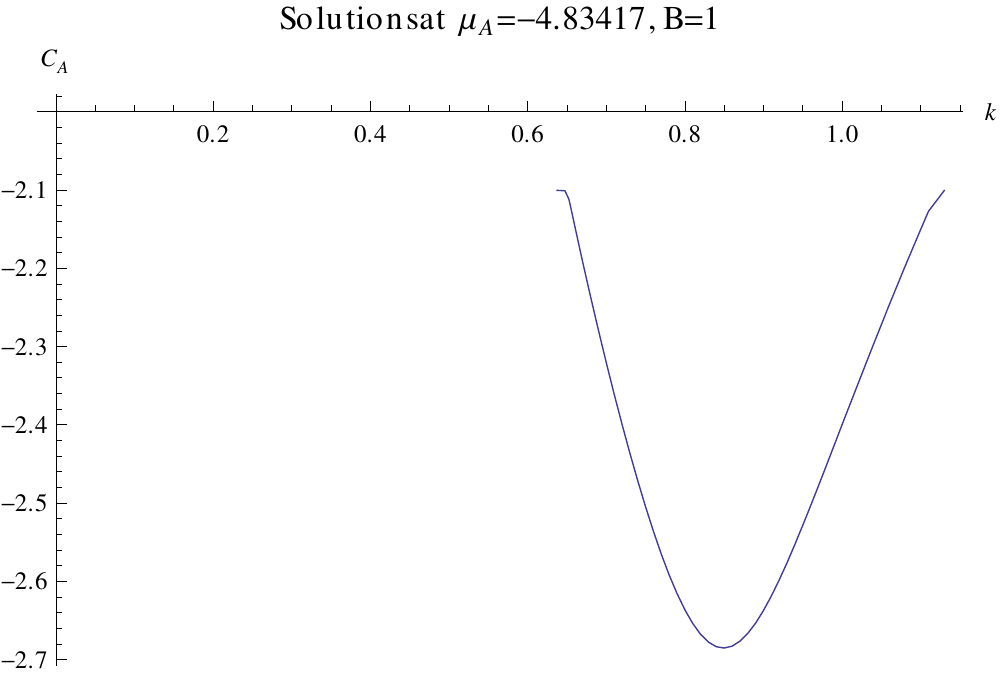}\quad
\includegraphics[width=.43\textwidth]{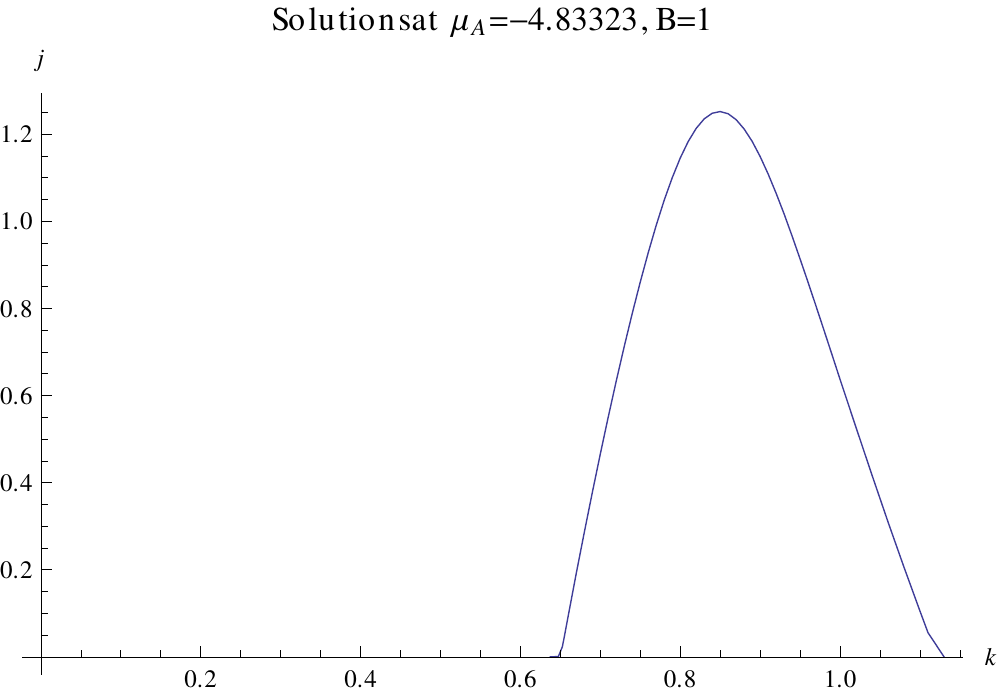}
\end{center}
  \caption{Data for solutions at $C_B=0$, for the (arbitrary) value
    $\mu=-4.83$. The second plot clearly shows that when $B\not=0$, constant $j$
    solutions require a scan through the full parameter space
    $\{h_0,k,C_A,C_B\}$.\label{f:cB0}}
\end{figure}

A first indication of the behaviour of the solutions can be obtained
by looking at the effect of increasing $C_B$ at non-zero value of $B$.
Remember from section~\ref{s:perturbative_stability} that there exists
a marginally stable mode in the spectrum, which follows the downward
bending curve in figure~\ref{f:physical_modes}. Along this curve there
might be a decay to a new ground state. That this indeed happens is
easily confirmed by looking for nonlinear solutions at non-zero $C_A$
and $C_B$. A number of physical solutions is depicted in
figure~\ref{f:varyCB}. They confirm that the marginal modes found in
the previous section in fact correspond to true instabilities and new
ground states. However, these figures say little about the actual
physics, as neither $C_A$ nor $C_B$ are physical parameters.

It is tempting to associate $C_B$ with a baryon chemical potential,
but as we have already mentioned, this is incorrect, as the parameter
$\mu_B$ completely decouples and does not influence the value of the
Hamiltonian. Instead, the correct interpretation is that the physical
parameters in our problem are $\mu_A$ and $j$, which happen to be
related in a non-trivial way to $C_A$ and $C_B$. Families of solutions
(parametrised by~$k$) at fixed $\mu_A$ and $j$ lie on curves in the
space spanned by $C_A$ and $C_B$, which only become straight lines at
$C_B=0$ when $B=0$. This is once more clearly visible in
figure~\ref{f:cB0}. This depicts a family of solutions at constant
$\mu_A$, parametrised by $k$, at $C_B=0$. These clearly do not have a
constant value of $j$. 

In order to find solutions with both $\mu_A$ and $j$ fixed, we need to
allow for a variation of both $C_A$ and $C_B$ as a function of
$k$. Even with the magnetic field fixed to a particular value, this
still means that the normalisable solutions lie on a curve in a
four-dimensional parameter space spanned by $\{h_0, k, C_A, C_B\}$.
This makes a brute force scan computationally
infeasible. Independent of the large dimensionality of this
  problem, we also found that the larger $\mu_A$ cases require
  substantial computational time because of the fact that the
  asymptotic value $h(-\infty)$ varies rather strongly as a function
  of $k$ and the other parameters. In other words, the valley of
  solutions is rather steep and the bottom of the valley, where $h(z)$
  is normalisable at $z\rightarrow-\infty$, is difficult to trace. In
  this respect, it is useful to note that a solution written
  in C\raisebox{.4ex}{\scriptsize ++} using odeint~\cite{Ahnert:2011a}
  and GSL~\cite{Galassi:2009a} outperformed our Mathematica
  implementations by two to three orders of magnitude (!)  in these
  computations. The details of the procedure which we followed can be
found in appendix~\ref{a:scanning}.

We first focus on solutions at fixed and rather arbitrary value $B=1$;
the dependence on $B$ will be discussed in the next section. We will
mainly discuss $j=0$ solutions. Our numerical investigations of $j\neq
0$ solutions show that all of these actually have higher free energy,
and are thus not real ground states of the system (see below).  In figure
\ref{f:energy_overview} we display a set of configurations at constant
$\mu_A$ and vanishing~$j$. Also depicted is the difference of the
Hamiltonian of the homogeneous solution for this pair of $\mu_A, j$
values, and the Hamiltonian of the non-homogeneous solution.

\begin{figure}[t]
\begin{center}
\vspace{-2ex}
\hbox{\hspace{-.6cm}\includegraphics[width=.6\textwidth]{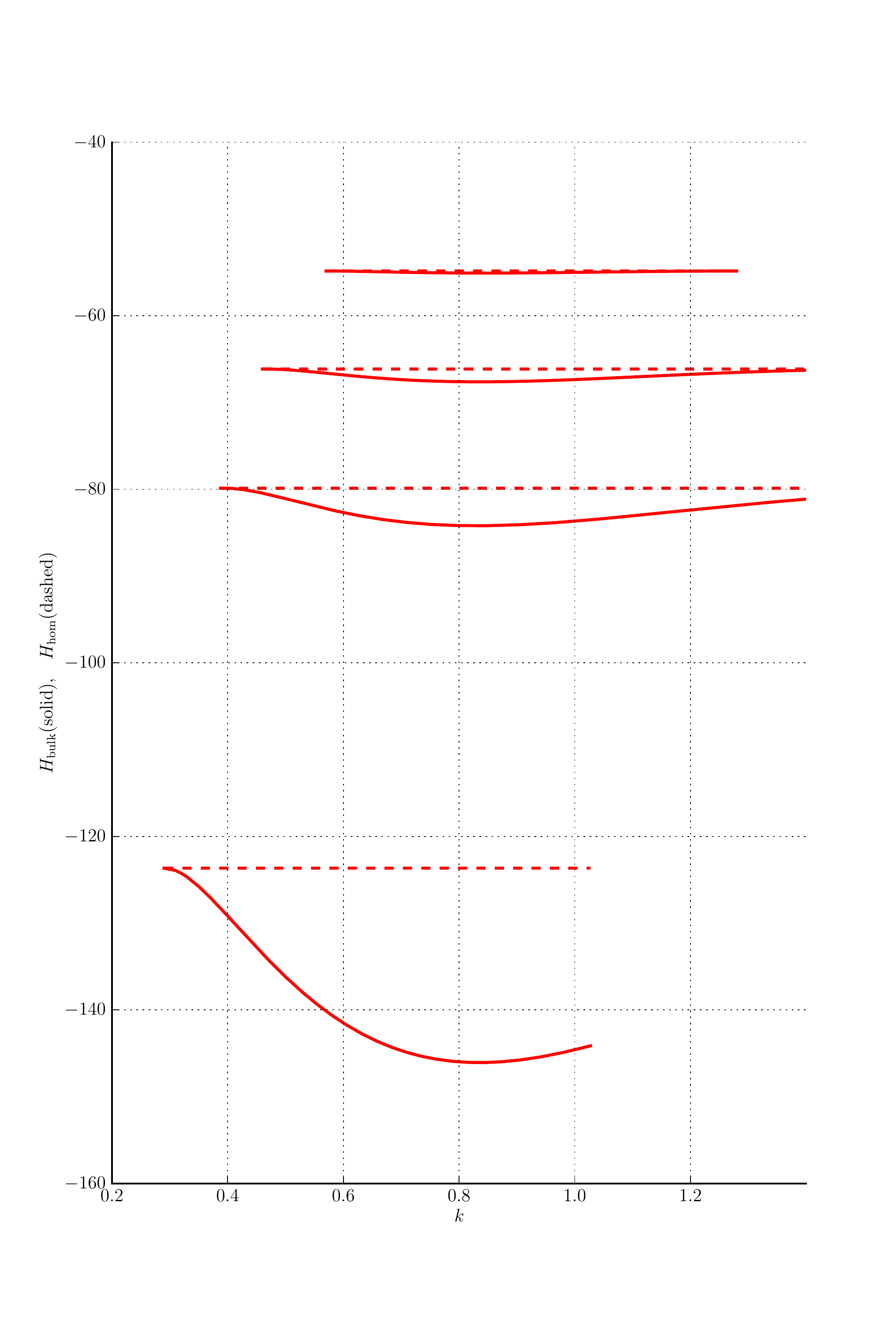}\hspace{-1.5cm}\includegraphics[width=.6\textwidth]{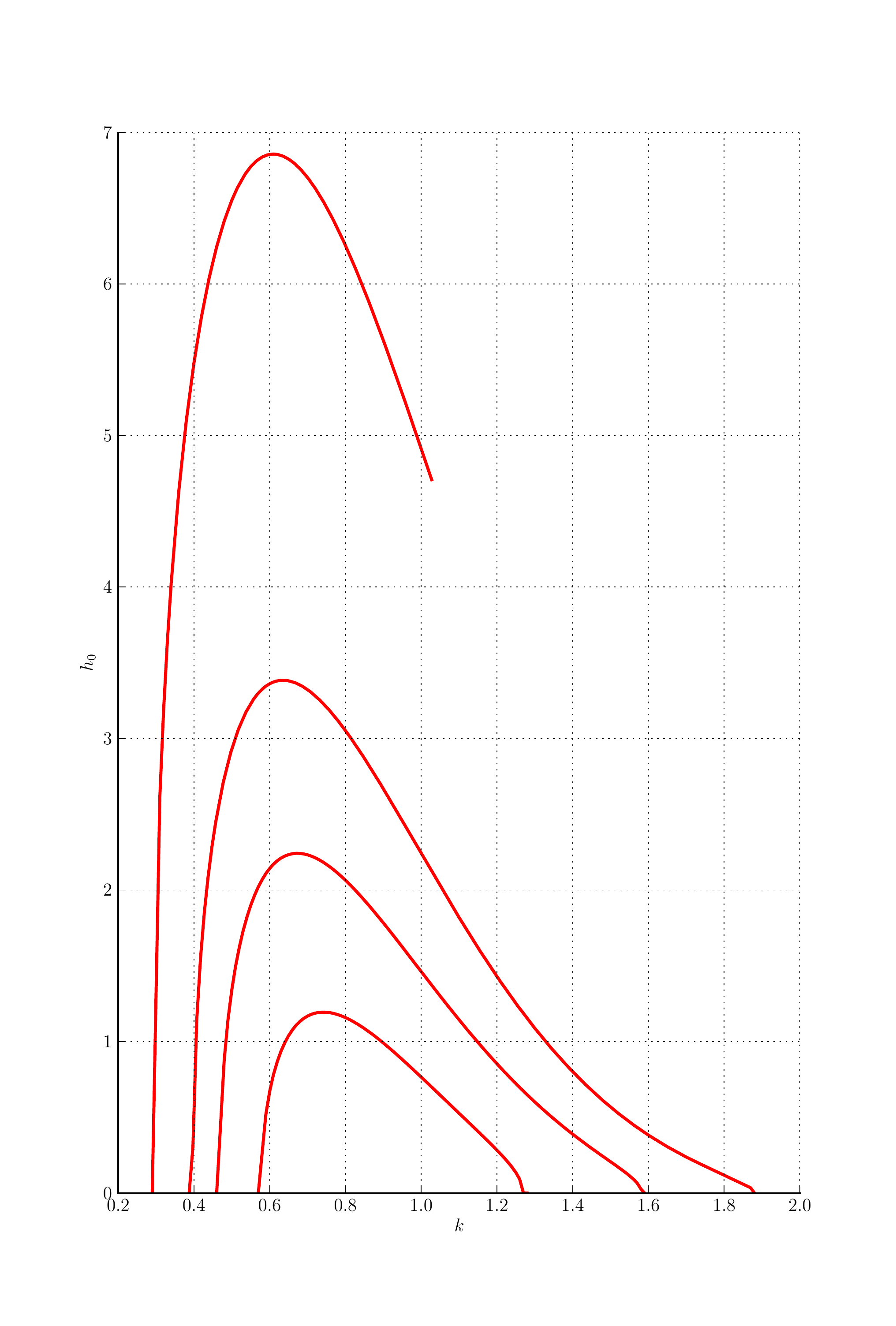}}
\vspace{-2cm}
\end{center}
\caption{Left: Comparison of the energy of the non-homogeneous ground state
  (solid curve) with the energy of the homogeneous state (dashed
  curves) for $j=0, B=1$ and four values of $\mu_A$; from top to
  bottom $\mu_A = -5.5, -6.0, -6.5$ and $-7.5$.
Right: $h_0$ versus $k$, for the same values of the parameters
(from bottom to top).\label{f:energy_overview}}
\end{figure}

As in earlier work without magnetic
field~\cite{Ooguri:2010xs,Bayona:2011ab}, there is a family of
solutions parametrised by the wave momentum $k$, and the physical
solution is the one for which the Hamiltonian is minimised.  Perhaps
somewhat surprisingly, for smaller values of the chemical potential
(we have done computations up to about $\mu_A=-8$) the momentum
$k_{\text{g.s.}}$ at which the non-homoge\-neous ground state attains
its minimum energy is only very weakly dependent on $\mu_A$. Our
numerics for the range of $\mu_A$ up to $-7.5$ show that the ground
state momentum for all these cases equals $0.83$ to within less than
one percent. Furthermore, this ground state momentum is also the same
(to within numerical accuracy) as the ground state momentum for the
$B=0$ case analysed in \cite{Bayona:2011ab}. This is despite the fact
that the actual solutions are quite different. It would be interesting
to understand this behaviour better.

If one evaluates the Hamiltonian of the A'-formalism~\eqref{e:HTheta},
one finds that all non-homogeneous configurations always have higher
energy than the corresponding homogeneous one at the same values of
$\mu_A$ and $j$. We take this as a strong sign that there is still
something un-understood about the A'-formalism, as the analysis of
section~\ref{s:perturbative_stability} clearly indicates a
perturbative instability. It is in principle possible that we are not
looking at the correct ansatz for the ground state, but we consider
this unlikely. A similar statements holds for the B-formalism.

\begin{figure}[t]
\begin{center}
\hbox{\hspace{-.8cm}\includegraphics[width=.6\textwidth]{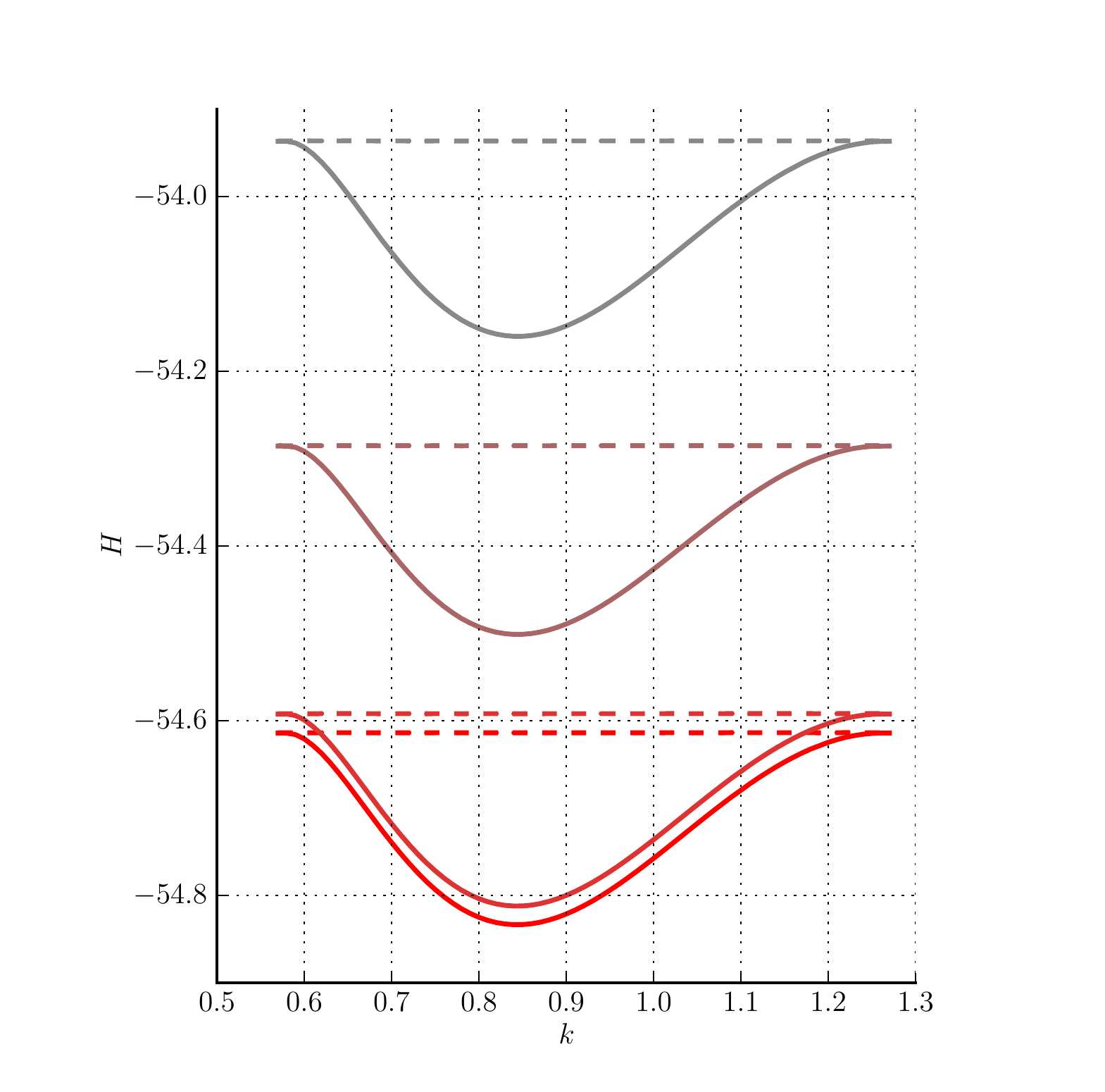}\hspace{-1cm}\includegraphics[width=.6\textwidth]{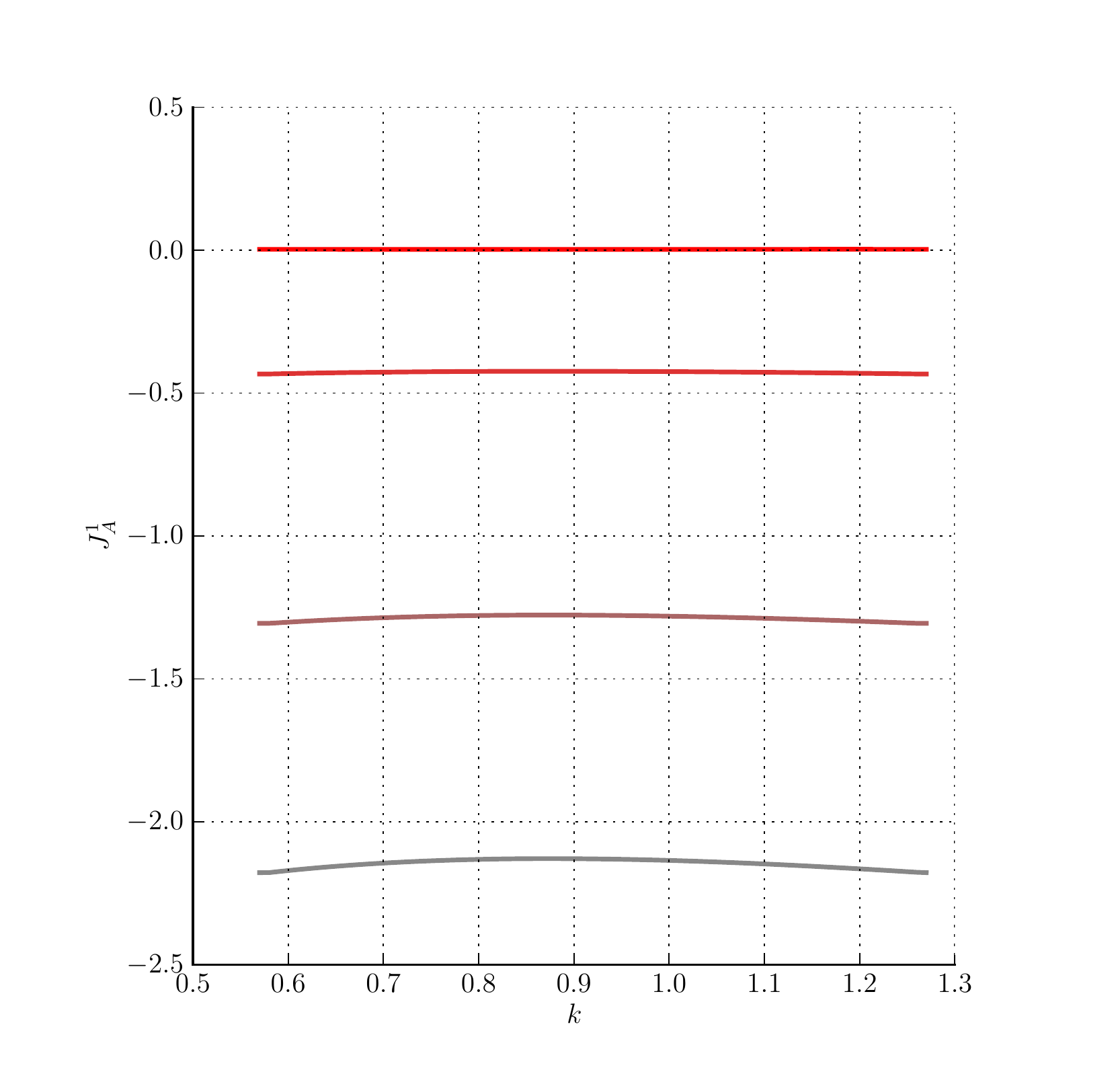}}\vspace{-1cm}
\end{center}
\caption{The dependence of the Hamiltonian and the current $J^1_A$ on
  the parameter $j$ (showing $j=0$, $0.1$, $0.3$ and $0.5$ bottom to
  top on the left, top to bottom on the right). This shows that minimisation with respect to~$j$ drives the
  system to $j=0$, both for the homogeneous and for the
  non-homogeneous state, and that this also drives the $\eta'$-gradient
  sitting in $J^1_A$ to zero. Under $j\rightarrow -j$ the Hamiltonian
  is invariant while $J^1_A$ changes sign. \label{f:vary_j}}
\end{figure}

The dependence of these solutions on $j$ can also be computed, and is
depicted in figure~\ref{f:vary_j}. From those plots one observes two
things. Firstly, a minimisation of the Hamiltonian with respect to $j$
drives the system to $j=0$. Secondly, a non-zero $j$ leads to a
non-vanishing $J^1_A$ current, which can be interpreted as an
$\eta'$-gradient. Together, these observations show that the preferred
non-homogeneous state is one with a vanishing $\eta'$-gradient. 

\begin{figure}[p!]
\begin{center}
\hbox{\hspace{-1.6cm}\includegraphics[width=.45\textwidth]{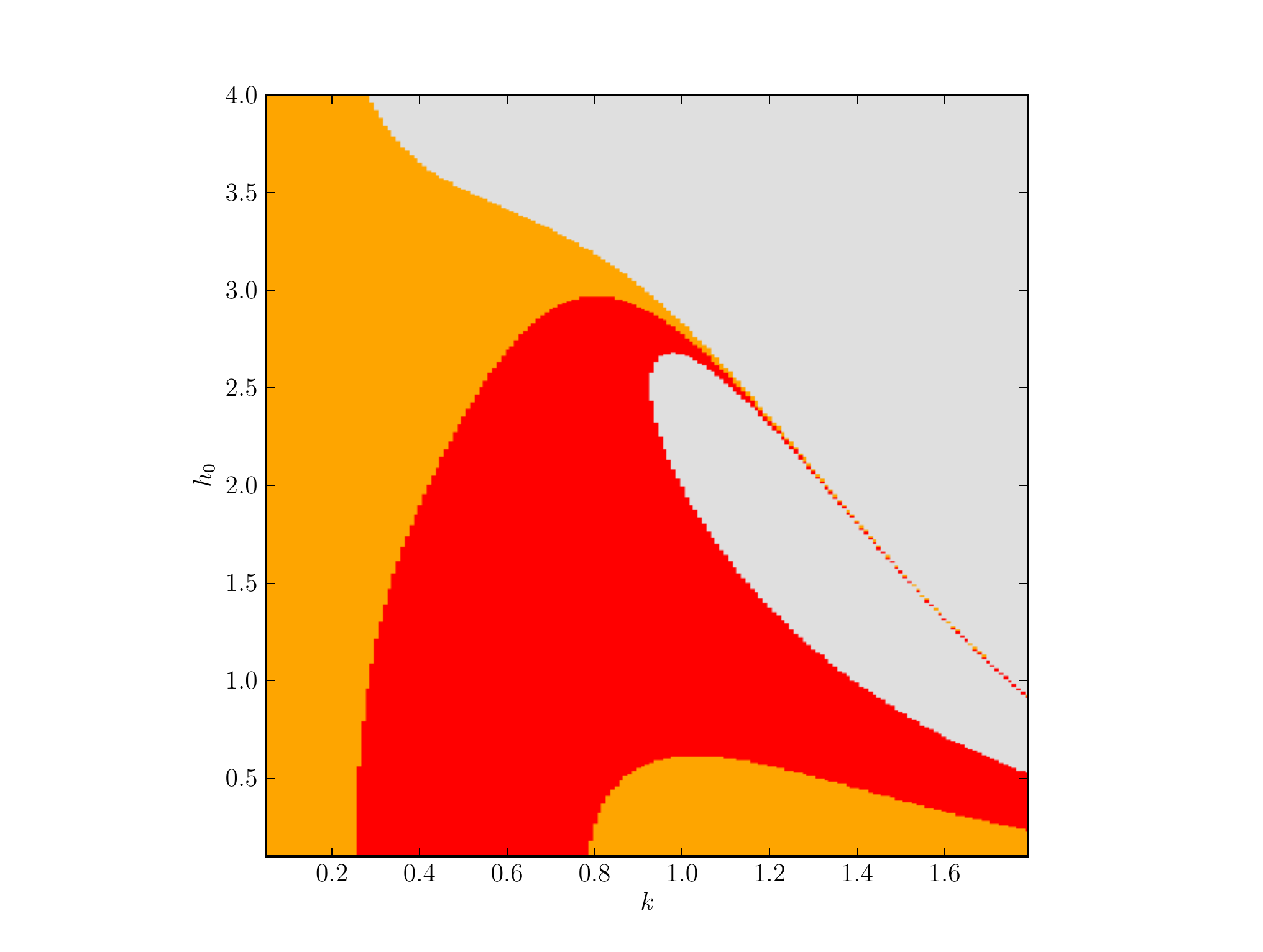}\hspace{-1.2cm}
\includegraphics[width=.45\textwidth]{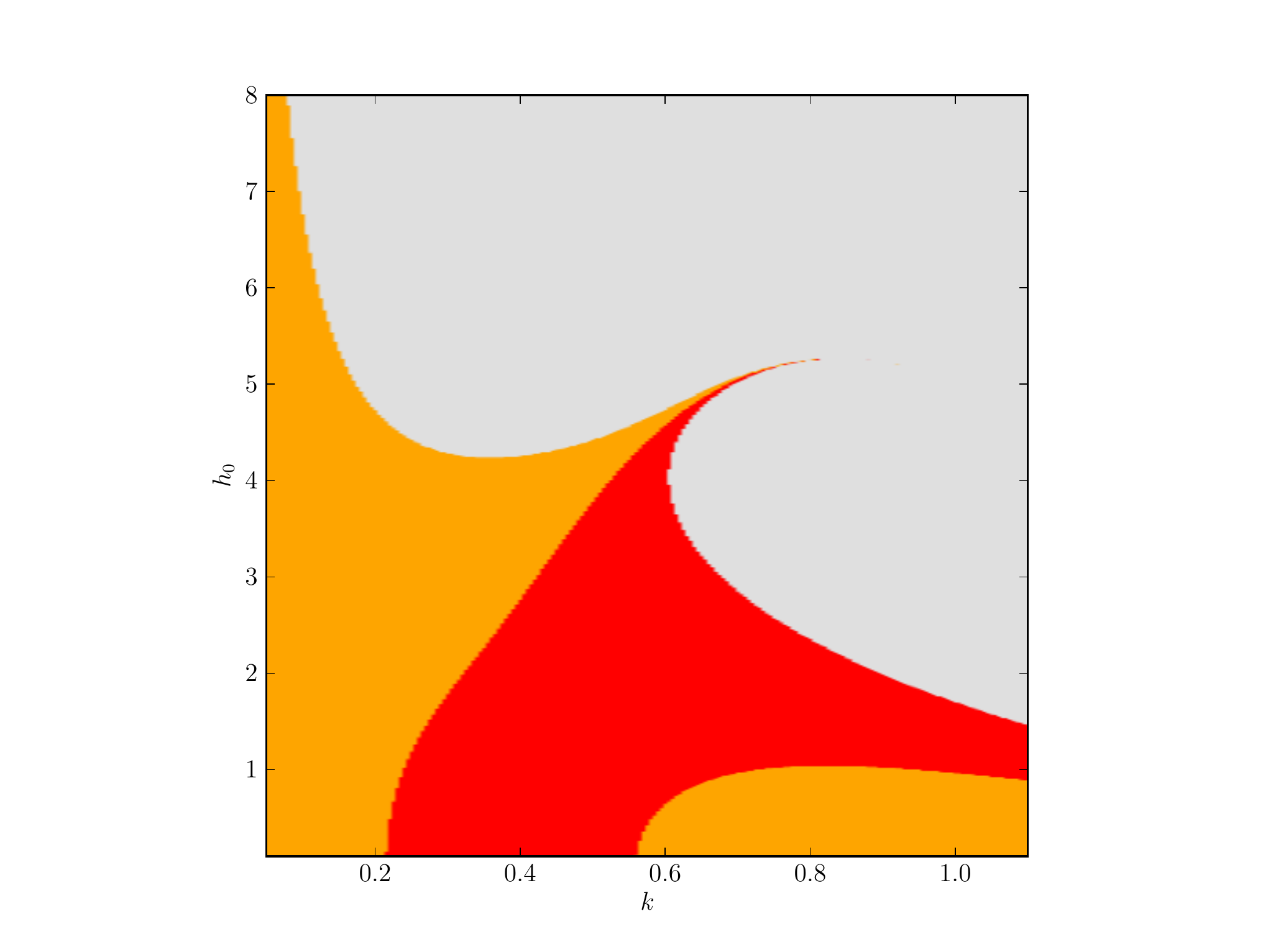}\hspace{-1.2cm}
\includegraphics[width=.45\textwidth]{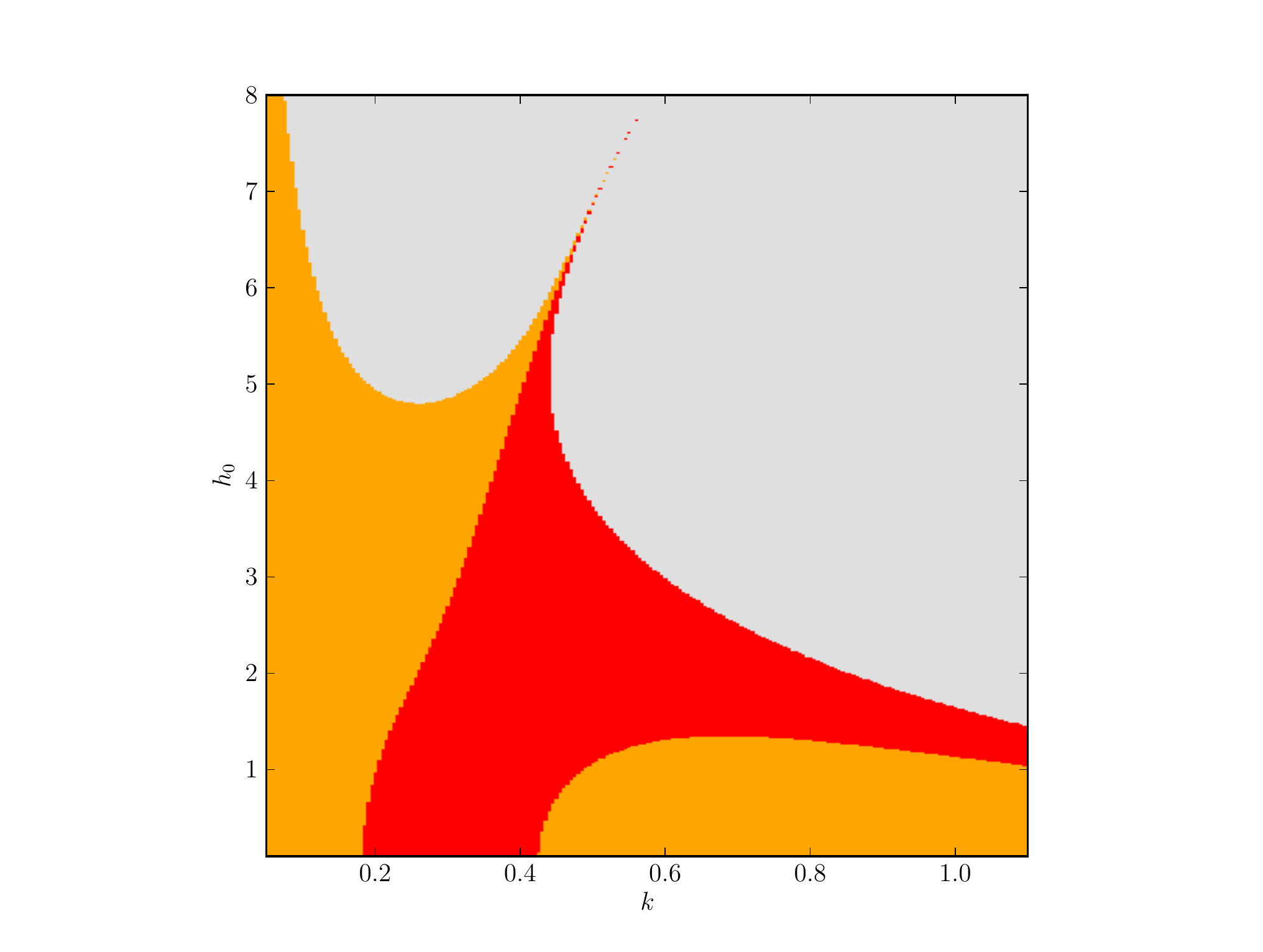}}
\vspace{-1cm}
\end{center}
\caption{The sign of $h(z\rightarrow-\infty)$, where red and orange
  regions denote minus and plus respectively, and grey regions
  correspond to solutions that diverge when
  $z\rightarrow-\infty$. Normalisable non-homogeneous solutions sit at
  the boundary between red and orange regions. The branch that starts
  near $k=0.2$ is the relevant one, as it has the lowest energy. It
  becomes increasingly difficult to find numerically as it is
  sandwiched in a steep valley as $k$ increases. All plots are at
  $C_A=-2.4$, with $B=1.9$, $2.1$ and $2.3$ respectively. Note the
  different axis scales on the first plot.\label{f:tricky_numerics}}
\vspace{1cm}
\begin{center}
\includegraphics[width=.49\textwidth]{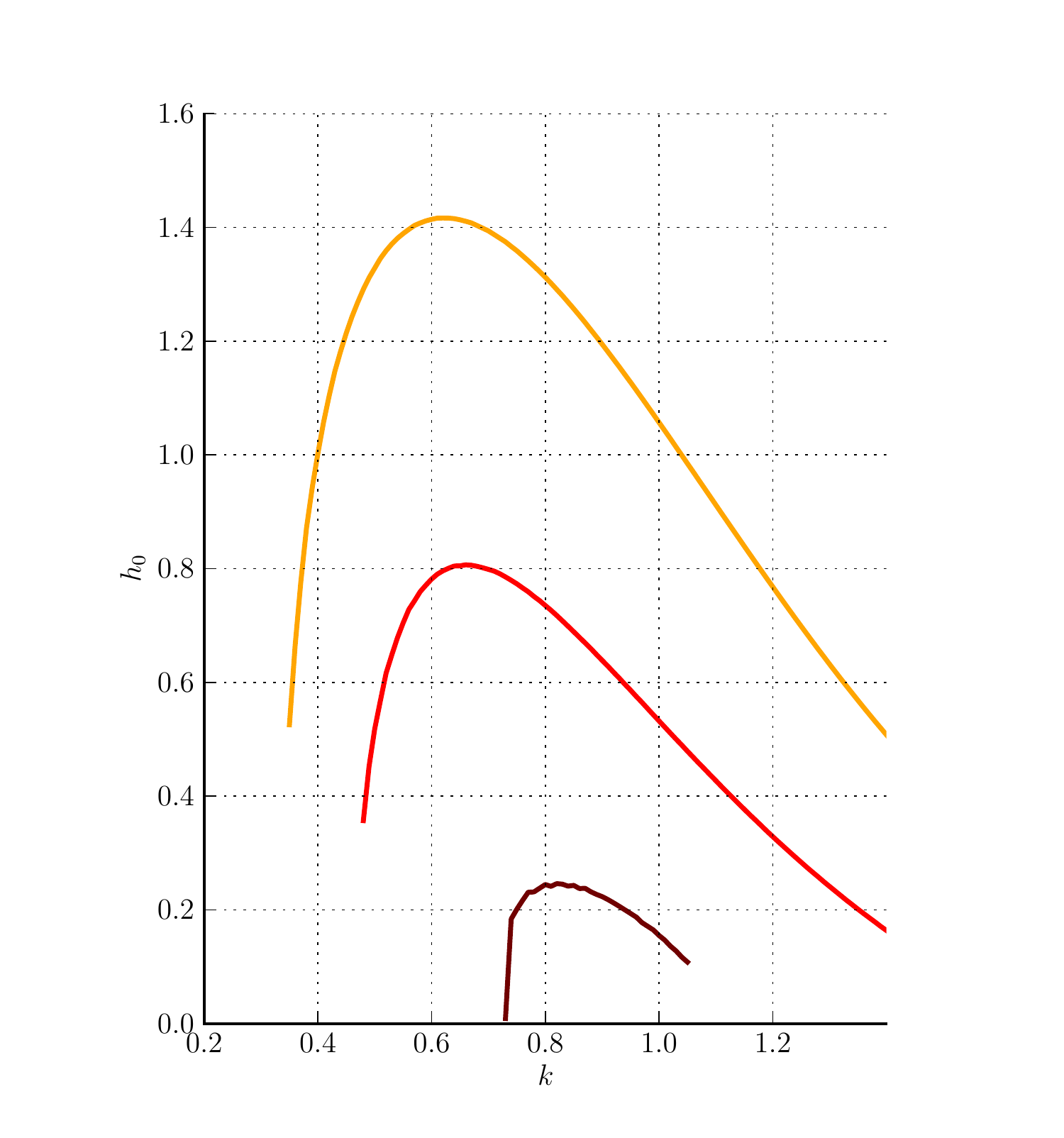}
\includegraphics[width=.49\textwidth]{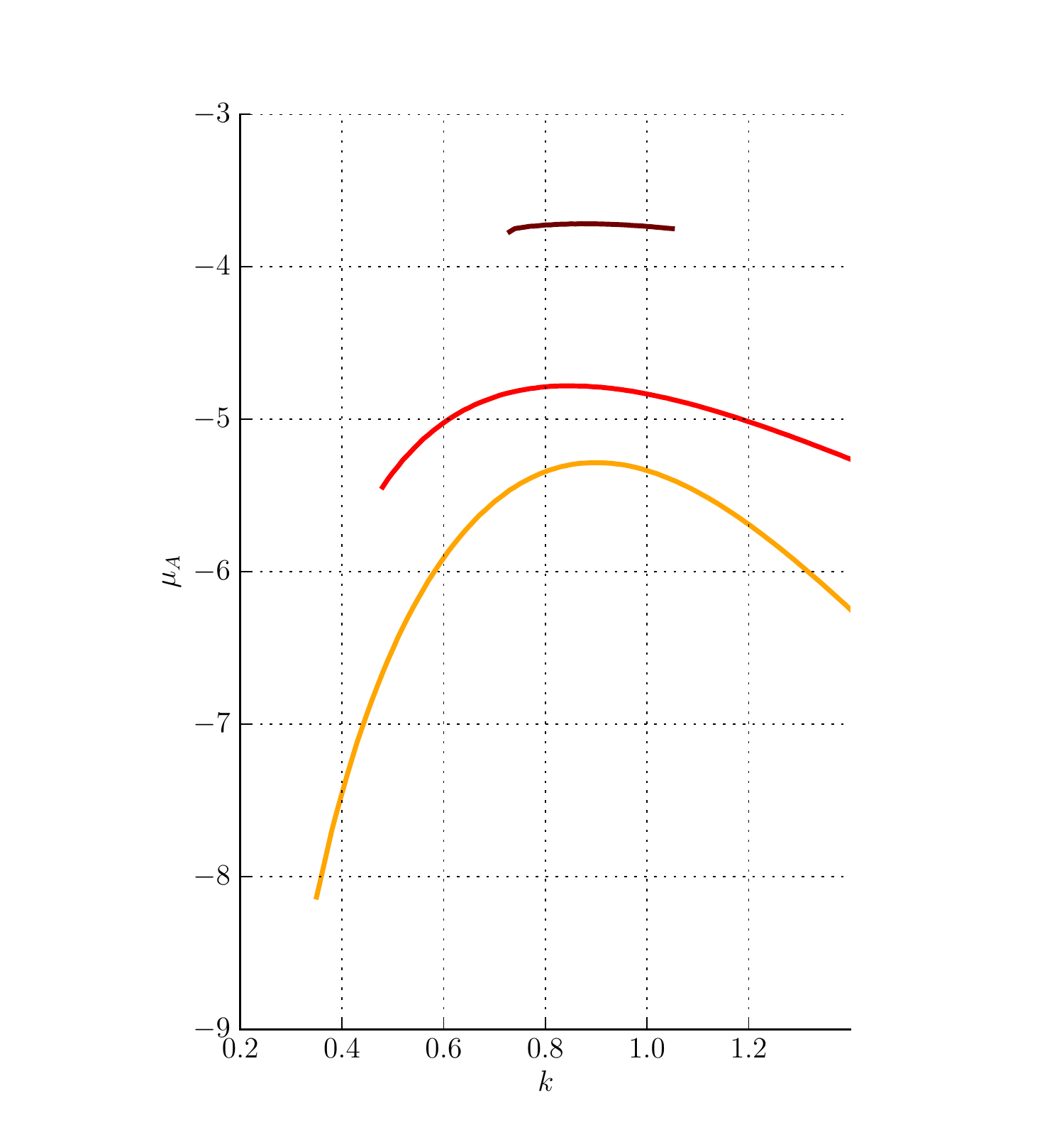}
\caption{Left: The increase of $B$ from $0$ to $1.0$ to $1.5$ (bottom
  black to top orange curves) at $C_B=0$ and fixed $C_A=-2.4$ has the
  effect of raising the curves of normalisable solutions. This implies
  that a larger magnetic field requires a smaller value of $|C_A|$ to trigger
  an instability. Right: However, one cannot conclude anything about the
  effect at constant $\mu_A$ from this, since this physical parameter
  varies over the curves in the figure on the left.
  \label{f:naiveBincrease}}
\end{center}
\end{figure}

\subsection{Critical magnetic field}

\begin{figure}[t]
\begin{center}
\hbox{\hspace{-.5cm}\includegraphics[width=.52\textwidth]{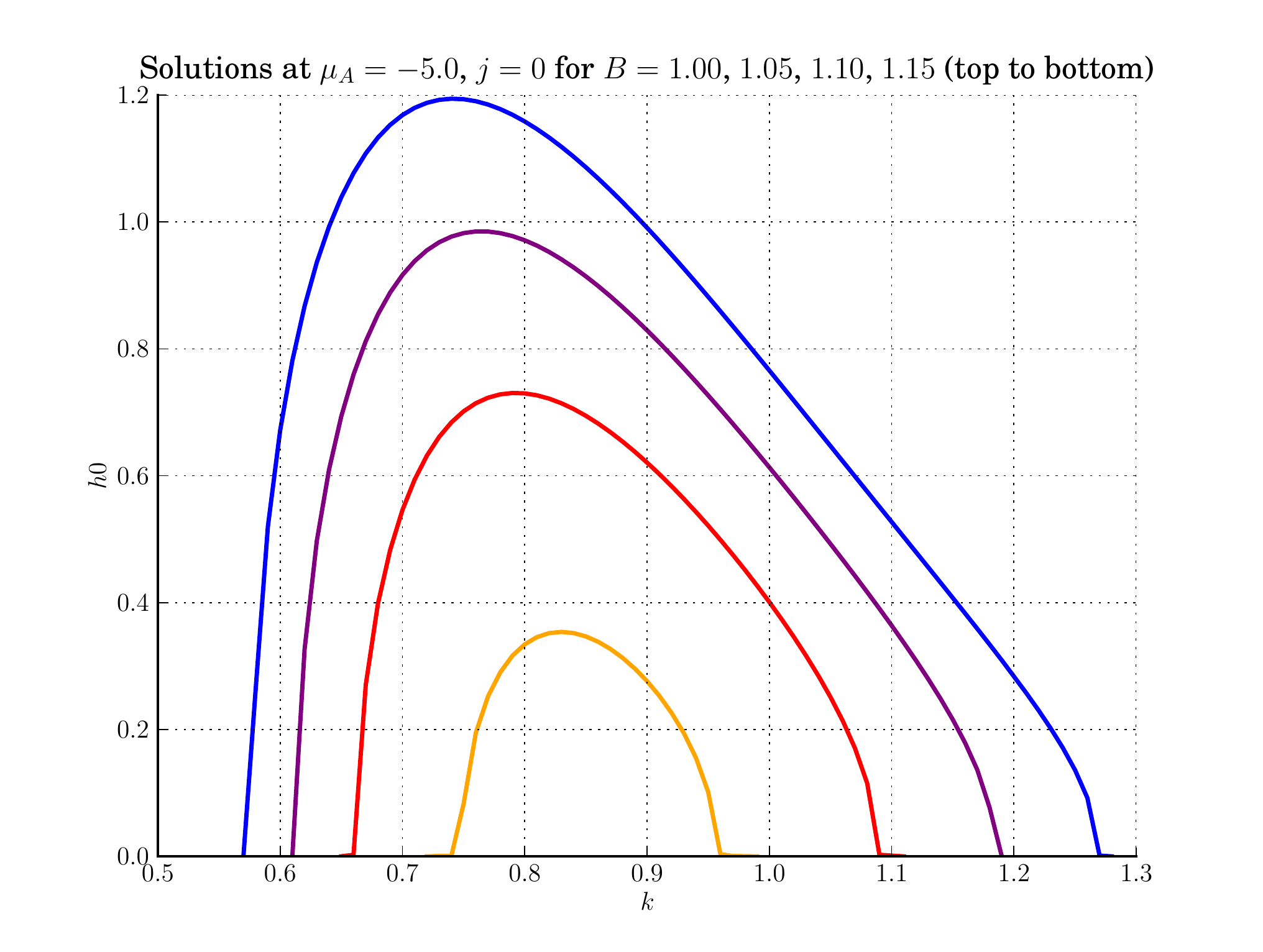}
  \includegraphics[width=.52\textwidth]{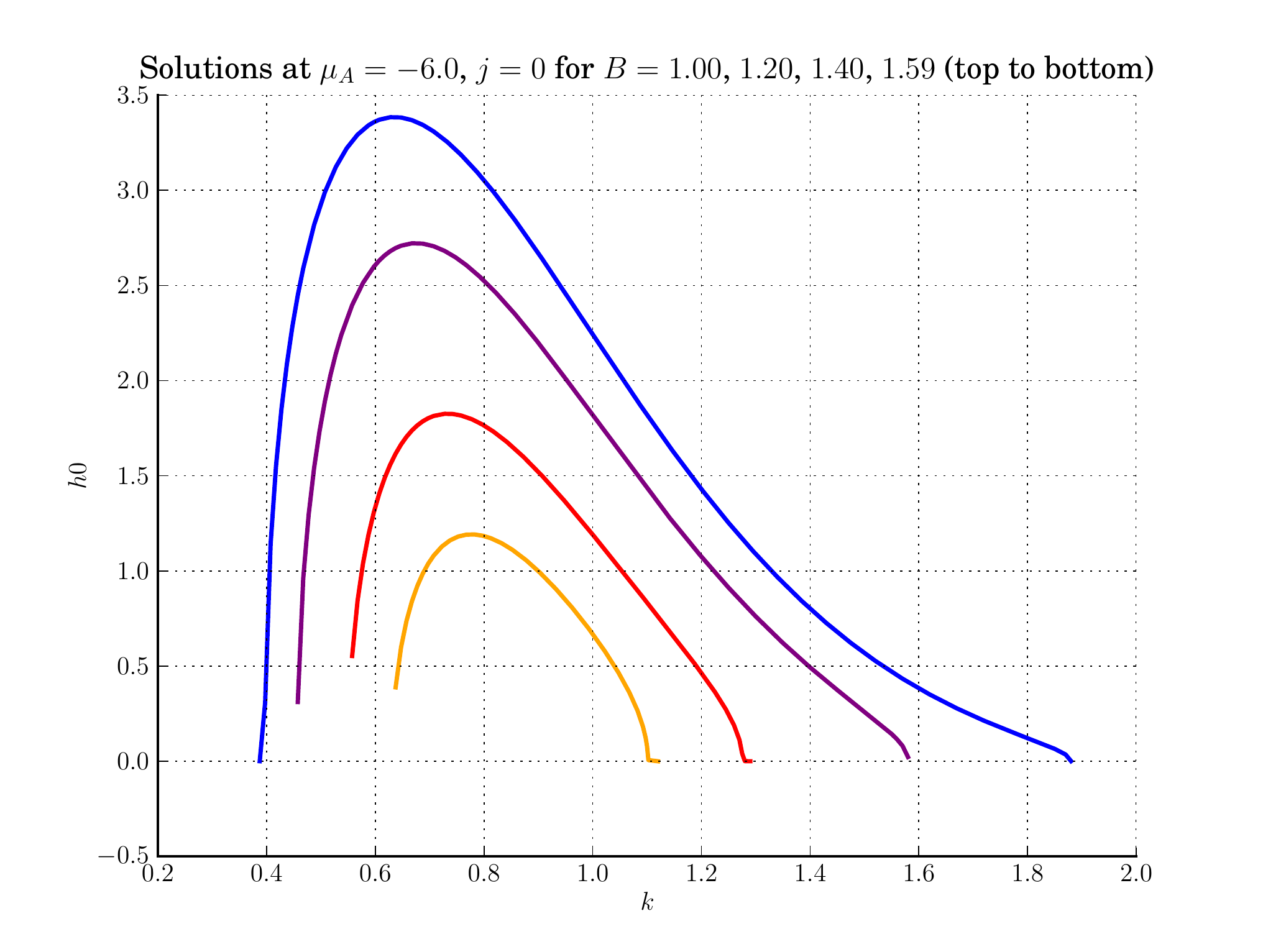}}
\vspace{-1cm}
\end{center}
\caption{Solutions at constant $\mu=-5.0$ (left) and $\mu=-6.0$
  (right) and $j=0$, for various values of the magnetic field. This
  shows that a magnetic field \emph{suppresses} the instability to a
  non-homogeneous ground state, and there is a critical magnetic field
  $B_{\text{crit}}(\mu_A)$ above which the non-homogeneous solution
  ceases to exist.\label{f:B_suppresses_instability}}
\vspace{.5cm}
\begin{center}
  \includegraphics[width=.45\textwidth]{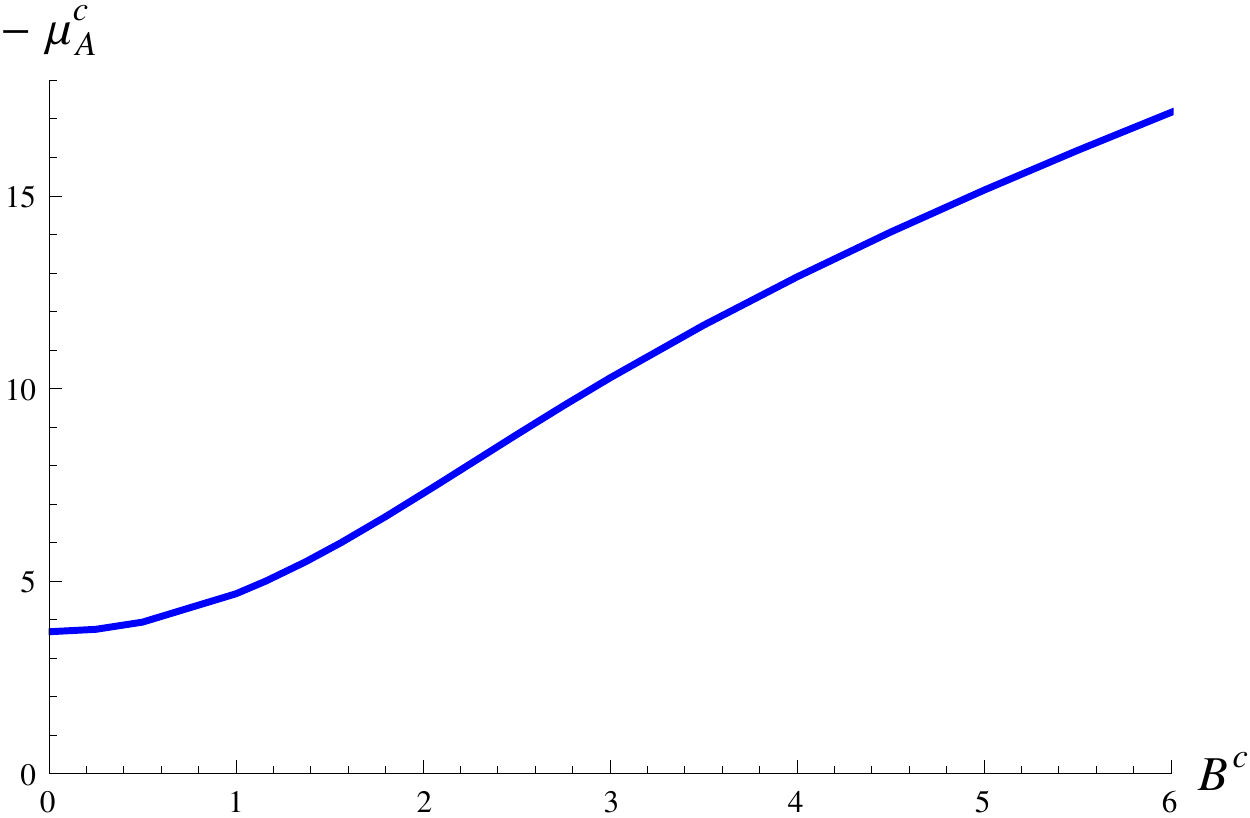}\hspace{1cm}
  \includegraphics[width=.45\textwidth]{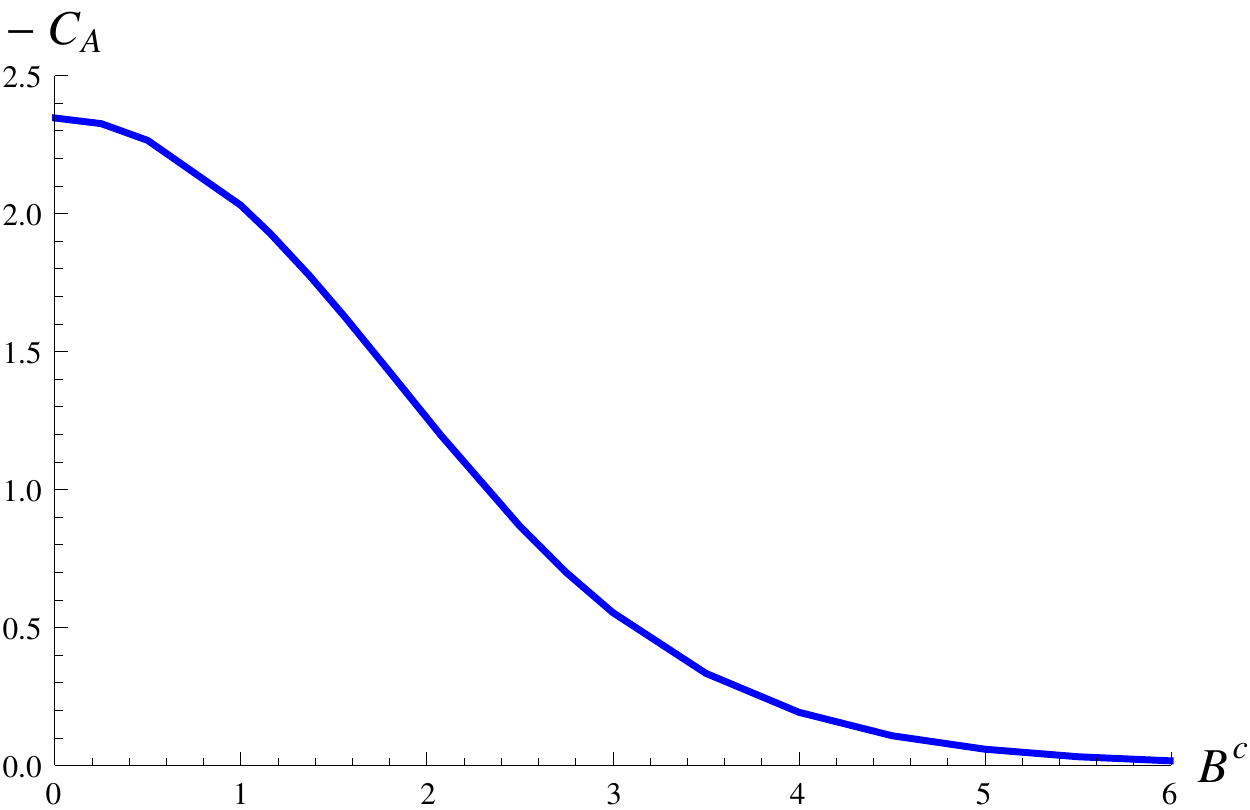}
\end{center}
\caption{Results from a linearised analysis close to the critical magnetic field. Left:
  critical chemical potential (i.e.~the potential for which a
  non-homogeneous ground state first develops) versus critical
  magnetic field (i.e.~the field for which the solution
  disappears). Right: as determined earlier for the non-linear case,
  the parameter $C_A$ goes to zero (asymptotically) as $B$ increases.
  \label{f:close_to_critical_mu_CA}}
\end{figure}

\begin{figure}[t]
\begin{center}
  \includegraphics[width=.45\textwidth]{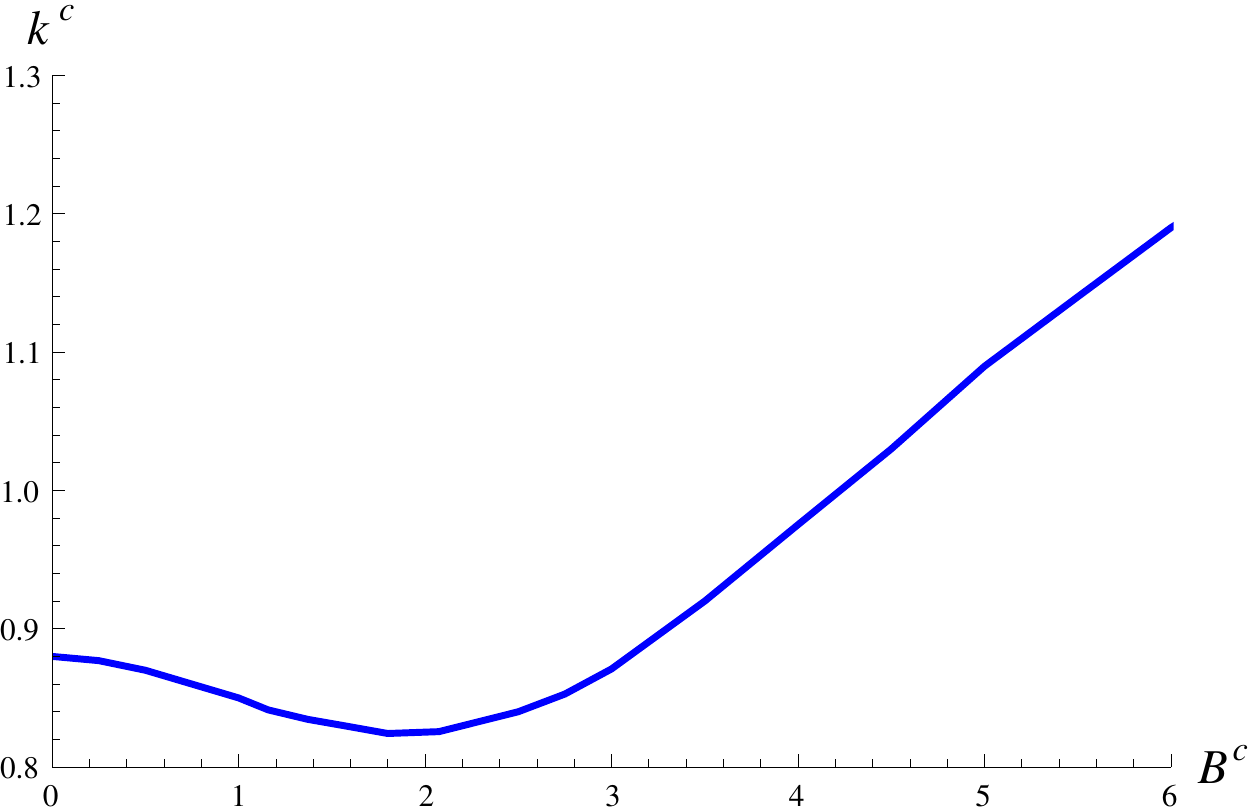}\hspace{1cm}
  \includegraphics[width=.45\textwidth]{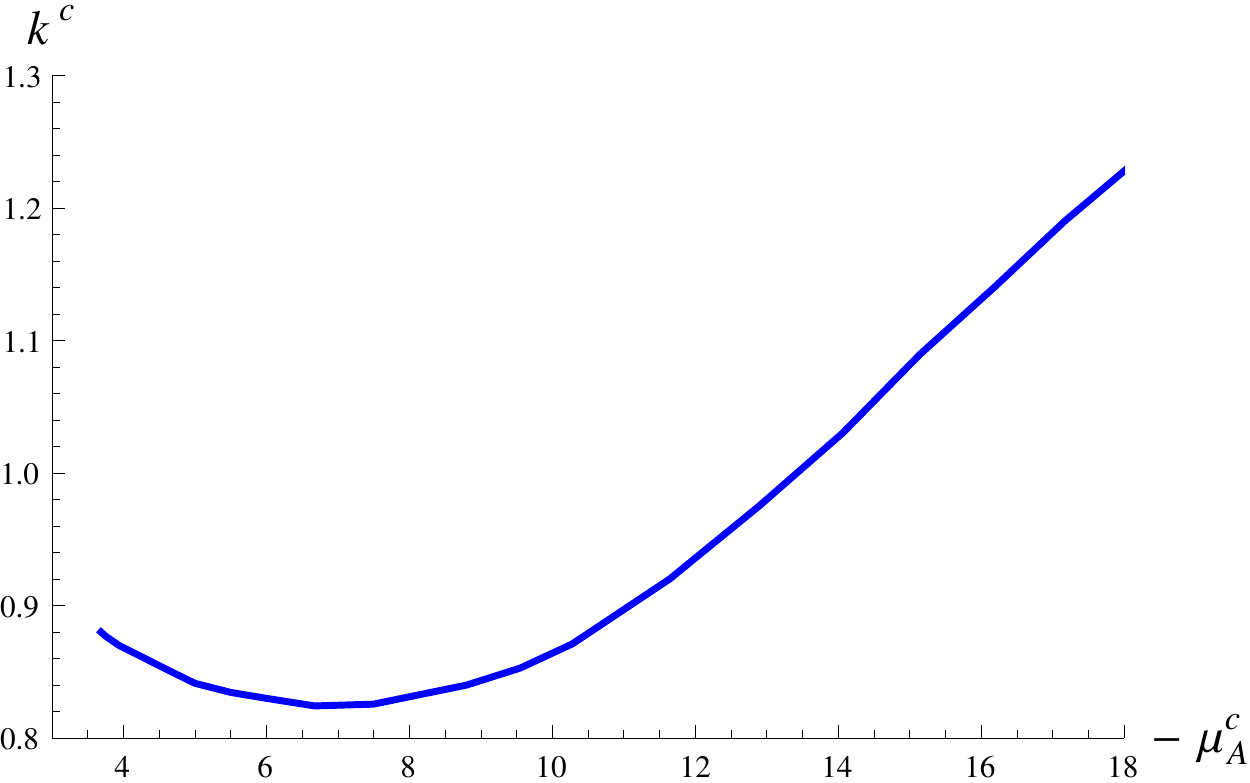}
\end{center}
\caption{Dependence of the ground state momentum $k_{\text{g.s.}}$ of
  the critical solution on the magnetic field or the chemical
  potential, computed in the linear approximation. The non-linear
  solutions which we computed cover the regime up to about
  $B_c=2$. One should keep in mind that the linear approximation
  will break down for sufficiently large values of $B_c$.\label{f:close_to_critical_k}}
\end{figure}

For larger magnitudes of the external magnetic field, the numerics
becomes increasingly expensive as the valleys of solutions to the
equations of motion rapidly become steeper and are more and more
closely approached by regions in parameter space in which no regular
solutions exist (i.e.~regions in which $h(z\rightarrow\infty)$
diverges). See figure~\ref{f:tricky_numerics} for an impression.

A first analysis which one can make is to simply scan for normalisable
solutions at fixed $C_A$ and $C_B$, for increasing values of $B$. This
leads to plots like in figure~\ref{f:naiveBincrease}. One observes
that a larger magnetic field requires a smaller $|C_A|$ for the chiral
spiral solution to exist. In \cite{Kim:2010pu} this effect was seen at
the level of the perturbative instability analysis, and led the
authors to conclude that a magnetic field enhances the
instability. However, this conclusion is premature and actually
incorrect. What one needs to do is to analyse the effect of the
magnetic field on solutions at fixed value of $\mu_A$ and $j$, not at
fixed values of the unphysical parameters $C_A$ and $C_B$. 

If one does this more elaborate analysis, the conclusion is actually
opposite: the magnetic field tends to stabilise the homogeneous
solution, and there exists a critical $B_{\text{crit}}(\mu_A)$ above
which the non-homogeneous solution ceases to exist. This can be seen
in figure~\ref{f:B_suppresses_instability}, where solutions at fixed
$\mu_A$ and $j$ are depicted. By increasing $B$ sufficiently slowly
while keeping the other physical parameters fixed, we can determine
this $B_{\text{crit}}$ in numerical form. 

It is difficult to get a good picture of $B_{\text{crit}}$ versus
$\mu_A$, as the numerics become expensive, for reasons we have
mentioned earlier. However, we can make use of the fact that near the
critical magnetic field, the parameter $h_0$ is small. Assuming that
this implies that the entire function $h(z)$ is small, we can then use
a linear approximation to the equations of motion, which is much
easier to solve. The result of this linear analysis is depicted in
figures~\ref{f:close_to_critical_mu_CA}
and~\ref{f:close_to_critical_k}. The latter shows that the ground
state momentum $k_{\text{g.s.}}$ is actually not as flat as the
non-linear computations suggests.

For larger $B$, one should keep in mind that the results may be
invalidated for a variety of reasons. Firstly, the linear
approximation may break down, as smallness of $h_0$ does not
necessarily imply smallness of the full function $h(z)$. Secondly, we
have seen that the valley near the non-linear solutions becomes very
steep for large values of $B$, and hence the linear solution may
deviate quite strongly from the non-linear one for large $B$. Thirdly,
when $B$ is large, DBI corrections to the equations of motion may
become relevant, as higher powers of the field strength are no longer
necessarily small. For these reasons, one should be careful
interpreting the large $B_c$ regime of
figure~\ref{f:close_to_critical_k}.

\section{Conclusions and open questions}

We have analysed in detail the instability of the Sakai-Sugimoto model
in the presence of a chiral chemical potential $\mu_A$, an
$\eta'$-gradient $j$ and a magnetic field $B$. We have shown that the
presence of marginally stable modes (overlooked in~\cite{Kim:2010pu})
is a signal for decay towards a non-homogeneous chiral spiral like
ground state. Minimising the Hamiltonian on a constant $\mu_A$,
constant $j$ curve leads to a unique ground state momentum
$k_{\text{g.s.}}$ (see figure~\ref{f:energy_overview}), which for
small $B$ is only weakly dependent on $\mu_A$. Increasing the magnetic
field to sufficiently large values suppresses the instability and
drives the system back to the homogeneous phase
(figure~\ref{f:close_to_critical_k}).  This result may be related to
the effective reduction of the Landau levels in a strong magnetic
field (see \cite{Basar:2012gm}). 

A linear analysis suggests that the ground state momentum may have a
non-trivial dependence on the chemical potential, and might be
compatible with a linear scaling at large $\mu_A$ (as in the
Deryagin-Grigoriev-Rubakov non-homogeneous \mbox{large-$N_c$} QCD ground
state~\cite{Deryagin:1992rw}).  However, substantial additional work is
necessary to determine whether this is indeed happening in the full
non-linear theory.

We have found that a recent proposal for the correction of the
currents~\cite{Landsteiner:2011tf,Gynther:2010ed}, while capable of
producing the chiral magnetic effect both in the homogeneous and
non-homogeneous ground states, is incomplete. It is a) not unique and
b) leads to a Hamiltonian which does not prefer the non-homogeneous
ground state. We emphasise that this is a problem with the currents
and boundary terms in the Hamiltonian, as the perturbative instability
analysis is not affected by these corrections, and neither are the
non-linear solutions.

\acknowledgments

We thank Karl Landsteiner for discussions
about~\cite{Gynther:2010ed,Landsteiner:2011tf}. This work was
partially supported by STFC grant ST/G000433/1. The work of MZ was
supported by an EPSRC Leadership Fellowship.

\vfill\eject
\appendix

\section{Appendix: Technical details}
\subsection{The Hamiltonian after the anomaly corrections}

Here we will give a detailed derivation of the Hamiltonian for the
Sakai-Sugimoto system taking special care about surface terms that are
present, as well as taking into account the corrections to the action
coming from the anomaly. 

After anomaly in the system have been corrected by  inclusion of
extra terms in the action, the full Langrangian density can be written
as 
\begin{multline}
 \tilde {\cal L} = - \kappa \sqrt{-g} \left [ {\cal
    F}^{0a}_V {\cal F}_{0a}^V + \frac12 {\cal F}^{ab}_V {\cal
    F}_{ab}^V + {\cal F}^{0a}_A {\cal F}_{0a}^A + \frac12 {\cal
    F}^{ab}_A {\cal F}_{ab}^A \right ] \\[1ex] 
+ \frac{\alpha}{4}
\epsilon^{0abcd} \Big [ {\cal A}_0^A ( 3 {\cal F}_{ab}^V {\cal
    F}_{cd}^V + {\cal F}_{ab}^A {\cal F}_{cd}^A ) + 4 {\cal A}_b^A ( 3
  {\cal F}_{0a}^V {\cal F}_{cd}^V + {\cal F}_{0a}^A {\cal F}_{cd}^A )
  \Big ] \,.  
\end{multline}

The conjugate momenta associated with the vector and axial gauge
fields take the form
\begin{equation}
\begin{aligned}
\tilde \Pi^a_V &= \frac{
  \partial \tilde L}{ \partial (\partial_0 {\cal A}_a^V )} = - 2 \kappa \sqrt{-g} {\cal F}^{0a}_V + 3 \alpha
\epsilon^{0abcd} {\cal A}_b^A {\cal F}_{cd}^V \,, \\[1ex]
 \tilde \Pi^a_A &= \frac{ \partial \tilde L}{ \partial (\partial_0 {\cal A}_a^A )} = - 2 \kappa \sqrt{-g} {\cal F}^{0a}_A + \alpha
\epsilon^{0abcd} {\cal A}_b^A {\cal F}_{cd}^A \,.  
\end{aligned}
\end{equation}
Hence, the on-shell Hamiltonian takes the form  $\tilde H = \tilde H_{\text{Bulk}} + \tilde H_{\text{Bdy}}$
\begin{eqnarray}
\tilde H_{\text{Bulk}}  &=& \kappa \int\! {\rm d}^3x {\rm d}z    \sqrt{-g} \left [ - {\cal F}^{0a} {\cal F}_{0a} + \frac12 {\cal F}^{ab} {\cal F}_{ab}  \right ]   \\ 
\tilde H_{\text{Bdy}} &=&  \int\! {\rm d}^3x {\rm d}z \, \partial_a \left [ \tilde \Pi^a_V {\cal A}_0^V + \tilde \Pi^a_A {\cal A}_0^A \right ] \, .
\end{eqnarray}
where  we have used the gauge field equations for the time component of gauge potential (generalised Gauss law).

For our inhomogeneous ansatz  the conjugate momenta simplify to
\begin{equation}
\begin{aligned}
\tilde \Pi^z_V &= 2 \kappa \sqrt{-g} g^{zz} g^{00} \partial_z f_V - 6 \alpha (a_A B + k h_A h_V) = 0 \,  \, ,\\[1ex]
\vec{\tilde \Pi}_V &= - 6 \alpha \left [ a_A (\partial_z h_V) - h_A (\partial_z a_V) \right ] \left [ \sin(k x_1) \hat x_2 + \cos(kx_1) \hat x_3 \right ] \, , \\[1ex]
\tilde \Pi^z_A &=  2 \kappa \sqrt{-g} g^{zz} g^{00} \partial_z f_A - 2 \alpha k h_A^2 = 6 \alpha  B a_V + 3 \alpha k h_V^2 + \alpha k h_A^2  - 2 \kappa \tilde \rho  \, , \\[1ex]
\vec{\tilde \Pi}_A  &=  -2 \alpha \left [ a_A (\partial_z h_A) - h_A (\partial_z a_A) \right ] \left [ \sin(k x_1) \hat x_2 + \cos(kx_1) \hat x_3 \right ] \, .
\end{aligned}
\end{equation}
Using this result and the boundary conditions we have imposed on the
fields, one gets the boundary term of the Hamiltonian
\begin{equation}
\tilde H_{\text{Bdy}} = - 2 \kappa \tilde \rho \int\! {\rm d}^3x {\rm d}z \partial_z f_A = 4 \kappa V_x \tilde \rho \mu_A = - V_x \tilde J^0_A \mu_A \,.
\end{equation}
The bulk term  simplifies for our ansatz to 
\begin{align}
\tilde H_{\text{Bulk}} &= \kappa \int {\rm d}^3 x {\rm d}z \,
\Big \{  - \sqrt{-g} g^{zz} g^{00} (\partial_z f)^2 +  \sqrt{-g} g^{zz} g^{xx} \left [ (\partial_z h)^2  + (\partial_z a)^2 \right ] \\[1ex]
&+ \sqrt{-g} (g^{xx})^2 \left[ B^2 + k^2  h^2 \right ] \Big \} \, .
\end{align}
In terms of the variables (\ref{hatdimless}) the bulk and boundary terms take the form 
\begin{align}
\tilde H_{\text{Bulk}} &=
{\cal H}_0 \int\! {\rm d}z\, \Big \{ \frac{1}{K_z} \left [ \hat b + \frac{\hat \alpha}{2} \hat k \hat h^2 \right ]^2 
+  K_z \left [ (\partial_z \hat h)^2  + \left (\frac{\partial_z \hat b}{\hat \alpha \hat B} \right )^2 \right ] \\[1ex] 
&+ K_z^{-1/3} \hat k^2 \hat h^2  \Big \}
+  \tilde H_{\text{Div}}   \,, \\[1ex]
\tilde H_{\text{Bdy}} &= 4 {\cal H}_0 \hat \rho \hat \mu_A \, ,
\end{align}
where ${\cal H}_0 = M_{\text{KK}}^4 V_x \bar \lambda^3 N_c /(8 \pi^2) $, $\hat b = \hat B \hat a - \hat \rho$ and $H_{\text{Div}} = {\cal H}_0 \hat B^2 \int\! {\rm d}z\, K_z^{-1/3}$.
For the homogeneous case we get simple analytic result 
\begin{align}
 \tilde H_{\text{Bulk}} &= 2 {\cal H}_0 \hat B \coth (\frac{\pi}{2}
 \hat B) (\hat \mu_A^2 + \hat j^2) + \tilde H_{\text{Div}}\nonumber\\
&\qquad\qquad  = {\cal H}_0 (\hat C_A^2 + \hat C_B^2 ) \frac{ \sinh(\hat B \pi)}{\hat B} + \tilde H_{\text{Div}} \, , \\[1ex]
\tilde H_{\text{Bdy}} &= -4 {\cal H}_0 \hat B \coth (\frac{\pi}{2} \hat B) \hat \mu_A^2 = - 2 {\cal H}_0 \hat C_A^2 \frac{ \sinh(\hat B \pi)}{\hat B} \,.
\end{align}

\subsection{The B-formalism ansatz}
\label{a:Bformalism}

In formalism B the time component of the chiral gauge field is zero at
the boundary. An ansatz that preserves the field strengths (and hence
the equations of motion) is given by
\begin{equation}
\begin{aligned}
{\cal A}^V_z &= - t \, \partial_z f_V  \quad , \quad {\cal A}^A_z = - t \, \partial_z f_A  \, , \\[1ex]
{\cal A}_0^V &= 0 \quad , \quad  {\cal A}_0^A = 0 \, , \\[1ex]
\vec{\cal A}_V &= \frac{B}{2} \left [ - x_3 \hat x_2 + x_2 \hat x_3 \right ] + h_V(z) \left [ \cos (k x_1) \hat x_2 - \sin (k x_1) \hat x_3 \right ] + a_V(z) \hat x_1 \, ,\\[1ex]
\vec{ \cal A}_A &= a_A (z) \hat x_1 +  h_A(z) \left [ \cos (k x_1) \hat x_2 - \sin (k x_1) \hat x_3 \right ] \, , \label{gaugeansatzformB}
\end{aligned}
\end{equation}
which should be used together with the boundary conditions 
\begin{equation}
\begin{aligned}
{\cal A}^V_z (z \to \pm \infty) &= {\cal A}^A_z (z \to \pm \infty) = 0 \, , \\[1ex]
h_V( z \to \pm \infty) &=  h_A( z \to \pm \infty) = 0 \, , \\[1ex]
a_V (z \to \pm \infty) &=  a_A( z \to \pm \infty ) = \mp j \,  \label{BCforansatzformB}.
\end{aligned}
\end{equation}
The components of the vector and axial currents take the same form as
those in the A-formalism, given in \eqref{allcurrents}, with the
exception of $\tilde{J}^1_V$ which reads
\begin{equation}
\tilde J^1_V =  - 4 \kappa \lim_{z \to \infty} \left [ \sqrt{-g} g^{zz} g^{xx} \partial_z a_V \right ]   
= 12 \alpha B \lim_{z \to \infty} f_A  = - 12 \alpha B \mu_A \,  .
\end{equation}
This in particular implies that there is a non-zero chiral magnetic
effect in formalism~B.

\subsection{Scanning parameter space for solutions}
\label{a:scanning}

We recall from the main text that we aim for the minimisation of the
Hamiltonian as a function of the condensate momentum $k$, on curves
which have fixed values of $\mu_A$ and $j$. Unfortunately, the latter
two parameters are determined only indirectly, after a solution has
been found; their dependence on the shooting parameters $h_0$, $C_A$
and $C_B$ is not known analytically. Minimising the Hamiltonian over
curves at fixed value of $C_A$ and $C_B$ would in general not be
equivalent (i.e.~wrong).

One could in principle make a fine-grained lattice scan for solutions
in the parameter space spanned by $\{ h_0, k, C_A, C_B\}$, and then
consider only those points which have a certain fixed value of $\mu_A$
and $j$. This, however, is computationally extremely expensive, and
wastes a lot of time on regions of parameter space which will never be
used. We therefore follow a different approach.

The general problem which we need to solve for an efficient
determination of the required curves is the following. We have a
$d$-dimensional parameter space of configurations, and physical
solutions are those in which $d-1$ functions of the configuration are
vanishing simultaneously. One of these functions is always the value
of the function $h(z)$ evaluated at $-\infty$, as this imposes
normalisability. Among the other functions we have, for instance, the
value of $\mu_A$ for this configuration minus some fixed reference
value, in case we want to scan for solutions at this fixed value of
the chemical potential. This `isocurve tracing' problem is most easily
implemented by starting from one known point on which the functions
are simultaneously zero, and then using a version of the
Newton-Raphson method to find the location of a neighbouring
simultaneous zero.\footnote{General purpose Mathematica and
  C\raisebox{.3ex}{\scriptsize ++} implementations of this isocurve
  tracing algorithm are available upon request from the authors.} If we
have $n$ lattice points in every direction of parameter space, this
brings the computational cost down from being order $O(n^d)$ to the
much more tractable $O(n)$.

\begin{figure}[t]
\begin{center}
\includegraphics[width=.6\textwidth]{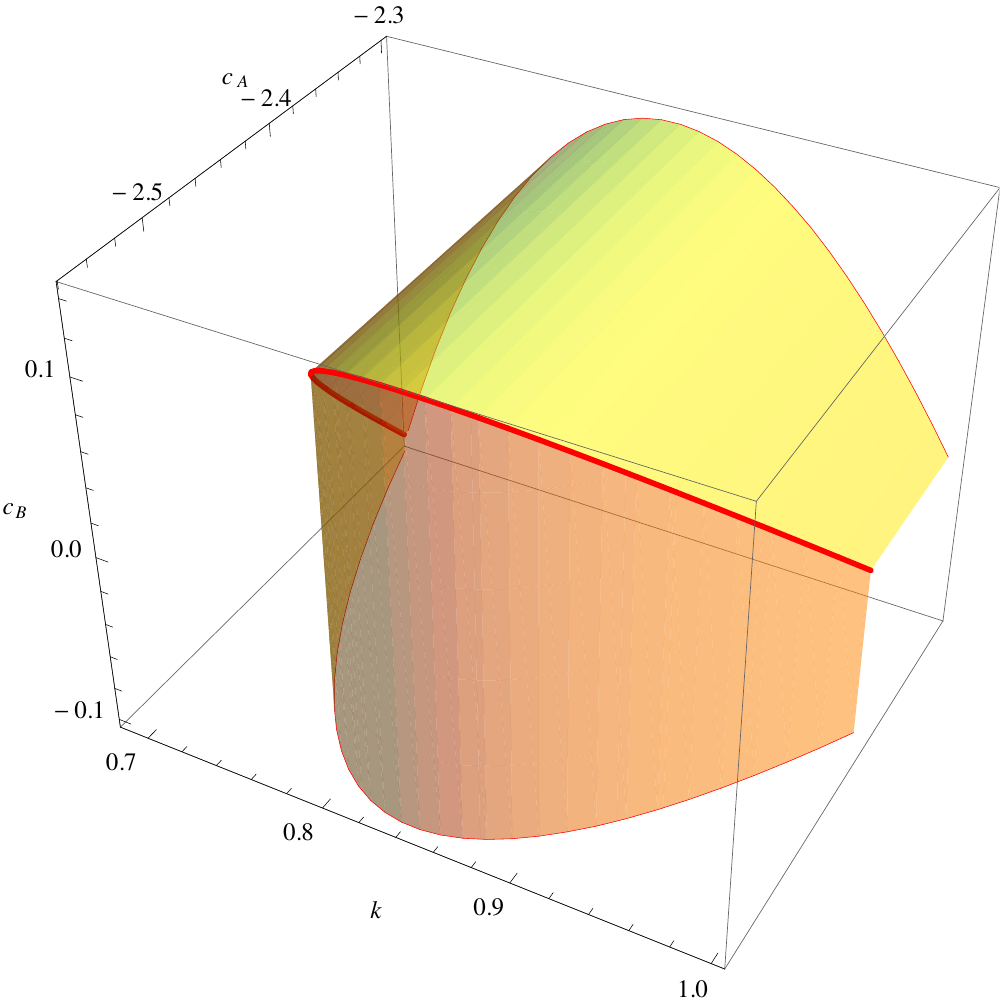}
\end{center}
\caption{Typical curve of solutions, here at constant $\mu_A=-4.83$
  and constant $j=0.64$ (thick curve) and for $B=1$. The orange/yellow
  surfaces show the projections onto the $\{k,C_A\}$ and $\{k,C_B\}$
  planes respectively. Only for vanishing magnetic does this curve lie
  completely inside the $\{k,C_A\}$ plane.\label{f:typical4d}}
\end{figure}

We start by scanning for physical solutions at $C_B=0$ and some fixed
value of $C_A$, which is a two-dimensional problem in $\{ h_0, k\}$
space and yields curves like those in
figure~\ref{f:cB0}. On these curves neither $\mu_A$ nor
$j$ will in general be constant. However, one can choose one point on
such a solution curve, and use this as a seed point $p^{\text{seed}}$
for the isocurve tracing described above. Denoting the value of
$\mu_A$ and $j$ by $\mu_A^{\text{seed}}$ and $j^{\text{seed}}$
respectively, we then trace the common zero of the three functions
$h(-\infty)$, $\mu_A - \mu_A^{\text{seed}}$ and $j-j^{\text{seed}}$ in
the four-dimensional parameter space $\{ h_0, k, C_A, C_B\}$ (only
when $B=0$ do these curves lie completely in the $\{h_0, k, C_A\}$
subspace).  A typical solution is depicted in figure~\ref{f:typical4d}. The
only remaining problem with these curves is now that in general
$j^{\text{seed}} \not=0$ (the only such points on the curves in
figure~\ref{f:cB0} are located on the endpoints of those
curves, where the coefficient $h_0$ goes to zero).

\begin{figure}[t]
\begin{center}
\includegraphics[width=.5\textwidth]{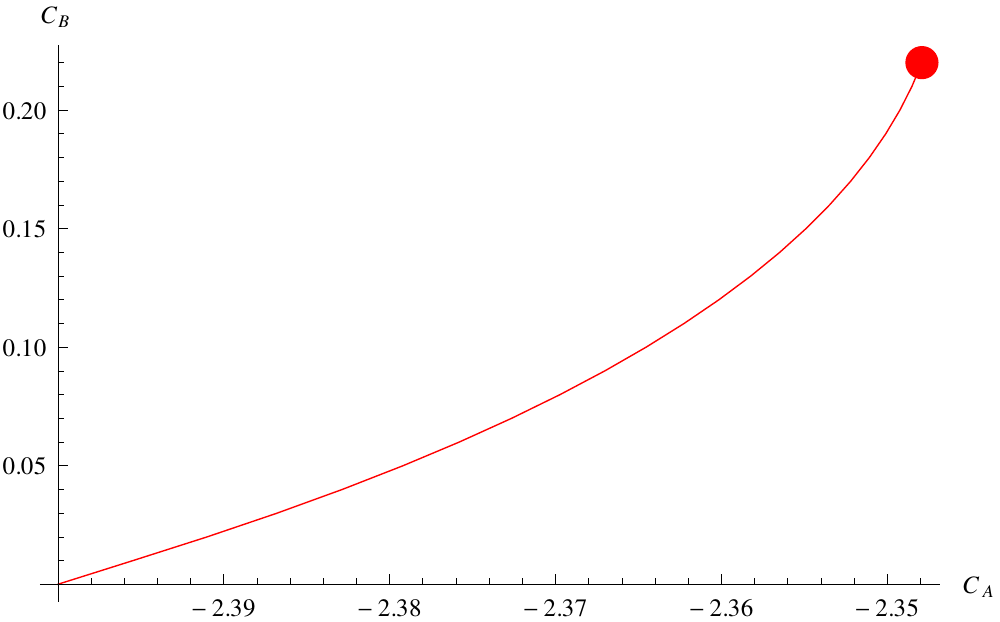}
\end{center}
\caption{Typical isocurve trace, here at the sample value of
  $\mu_A=-4.83$, starting from the seed point $p^{\text{seed}}$ with
  $C_B=0$ to the new seed point $\tilde{p}^{\text{seed}}$ (indicated
  by the dot) at which $j=0$ and both $C_A$ and $C_B$ are
  non-zero.\label{f:flowtoj0}}
\end{figure}
In order to find curves at $j=0$, we start again from
$p^{\text{seed}}$, but first do an isocurve trace at fixed $k$ in the
$\{h_0, C_A, C_B\}$ space until we find a point at which $j=0$. An
example is given in figure~\ref{f:flowtoj0}. This
point is then used as our new seed point $\tilde{p}^{\text{seed}}$ for
the four-dimensional scan. Repeating the whole process using a seed
point obtained for different initial values of $C_A$ then produces a
set of curves at $j=0$ for various constant values of $\mu_A$. 

On each of these curves we can now finally compute the Hamiltonian as
a function of $k$, and find the value $k_{\text{min}}$ for which it is
minimised. For a sample set of values of $\mu_A$ the result is
depicted in figure~\ref{f:energy_overview}.

%\bibliographystyle{kasper}
%\bibliography{kasbib}

\begin{thebibliography}{32}
\expandafter\ifx\csname natexlab\endcsname\relax\def\natexlab#1{#1}\fi

\bibitem[Son and Stephanov(2008)]{Son:2007ny}
D.~T. Son and M.~A. Stephanov, ``{Axial anomaly and magnetism of nuclear and
  quark matter}'', {\em Phys.\ Rev.} {\bfseries D77} (2008) 014021,
 \href{http://arxiv.org/abs/0710.1084}{{\ttfamily arXiv:0710.1084}}.
%%CITATION = 0710.1084;%%.

\bibitem[Kharzeev et~al.(2008)Kharzeev, McLerran, and
  Warringa]{Kharzeev:2007jp}
D.~E. Kharzeev, L.~D. McLerran, and H.~J. Warringa, ``{The Effects of
  topological charge change in heavy ion collisions: 'Event by event P and CP
  violation'}'', {\em Nucl.\ Phys.} {\bfseries A803} (2008) 227--253,
 \href{http://arxiv.org/abs/0711.0950}{{\ttfamily arXiv:0711.0950}}.
%%CITATION = ARXIV:0711.0950;%%.

\bibitem[Fukushima et~al.(2008)Fukushima, Kharzeev, and
  Warringa]{Fukushima:2008xe}
K.~Fukushima, D.~E. Kharzeev, and H.~J. Warringa, ``{The Chiral Magnetic
  Effect}'', {\em Phys.\ Rev.} {\bfseries D78} (2008) 074033,
 \href{http://arxiv.org/abs/0808.3382}{{\ttfamily arXiv:0808.3382}}.
%%CITATION = 0808.3382;%%.

\bibitem[Son and Zhitnitsky(2004)]{Son:2004tq}
D.~Son and A.~R. Zhitnitsky, ``{Quantum anomalies in dense matter}'', {\em
  Phys.\ Rev.} {\bfseries D70} (2004) 074018,
 \href{http://arxiv.org/abs/hep-ph/0405216}{{\ttfamily arXiv:hep-ph/0405216}}.
%%CITATION = HEP-PH/0405216;%%.

\bibitem[Metlitski and Zhitnitsky(2005)]{Metlitski:2005pr}
M.~A. Metlitski and A.~R. Zhitnitsky, ``{Anomalous axion interactions and
  topological currents in dense matter}'', {\em Phys.\ Rev.} {\bfseries D72}
  (2005) 045011,
 \href{http://arxiv.org/abs/hep-ph/0505072}{{\ttfamily arXiv:hep-ph/0505072}}.
%%CITATION = HEP-PH/0505072;%%.

\bibitem[Son and Surowka(2009)]{Son:2009tf}
D.~T. Son and P.~Surowka, ``Hydrodynamics with triangle anomalies'', {\em
  Phys.\ Rev.\ Lett.} {\bfseries 103} (2009) 191601,
 \href{http://arxiv.org/abs/0906.5044}{{\ttfamily arXiv:0906.5044}}.
%%CITATION = ARXIV:0906.5044;%%.

\bibitem[Kharzeev and Yee(2011)]{Kharzeev:2010gd}
D.~E. Kharzeev and H.-U. Yee, ``{Chiral Magnetic Wave}'', {\em Phys.\ Rev.}
  {\bfseries D83} (2011) 085007,
 \href{http://arxiv.org/abs/1012.6026}{{\ttfamily arXiv:1012.6026}}.
%%CITATION = ARXIV:1012.6026;%%.

\bibitem[Bergman et~al.(2009)Bergman, Lifschytz, and Lippert]{Bergman:2008qv}
O.~Bergman, G.~Lifschytz, and M.~Lippert, ``{Magnetic properties of dense
  holographic QCD}'', {\em Phys.\ Rev.} {\bfseries D79} (2009) 105024,
 \href{http://arxiv.org/abs/0806.0366}{{\ttfamily arXiv:0806.0366}}.
%%CITATION = ARXIV:0806.0366;%%.

\bibitem[Thompson and Son(2008)]{Thompson:2008qw}
E.~G. Thompson and D.~T. Son, ``{Magnetized baryonic matter in holographic
  QCD}'', {\em Phys.\ Rev.} {\bfseries D78} (2008) 066007,
 \href{http://arxiv.org/abs/0806.0367}{{\ttfamily arXiv:0806.0367}}.
%%CITATION = 0806.0367;%%.

\bibitem[Bergman et~al.(2012)Bergman, Erdmenger, and Lifschytz]{Bergman:2012na}
O.~Bergman, J.~Erdmenger, and G.~Lifschytz, ``A review of magnetic phenomena in
  probe-brane holographic matter'',
 \href{http://arxiv.org/abs/1207.5953}{{\ttfamily arXiv:1207.5953}}.
%%CITATION = ARXIV:1207.5953;%%.

\bibitem[Kim et~al.(2010)Kim, Sahoo, and Yee]{Kim:2010pu}
K.-Y. Kim, B.~Sahoo, and H.-U. Yee, ``Holographic chiral magnetic spiral'',
  {\em JHEP\,} {\bfseries 10} (2010) 005,
 \href{http://arxiv.org/abs/1007.1985}{{\ttfamily arXiv:1007.1985}}.
%%CITATION = 1007.1985;%%.

\bibitem[Hoyos et~al.(2011)Hoyos, Nishioka, and O'Bannon]{Hoyos:2011us}
C.~Hoyos, T.~Nishioka, and A.~O'Bannon, ``{A Chiral Magnetic Effect from
  AdS/CFT with Flavor}'', {\em JHEP} {\bfseries 1110} (2011) 084,
 \href{http://arxiv.org/abs/1106.4030}{{\ttfamily arXiv:1106.4030}}.
%%CITATION = ARXIV:1106.4030;%%.

\bibitem[Gynther et~al.(2011)Gynther, Landsteiner, Pena-Benitez, and
  Rebhan]{Gynther:2010ed}
A.~Gynther, K.~Landsteiner, F.~Pena-Benitez, and A.~Rebhan, ``Holographic
  anomalous conductivities and the chiral magnetic effect'', {\em JHEP\,}
  {\bfseries 1102} (2011) 110,
 \href{http://arxiv.org/abs/1005.2587}{{\ttfamily arXiv:1005.2587}}.
%%CITATION = ARXIV:1005.2587;%%.

\bibitem[Aharony et~al.(2007)Aharony, Peeters, Sonnenschein, and
  Zamaklar]{Aharony:2007uu}
O.~Aharony, K.~Peeters, J.~Sonnenschein, and M.~Zamaklar, ``{Rho meson
  condensation at finite isospin chemical potential in a holographic model for
  QCD}'', {\em JHEP\,} {\bfseries 082} (2007) 1007,
 \href{http://arxiv.org/abs/0709.3948}{{\ttfamily arXiv:0709.3948}}.
%%CITATION = ARXIV:0709.3948;%%.

\bibitem[Bayona et~al.(2011)Bayona, Peeters, and Zamaklar]{Bayona:2011ab}
C.~B. Bayona, K.~Peeters, and M.~Zamaklar, ``A non-homogeneous ground state of
  the low-temperature {Sakai-Sugimoto model}'', {\em JHEP\,} {\bfseries 1106}
  (2011) 092,
 \href{http://arxiv.org/abs/1104.2291}{{\ttfamily arXiv:1104.2291}}.
%%CITATION = ARXIV:1104.2291;%%.

\bibitem[Landsteiner et~al.(2012)Landsteiner, Megias, Melgar, and
  Pena-Benitez]{Landsteiner:2011tf}
K.~Landsteiner, E.~Megias, L.~Melgar, and F.~Pena-Benitez, ``Gravitational
  anomaly and hydrodynamics'', {\em J.Phys.Conf.Ser.} {\bfseries 343} (2012)
  012073,
 \href{http://arxiv.org/abs/1111.2823}{{\ttfamily arXiv:1111.2823}}.
%%CITATION = ARXIV:1111.2823;%%.

\bibitem[Rebhan et~al.(2010)Rebhan, Schmitt, and Stricker]{Rebhan:2009vc}
A.~Rebhan, A.~Schmitt, and S.~A. Stricker, ``{Anomalies and the chiral magnetic
  effect in the Sakai-Sugimoto model}'', {\em JHEP} {\bfseries 1001} (2010)
  026,
 \href{http://arxiv.org/abs/0909.4782}{{\ttfamily arXiv:0909.4782}}.
%%CITATION = ARXIV:0909.4782;%%.

\bibitem[Yamamoto(2012)]{Yamamoto:2012bi}
A.~Yamamoto, ``{Chiral Magnetic Effect} on the lattice'',
 \href{http://arxiv.org/abs/1207.0375}{{\ttfamily arXiv:1207.0375}}.
%%CITATION = ARXIV:1207.0375;%%.

\bibitem[Fukushima and Mameda(2012)]{Fukushima:2012fg}
K.~Fukushima and K.~Mameda, ``{Wess-Zumino-Witten action and photons from the
  Chiral Magnetic Effect}'',
 \href{http://arxiv.org/abs/1206.3128}{{\ttfamily arXiv:1206.3128}}.
%%CITATION = ARXIV:1206.3128;%%.

\bibitem[Gorsky et~al.(2011)Gorsky, Kopnin, and Zayakin]{Gorsky:2010xu}
A.~Gorsky, P.~Kopnin, and A.~Zayakin, ``On the {Chiral Magnetic Effect} in
  soft-wall {AdS/QCD}'', {\em Phys.\ Rev.} {\bfseries D83} (2011) 014023,
 \href{http://arxiv.org/abs/1003.2293}{{\ttfamily arXiv:1003.2293}}.
%%CITATION = ARXIV:1003.2293;%%.

\bibitem[Rubakov(2010)]{Rubakov:2010qi}
V.~Rubakov, ``{On chiral magnetic effect and holography}'',
 \href{http://arxiv.org/abs/1005.1888}{{\ttfamily arXiv:1005.1888}}.
%%CITATION = ARXIV:1005.1888;%%.

\bibitem[Sakai and Sugimoto(2005)]{Sakai:2004cn}
T.~Sakai and S.~Sugimoto, ``Low energy hadron physics in holographic {QCD}'',
  {\em Prog.\ Theor.\ Phys.} {\bfseries 113} (2005) 843--882,
 \href{http://arxiv.org/abs/hep-th/0412141}{{\ttfamily hep-th/0412141}}.
%%CITATION = HEP-TH 0412141;%%.

\bibitem[Sakai and Sugimoto(2006)]{Sakai:2005yt}
T.~Sakai and S.~Sugimoto, ``More on a holographic dual of {QCD}'', {\em Prog.\
  Theor.\ Phys.} {\bfseries 114} (2006) 1083--1118,
 \href{http://arxiv.org/abs/hep-th/0507073}{{\ttfamily hep-th/0507073}}.
%%CITATION = HEP-TH/0507073;%%.

\bibitem[Rebhan et~al.(2010)Rebhan, Schmitt, and Stricker]{Rebhan:2010ax}
A.~Rebhan, A.~Schmitt, and S.~Stricker, ``Holographic chiral currents in a
  magnetic field'', {\em Prog.Theor.Phys.Suppl.} {\bfseries 186} (2010)
  463--470,
 \href{http://arxiv.org/abs/1007.2494}{{\ttfamily arXiv:1007.2494}}.
%%CITATION = ARXIV:1007.2494;%%.

\bibitem[Bardeen(1969)]{Bardeen:1969md}
W.~A. Bardeen, ``{Anomalous Ward identities in spinor field theories}'', {\em
  Phys.\ Rev.} {\bfseries 184} (1969)
1848--1857.
%%CITATION = PHRVA,184,1848;%%.

\bibitem[Peskin and Schroeder(1995)]{Peskin:1995a}
M.~Peskin and D.~Schroeder, ``An introduction to quantum field theory'',
  Perseus, 1995.

\bibitem[Ooguri and Park(2011)]{Ooguri:2010xs}
H.~Ooguri and C.-S. Park, ``Spatially modulated phase in holographic
  quark-gluon plasma'', {\em Phys.\ Rev.\ Lett.} {\bfseries 106} (2011) 061601,
 \href{http://arxiv.org/abs/1011.4144}{{\ttfamily arXiv:1011.4144}}.
%%CITATION = 1011.4144;%%.

\bibitem[Basar et~al.(2010)Basar, Dunne, and Kharzeev]{Basar:2010zd}
G.~Basar, G.~V. Dunne, and D.~E. Kharzeev, ``{Chiral Magnetic Spiral}'', {\em
  Phys.Rev.Lett.} {\bfseries 104} (2010) 232301,
 \href{http://arxiv.org/abs/1003.3464}{{\ttfamily arXiv:1003.3464}}.
%%CITATION = ARXIV:1003.3464;%%.

\bibitem[Ahnert and Mulansky(2011)]{Ahnert:2011a}
K.~Ahnert and M.~Mulansky, ``Odeint - solving ordinary differential equations
  in {C++}'',  \href{http://arxiv.org/abs/1110.3397}{{\ttfamily
  arXiv:1110.3397}}, \url{http://www.odeint.com}.

\bibitem[et~al.(2009)]{Galassi:2009a}
M.~G. et~al., ``{GNU Scientific Library Reference Manual}, 3rd edition'', 2009.

\bibitem[Basar and Dunne(2012)]{Basar:2012gm}
G.~Basar and G.~V. Dunne, ``The {Chiral Magnetic Effect} and axial anomalies'',
 \href{http://arxiv.org/abs/1207.4199}{{\ttfamily arXiv:1207.4199}}.
%%CITATION = ARXIV:1207.4199;%%.

\bibitem[Deryagin et~al.(1992)Deryagin, Grigoriev, and
  Rubakov]{Deryagin:1992rw}
D.~V. Deryagin, D.~Y. Grigoriev, and V.~A. Rubakov, ``Standing wave ground
  state in high density, zero temperature {QCD} at {large-$N_c$}'', {\em Int.\
  J.\ Mod.\ Phys.} {\bfseries A7} (1992)
659--681.
%%CITATION = IMPAE,A7,659;%%.

\end{thebibliography}

\begingroup\raggedright\endgroup

\end{document}